%% file: main.tex
\documentclass[prb, floatfix, english amsmath,amssymb,aps,twocolumn,superscriptaddress, longbibliography, reprint]{revtex4-2}

\usepackage{graphicx}
\usepackage{dcolumn}
\usepackage{bm}
\usepackage{hyperref}
\usepackage[dvipsnames, svgnames]{xcolor}
\usepackage{amsmath}
\usepackage{amssymb}
\usepackage{booktabs}
\usepackage{physics}
\usepackage{tikz}
\usepackage{quantikz}
\usepackage{comment}
\usepackage{makecell, array}
\usepackage{helvet}  
\hypersetup{colorlinks=true, citecolor=blue, urlcolor=blue,
linkcolor=blue, breaklinks=true}

\newcommand{\ZZ}{\mathbb{Z}}
\newcommand{\bigO}[1]{\mathcal{O}\!\left( #1 \right)}
\newcommand{\qcode}[1]{[\![#1]\!]}

\newcommand{\fullSpaceReduction}{$\sim\!5.5\times$}
\newcommand{\headlineQubits}{$2,240$}
\newcommand{\headlineTime}{$\sim\!200$}
\newcommand{\headlineQubitsFH}{$\sim\!6{,}300$}
\newcommand{\headlineTimeFH}{$\sim\!200$}

\begin{document}

\title{Fast and Parallel High-Rate STAR Architecture for Megaquop Quantum Simulation}
\author{Refaat Ismail}
\email{rismail@quera.com}
\affiliation{QuEra Computing Inc., 1380 Soldiers Field Road, Boston, MA 02135, USA}
\author{Milan Kornja\v{c}a}
\email{mkornjaca@quera.com}
\affiliation{QuEra Computing Inc., 1380 Soldiers Field Road, Boston, MA 02135, USA}
\author{Hong-Ye Hu}
\affiliation{Department of Physics, Harvard University, Cambridge, MA 02138, USA}
\author{Nishad Maskara}
\affiliation{QuEra Computing Inc., 1380 Soldiers Field Road, Boston, MA 02135, USA}
\affiliation{Center for Theoretical Physics - a Leinweber Institute, Massachusetts Institute of Technology, Cambridge, MA 02139, USA}
\author{Sheng-Tao Wang}
\affiliation{QuEra Computing Inc., 1380 Soldiers Field Road, Boston, MA 02135, USA}
\author{Hengyun Zhou}
\affiliation{QuEra Computing Inc., 1380 Soldiers Field Road, Boston, MA 02135, USA}
\affiliation{Department of Electrical Engineering and Computer Science, Massachusetts Institute of Technology, Cambridge, MA 02138, USA}
\author{Chen Zhao}
\email{czhao@quera.com}
\affiliation{QuEra Computing Inc., 1380 Soldiers Field Road, Boston, MA 02135, USA}
\date{\today}
\begin{abstract}
  Fault-tolerant quantum simulation is approaching a development phase where encoding overhead, logical Clifford operations, magic-state preparation, and rotation synthesis must be optimized together for efficient implementation. Space-Time efficient Analog Rotation (STAR) architectures reduce two of these costs by preparing small-angle rotation magic states directly, and the transversal STAR variant further lowers the Clifford overhead. Existing concrete implementations, however, largely inherit the low $O(1/d^2)$ encoding rate of the surface code, while high-rate codes have not yet been integrated into comparably explicit computational architectures. Here, we introduce a high-rate STAR architecture for local lattice Hamiltonian simulation based on a symmetry-driven co-design of the algorithm, QEC code, and neutral-atom hardware. Translation symmetries of the target lattice determine the choice of bicycle chain codes, a tunable family of self-dual bivariate bicycle codes that natively implement Clifford gates required for lattice simulation. Disjoint logical representatives allow STAR injections to be performed in parallel on all $k$ logical qubits in a code block, amortizing resource state preparation and enabling practical post-selection rates. On neutral-atom platform, the same translation symmetry compiles the key logical operations into low-depth, hardware-native acousto-optic-deflector shifts. End-to-end estimates show that an $8\times 8$ transverse-field Ising simulation to $T^*\approx 8\,(zJ)^{-1}$ requires \headlineQubits{} physical qubits and \headlineTime{}\,s per shot, a \fullSpaceReduction{} space reduction relative to a surface code STAR baseline at comparable speed; for Fermi--Hubbard dynamics to $T^*\approx 4\,(zt)^{-1}$, the corresponding estimates are \headlineQubitsFH{} physical qubits and \headlineTimeFH{}\,s per shot. These results provide a concrete route toward early fault-tolerant quantum simulation with high-rate codes.
\end{abstract}
\maketitle

\section{Introduction}
\label{sec:intro}

Quantum computers are widely believed to offer computational advantages beyond
classical reach for important classes of problems~\cite{feynman1982simulating,lloyd1996universal}.
Realizing this potential, however, requires quantum error correction (QEC)
to protect logical information from hardware noise, and the overhead imposed
by QEC remains a central obstacle to practical fault-tolerant quantum
computing (FTQC).

The conventional approach to FTQC is code-driven: one first selects a
general-purpose error-correcting code, then engineers fault-tolerant
implementations of a universal gate set---typically Clifford gates supplemented
by a non-Clifford gate such as the $T$ gate~\cite{bravyi2005universalquantum}. This
pipeline is powerful because of its generality: by composing error-corrected
Clifford operations with magic states produced through cultivation~\cite{gidney2024magicstate,vaknin2025magicstate,chen2025efficientmagic,sahay2025foldtransversalsurface}
or distillation~\cite{bravyi2005universalquantum,knill2004faulttolerant,bravyi2012magicstate},
it can in principle execute any quantum algorithm to arbitrarily low logical
error rate. In practice, however, it incurs three compounding layers of overhead.
First, the \emph{encoding overhead}: the celebrated surface code architectures
encode a single logical qubit in $d^2$ physical qubits~\cite{fowler2012surfacecodes,zhou2025resourceanalysis}, where $d$ is the code distance ranging from around $10$ to $30$.
Second, the \emph{gate overhead}: implementing logical gates
fault-tolerantly within the chosen code introduces additional space-time cost.
Third, the \emph{synthesis overhead}: algorithms whose gate structure does not
naturally decompose into Clifford$+T$ must pay an additional compilation
overhead. Trotterized Hamiltonian simulation, for instance, is dominated by
small-angle $R_Z$ rotations ($\theta \sim 10^{-2}$--$10^{-3}$\,rad), each of which may
require tens of $T$ gates when synthesized into Clifford$+T$ operations~\cite{selinger2014efficient,kliuchnikov2016practical,ross2016optimal}.
For near-term devices with limited resources, these combined overheads
render the general-purpose pipeline prohibitively expensive.

Yet not every algorithm demands the full generality of the code-driven approach.
Quantum simulation of correlated many-body systems is widely regarded as one of
the first important applications for quantum computers, both because quantum
mechanical correlations are essential to understanding many-body physics and
materials, and because the required precision is modest: megaquop Trotterized
circuits are expected to surpass classical accuracy at logical error rates of order
$10^{-6}$ per gate~\cite{akahoshi2024partiallyfaulttolerant,campbell2022earlyfaulttolerant,ismail2025transversalstar}, far above the $10^{-9}$--$10^{-15}$ targets of, e.g.,
arithmetic and quantum chemistry~\cite{gidney2021howfactor,gidney2025howfactor,zhou2025resourceanalysis,cain2026shors,webster2026pinnaclearchitecture,babbush2026securing,litinski2023howcompute,beverland2022assessingrequirements}.
This observation has motivated the search for early fault-tolerant strategies
that trade generality for efficiency by exploiting algorithmic structures, such as the STAR (Space-Time efficient Analog Rotation) architecture~\cite{akahoshi2024partiallyfaulttolerant, toshio2024practicalquantum, ismail2025transversalstar}, the phantom codes~\cite{koh2026entangling}, or the
Reaction-time-limited Architecture for Space-time-efficient
Complex qLDPC Logic (RASCqL)~\cite{yang2026spacetimeefficienthardwarecompatiblecomplexquantum}.

The main advantage of the STAR architecture is that, instead of synthesizing each small-angle rotation
from many $T$ gates, it prepares analog magic states $\ket{\theta} \propto \ket{0} +
e^{i\theta}\ket{1}$ directly within the code block using a post-selected protocol,
and then teleports these states into the computation to implement the corresponding rotation gates.
The resulting $R_Z$ logical error rate scales as $\sim\!p \cdot |\theta|$,
so that for physical error rates $p \sim 10^{-3}$ and angles
$\theta \sim 10^{-3}$, this yields a logical error rate of order $10^{-6}$
per gate---sufficient for the megaquop regime with drastically reduced
synthesis overhead.
The transversal STAR extension~\cite{ismail2025transversalstar} goes
further, replacing lattice surgery for the Clifford backbone with
transversal gates on surface codes, thereby reducing the Clifford spacetime overhead
from $\bigO{d^3}$ to $\bigO{d^2}$~\cite{zhou2025lowoverheadtransversal}. Together, these advances reduce the physical qubit
requirements of megaquop simulation to $\sim\!10^4$ and bring time-cost down $\gtrsim 100\times $ over fixed connectivity $T$-based architectures.

These advances exploit the gate-level structure of Hamiltonian
simulation, i.e., the abundance of small-angle rotations. However, the surface code's low encoding rate,
$\bigO{1/d^2}$,  still leads to substantial physical-qubit overhead.
High-rate quantum low-density parity-check (qLDPC) codes could provide encoding rates orders of magnitude higher~\cite{breuckmann2021quantumlowdensity,tillich2009quantumldpc,panteleev2022asymptoticallygood,leverrier2022quantumtanner,bravyi2024highthresholdlowoverhead,kasai2026breakingorthogonality,zhao2026ultrahighrate},
but developing a computationally efficient high-rate code architecture remains a major open challenge in the field.
Extending STAR to qLDPC codes requires further exploiting the global structure of the target problem
for efficient computation, thus raising two key challenges: First, it is necessary for the native logical gates of the high-rate code family
to match the structure of the target algorithm.
Second, naive sequential magic-state injection, in an $\qcode{n,k,d}$ code block, will keep the remaining logical qubits in the same block idling, which could create a significant bottleneck in the computational speed; in addition, sequential injections would result in an exponentially decaying
post-selection rate in the number of logical qubits $k$, posing significant scalability challenges.

As described in Ref.~\cite{ismail2025transversalstar}, translation symmetry of a local lattice Hamiltonian reduces the required native
logical gate set to global $H$, global $S$, and logical automorphisms on a
cyclic chain. Ref.~\cite{ismail2025transversalstar} further proposed cyclic
hypergraph product (HGP)~\cite{aydin2025cyclichypergraph} codes and self-dual bivariate bicycle (BB)~\cite{xu2025batchedhighrate,liang2025selfdualbivariate} codes as possible candidates. However, that work did not identify a concrete code instance, provide factory-level performance estimates, or provide a route to efficient, fully parallel injection. Separately, Ref.~\cite{xu2025batchedhighrate}
proposed a fully fault-tolerant architecture for local lattice Hamiltonian simulation
using a family of self-dual BB codes with a disjoint
logical basis. However, the proposal envisions either lattice surgery or distilled Clifford resource states to couple left and right logical qubit sets, which substantially slows down the computation. More fundamentally, it also relies on operationally complex magic-state cultivation and escape-adaptation protocols with significant synthesis and space-time overhead.

\begin{figure*}[t]
  \centering
  \includegraphics[width=\textwidth]{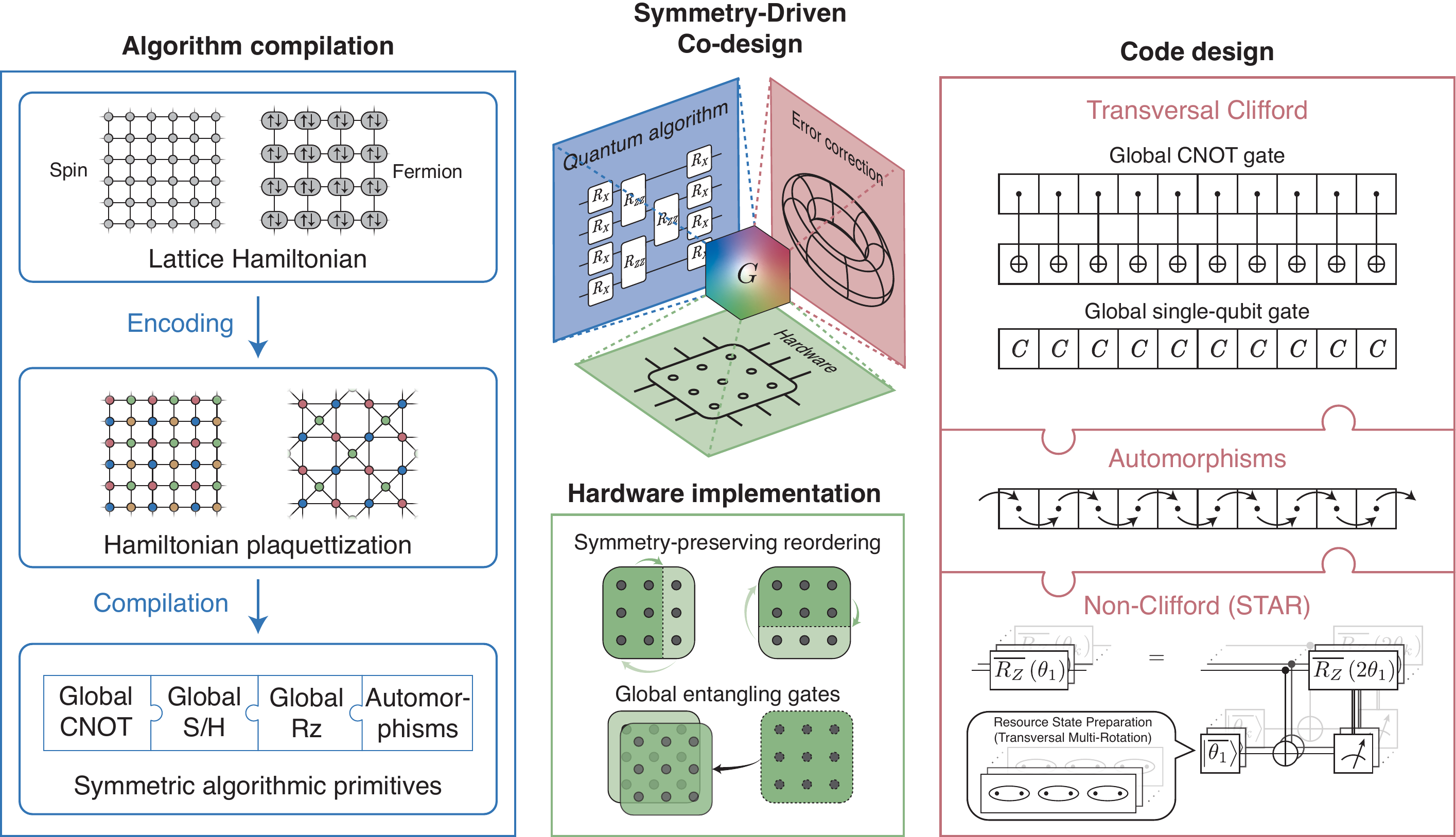}
  \caption{
    \textbf{A high-rate STAR architecture for local lattice Hamiltonian simulation,
    following the symmetry-driven co-design principle.}
    (Left) the translation symmetry $\ZZ_{\ell}$ of the target
    local lattice Hamiltonian determines the required algorithmic gadgets: the Hamiltonian can be grouped into sublattices (plaquettization) such that the interactions of the Hamiltonian are parallel between sublattices, resulting in global algorithmic primitives that preserve the translation symmetry. The Hamiltonian simulation could thus be compiled into purely plaquette parallel (global) Cliffords, global magic gates, and automorphisms that realize the primitive shifts of the plaquette lattice. While the simulation of local lattice spin Hamiltonians is direct, fermionic local lattice Hamiltonians are converted to the spin case via local compact encodings, such as Derby-Klassen~\cite{derby2021compact}, at the cost of mediating ancillae.
    (Right) the bicycle chain code realizes all the neccesary algorithmic primitives using native global transversal Clifford gates, $\ZZ_{\ell}$ shift automorphisms, and STAR injections. STAR injections can be performed in a fully parallel fashion due to the existence of non-overlapping logical basis in the bicycle chain codes, thus amortizing state preparation cost.
    (Middle) the translation symmetry translated into efficient parallel
    AOD moves -- cycles on a 2D-grid layout of qubits -- making all the error-corrected components of the architecture readily implementable with neutral atoms.
  }
  \label{fig:codesign}
\end{figure*}

In this work, we close the above gaps by proposing and verifying in detailed simulations a high-rate, parallel, and transversal STAR architecture
that threads the translation symmetry through
every layer of the stack---algorithmic gadgets, error correction codes,
and efficient hardware implementations (Fig.~\ref{fig:codesign}) ~\cite{bluvstein2022quantumprocessora,bluvstein2023logicalquantum,salesrodriguez2025experimentaldemonstration}, resulting in a maximally parallel algorithm execution.
The circuit is compiled into purely global, parallel Cliffords and rotation gates. We identify a
natural family of codes, \emph{bicycle chain codes}, whose $d$ and
$k$ can be tuned independently, while keeping all encoded logical qubits disjoint. This generalizes the
construction of Refs.~\cite{xu2025batchedhighrate,liang2025selfdualbivariate} to support global native gates
at arbitrary lattice sizes. Crucially, small-angle resource states can be
prepared fully in parallel across all $k$ disjoint logical qubits within a code block, reducing the factory circuit depth and enabling practical post-selection rates. The expected circuit depth for
repeat-until-success (RUS) teleportation scales as $\log_2 k$, yielding an average cost of
$\log_2(k)/k$ injections per magic rotation; this is in constrast to the cost of
$2$ injections per rotation for surface code STAR, further utilizing the encoding capacity of the high-rate codes.
Finally, on neutral-atom hardware, all key operations
correspond to hardware-native, highly-parallel acousto-optic deflector (AOD) cyclic-shifting moves.
The codesigned architecture simultaneously addresses all three overhead layers, converting them to simple, parallel operations:
encoding (via high-rate bicycle chain codes with cyclic-shift syndrome extraction), fault-tolerant gates (via
parallel transversal gates and shift automorphism), and small-angle magic synthesis (via direct and fully-parallel analog injection).

With end-to-end resource estimates supported by detailed circuit-level simulations with noise,
we find that this architecture yields a \fullSpaceReduction{} reduction in
physical qubit overhead relative to a surface code transversal STAR
baseline~\cite{ismail2025transversalstar} for an $8\times 8$ transverse-field Ising model.
In concrete terms, \headlineQubits{} physical qubits and \headlineTime{} seconds
of per-shot runtime suffice for megaquop-scale quantum simulation, providing a
low-resource pathway into the early fault-tolerant era. The framework extends
naturally to arbitrary local spin Hamiltonians and  fermionic systems via geometrically local fermion-to-qubit
encodings~\cite{derby2021compact,jafarizadeh2024recipelocal}. We exemplify this by showing that $8\times 8$ single-band Fermi-Hubbard model megaquop-scale quantum simulation could be performed with \headlineQubitsFH{} physical qubits and \headlineTimeFH{} seconds per shot.

The remainder of this paper is organized as follows.
Section~\ref{sec:codesign} presents the codesigned architecture for
local lattice Hamiltonian simulation: the algorithmic gadgets demanded by
translation symmetry, the bicycle chain codes that natively realize them,
and the efficient neutral-atom hardware implementation.
Section~\ref{sec:results} provides detailed logical error models for
each gadget, including transversal Clifford gates and the transversal multi-rotation (TMR) protocol for
STAR injections, and subsequently combines them into end-to-end resource estimates
benchmarked against alternative approaches including the surface code STAR baseline.
Section~\ref{sec:discussion} discusses limitations and future directions.

\section{Architecture Co-Design for local lattice Hamiltonian Simulation}
\label{sec:codesign}

In this section, we present the high-rate bicycle chain STAR architecture for translation-invariant local lattice Hamiltonian simulation. We first reduce the
Trotterized evolution to a small set of global algorithmic gadgets, then
construct the bicycle chain codes whose native operations realize these
gadgets, and finally describe how the resulting operations can be efficiently implemented with
neutral-atom AOD moves.

As a representative target, consider the following lattice spin Hamiltonian on an
$L\times L$ square lattice with periodic boundary conditions,
\begin{equation}
  H = \sum_{\langle i,j\rangle}
  \bigl( J_x X_i X_j + J_y Y_i Y_j + J_z Z_i Z_j \bigr)
  + \sum_i h_i X_i,
  \label{eq:Hamiltonian}
\end{equation}
where $L$ is even and the first sum is over nearest-neighbor bonds.
This family includes the transverse-field Ising model ($J_x = J_y = 0$),
the $XXZ$ model ($J_x = J_y \neq J_z$), and the isotropic Heisenberg model
($J_x = J_y = J_z$) as special cases.
Note that this architecture applies to broader families of geometrically local Hamiltonians. For example, Fermi-Hubbard Hamiltonians can also be reduced to a local spin Hamiltonian on a modified lattice, using geometrically local fermion-to-qubit encodings as described in Appendix~\ref{app:fermion-new}. Here, we specifically consider the implementation of a single-band, spin-full Fermi-Hubbard model on the square lattice:
\begin{equation}
  H_\mathrm{FH} = -t\!\!\sum_{\langle i,j\rangle,\sigma}
  \bigl(c^\dagger_{i\sigma} c_{j\sigma} + \mathrm{h.c.}\bigr)
  + U\sum_i n_{i\uparrow} n_{i\downarrow},
  \label{eq:FH_main}
\end{equation}
where $t$ denotes nearest-neighbor fermion hopping and $U$ on-site Coulomb repulsion.

\subsection{Algorithmic gadgets}
\label{sec:gadgets}

\begin{figure}[t]
  \centering
  \includegraphics[width=\columnwidth]{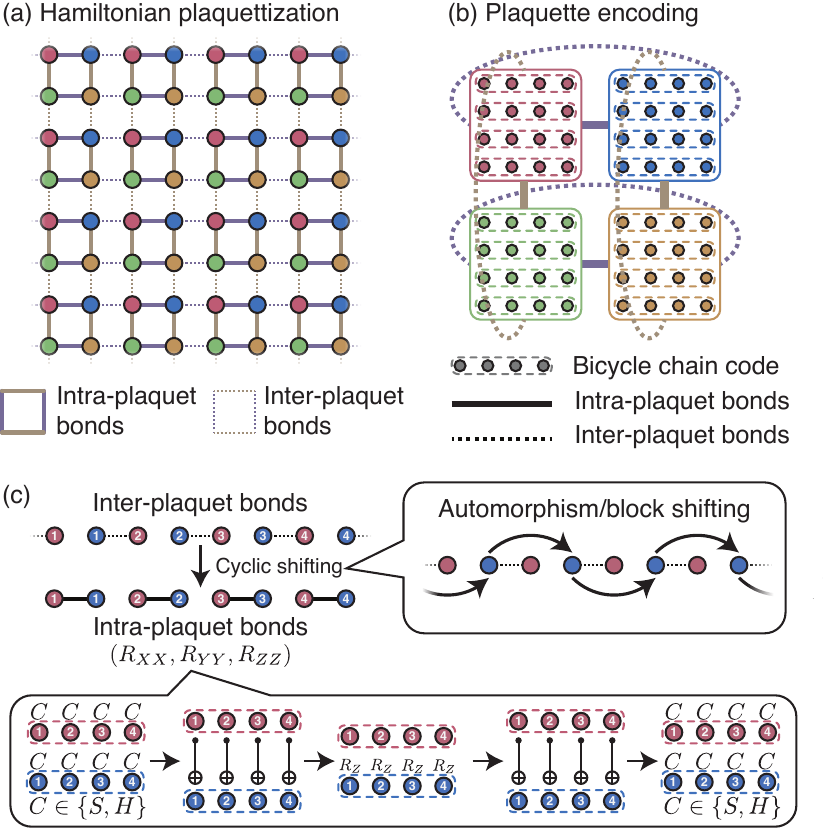}
  \caption{
    \textbf{Compilation of local lattice Hamiltonian simulation.}
    (a) A local lattice Hamiltonian is grouped into sublattices via plaquettization, then encoded using multiple code blocks of (b) bicycle chain codes that support native transversal CNOT, $H$, $S$, cyclic shift automorphisms, and disjoint logical qubits for parallel STAR injection. (c) Inter- and intra-plaquette interactions that support the Trotterized simulation can be implemented using native global gates.
  }
  \label{fig:compilation}
\end{figure}

Partitioning the lattice by coordinate parity (plaquettization) splits it into four disjoint
$L/2 \times L/2$ sublattices~\cite{ismail2025transversalstar}, as illustrated
in Fig.~\ref{fig:compilation}(a). The nearest-neighbor terms then form a global
interaction layer between two paired sublattices with inter- or
intra-plaquette bonds [Fig.~\ref{fig:compilation}(b)].
One Trotter step can therefore be organized into global $XX$, $YY$, and
$ZZ$ rotation layers, plus a layer of single-site $X$ rotations.
The $XX$ and $YY$ layers are reduced to $ZZ$ layers by conjugation with
global single-qubit Clifford gates ($H$ and $S$).
The inter-plaquette $ZZ$ rotations can be converted into intra-plaquette rotations via cyclic
shifts in two directions of the sublattice. The global intra $ZZ$ rotations
can then be implemented using global CNOTs and a parallel layer of $R_Z(\theta)$
rotations [Fig.~\ref{fig:compilation}(c)]:
\begin{equation}
  e^{-i\frac{\theta}{2} Z_A Z_B} = \mathrm{CNOT}_{AB}\; [I_A \otimes R_{Z_B}(\theta)]
  \;\mathrm{CNOT}_{AB}.
  \label{eq:zz_decomp}
\end{equation}
Thus, the gate content of the Trotterized circuit reduces to global CNOT, global
$H$ and $S$, row/column shift automorphisms, and parallel $R_Z$ operations. The required global Clifford structure is thus determined only by the Hamiltonian connectivity, not the interaction strengths.

We choose to encode each row of a sublattice into a different code block. This
further reduces the requirements of the automorphisms to cyclic shifts on a 1D
chain because the shifts in the other direction correspond to code block
shifting. This not only simplifies the code design complexity, but it also produces
more manageable resource state preparation post-selection rates. Therefore, the required native logical gadgets are:
\begin{enumerate}
  \item global CNOT;
  \item global $S$ and $H$;
  \item parallel $R_Z$;
  \item cyclic shifts on a 1D chain.
\end{enumerate}

\subsection{Bicycle chain codes}
\label{sec:bicycle_chain_code}

To search for a code family that natively provides the gadgets identified
above, we first explore general code properties that support those gadgets. Global CNOT is transversal between two code blocks in any Calderbank-Shor-Steane (CSS)
code~\cite{gottesman1997stabilizercodes}. A self-dual code, whose $X$ and $Z$ parity-check matrices are identical, would support transversal $H$~\cite{calderbank1997quantumerror}; and if all stabilizers have doubly-even weights, the code would further provide transversal $S$~\cite{calderbank1997quantumerror,zeng2007transversalityuniversality}. For parallel $R_Z$, the STAR
protocol prepares logical rotation states and teleports them via a
RUS gadget~\cite{akahoshi2024partiallyfaulttolerant,toshio2024practicalquantum,ismail2025transversalstar}; but
a \emph{complete set of disjoint logical} is required to support
parallel preparation and teleportation~\cite{xu2025batchedhighrate,gu2026qgpu}.
An odd-weight disjoint basis would also make the transversal $H$ and $S$ action
uniform across the basis. Finally, cyclic shifts on a 1D chain require code automorphisms that cyclically permute the logical qubits while preserving the
stabilizers, which suggests translation-invariant codes~\cite{haah2017algebraicmethods} as potential candidates such as BB codes or generalized bicycle codes~\cite{lin2023quantumtwoblock,webster2026pinnaclearchitecture}.

\begin{figure}[t]
  \centering
  \includegraphics[width=\columnwidth]{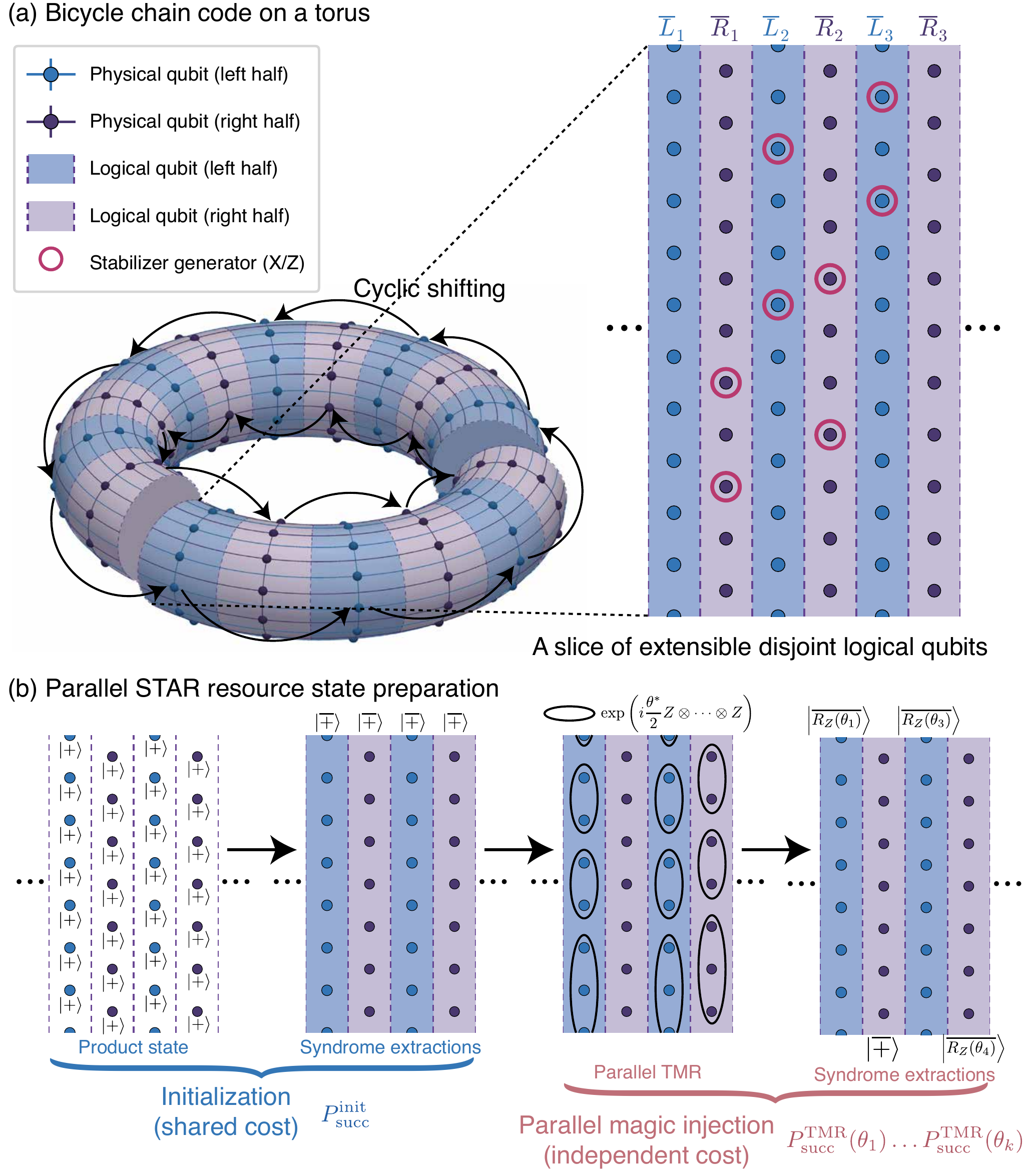}
  \caption{
    \textbf{Bicycle chain codes and parallel STAR magic generation.}
    (a) A bicycle chain code with parameters $[[n=2l\times m, k=2l, d \lesssim m]]$ is defined on an $l \times m$ torus. The torus circumference ($m$) limits the code distance, and each of the $2l$ slices of the torus hosts a disjoint logical qubit representative ($\bar{L}_i, \bar{R}_i$). The logical qubits are chained together via the weight-8 stabilizers -- the support of a single stabilizer generator is shown (top right), with the rest of the generator set related to it by torus translations. Finally, translations of the torus by two slices along its length generate $Z_l$ cyclic shift automorphism group action on the logical basis. (b) STAR injection can be executed in parallel on arbitrary subsets of logical qubits, due to the disjoint logical basis, thus amortizing the postselection cost of the injection. Each STAR injection proceeds by preparing an initial $\ket{+}^{\otimes n}$ state, and initializing the code by measuring $Z$ syndrome (half a syndrome round). After performing transversal multi-qubit rotation with $M$ disjoint multi-qubit rotations per logical qubit representative, a full round of syndrome measurement is performed, projecting the code to the logical subspace and implementing the desired logical rotation by post-selecting on the inconsistent stabilizers.
  }
  \label{fig:bicycle-chain-code}
\end{figure}

We realize all of these requirements inside the self-dual BB code family~\cite{bravyi2024highthresholdlowoverhead,xu2025batchedhighrate,liang2025selfdualbivariate} as bicycle chain codes (Appendix~\ref{app:bicycle-chain-codes}).
We note that not all self-dual BB codes discovered in Refs.~\cite{xu2025batchedhighrate,liang2025selfdualbivariate}
satisfy all the above properties, because supporting a disjoint logical basis and
the cyclic shifting on a chain are not the primary goals of those
constructions. Each bicycle chain code is defined on two $\ell \times m$
lattices with periodic boundary conditions (a regular torus), referred to as the left and right
data-qubit halves. The resulting code parameters are $\qcode{n=2l\times m, k=2l, d \lesssim m}$ -- with the torus circumference ($m$) limiting the code distance, and torus length ($l$) determining the number of encoded qubits. For the code instances used here, a complete set of disjoint
logical representatives can be simply chosen as odd-weight length-$\ell$ columns; these
columns are depicted as torus slices in Fig.~\ref{fig:bicycle-chain-code}(a).
The CSS structure, self-duality, and double-evenness directly realize the global transversal
CNOT, $H$, and $S$ gates. In addition, the fully disjoint logical basis and
the translation-invariant stabilizer pattern provide logical 1D cyclic shift automorphisms, separately
on the left and right halves of logical representatives. Furthermore, the chain length can be
increased by lengthening the torus (increasing $l$), supporting arbitrary chain
lengths for the left/right logical blocks, and thus arbitrary even $k$, enabling the code size to be adapted to the size of the simulation problem without affecting encoding rate.
Finally, enabled by the disjoint logical basis, parallel STAR magic can be introduced selectively on arbitrary subsets of
logical qubits with subset-parallel $R_Z$ injection and parallel RUS teleportation
[Fig.~\ref{fig:bicycle-chain-code}(b)].

The teleportation attempt on each
logical qubit succeeds with probability $1/2$, so resource state preparation is
repeated selectively on the failed ones until all rotations succeed. The
expected RUS attempts for injecting $k$ logical resource states are
$\sim\!\log_2(k)$ (Appendix~\ref{sec:magic}). Therefore, the per-logical RUS
attempts can be better than the surface code STAR RUS cost, which is $\sim\!2$ attempts on average.
To fully unlock this advantage, the post-selection rate per RUS should be
comparable to the surface code. The post-selection rate of parallel STAR
injections can be
decomposed into two parts --- the initialization cost for initializing a bicycle
chain to all $\ket{\overline{+}}$ states in the code block, and the injection cost per logical
TMR. The initialization cost is shared across all $k$ logical
qubits and the post-selection rate is $\sim\!(1-p)^{\mathcal{O}(n)}$. As verified by numerical simulations in the later section, in the practical regime where $\ell\leq 16$ ($n\leq 224$), we show that the bicycle chain code block with multiple logical qubits incurs comparable initialization cost as the surface code case with a single logical qubit.
Therefore, the average initialization cost per-logical can be lower in practice.
Secondly, the TMR cost is mostly characterized by the overlap between
the raw TMR output and the ideal logical state. The per-logical TMR cost is
independent across different logical qubits, so the per-logical injection cost
should be similar
to that of the surface code.

The main code instances we use as examples in this manuscript are $\qcode{14l, 2l, 6}$, summarized in Table~\ref{tab:bicycle-chain-instances}.
Note that the disjoint logical basis gives an encoding rate upper bound $1/d$, and instances at $d=6$ are almost optimal.
Moreover, based on our performance evaluation in Sec.~\ref{sec:results}, $d=6$ is enough for megaquop simulation.
Taking $\ell=3$, this code instance is identical to the $\qcode{42,6,6}$ code in Table~I of Ref.~\cite{xu2025batchedhighrate}.

\begin{table}[htbp]
  \centering
  \begin{ruledtabular}
    \begin{tabular}{cccccc}
      $a(x,y)$  & $m$ & $n$ & $k$ & $d$ & $k/n$ \\
      \midrule
      $1 + y^3 + xy^2 + xy^4$ & 7 & $14\ell$ & $2\ell$ & 6 & $1/7$ \\
    \end{tabular}
  \end{ruledtabular}
  \caption{
    Bicycle chain code instances used in this work. The number of logical qubits is proportional to one
    direction of the lattice size $\ell$. Codes with higher distances are available in Appendix~\ref{app:bicycle-chain-codes}.
  }
  \label{tab:bicycle-chain-instances}
\end{table}

\subsection{Hardware-efficient implementation}
\label{sec:hardware}

The translation symmetry of the bicycle chain codes in both stabilizer and logical
structure significantly simplifies the architectural design onto hardware-native
2D-grid moves in neutral-atom platforms using acousto-optic deflector (AOD) control~\cite{bluvstein2022quantumprocessora}.

For each bicycle chain code block, the left and right halves of data qubits are
arranged as two $\ell \times m$ arrays of atoms. Syndrome extraction uses two
additional $\ell \times m$ arrays for the $X$- and $Z$-check ancillas, with the
same periodic layout. We now describe the hardware implementation of logical
gadgets under this layout.

\paragraph{Global transversal Clifford gates.}
As described in Appendix~\ref{app:bicycle-chain-codes}, global transversal $H$
and $S$ are composed of single-qubit gates applied uniformly across the block;
they can therefore be implemented by global pulses and require no atom
shuttling.
A global transversal CNOT between two encoded blocks consists of pairwise
physical entangling gates between matched data qubits.
This can be implemented by AOD shuttling that aligns the corresponding data
arrays.
Once matched sites are brought into the entangling zone, the physical
entangling gates are applied in parallel.

\paragraph{Cyclic shift automorphisms.}
The shift automorphism is realized by shifting both the left and right halves
horizontally by one site in their $\ell \times m$ grids
[Fig.~\ref{fig:bicycle-chain-code}(a)].
This can be implemented geometrically by picking up the relevant data array,
translating it by one lattice spacing with periodic wraparound, and depositing
it back.
With one AOD array, this reshuttling takes two moves; with two AOD arrays,
it can be done in one.

\paragraph{Syndrome extraction.}
For the bicycle chain code instances considered in this manuscript, the syndrome
extraction schedule induced by the polynomial operation has CNOT depth 8. We
enumerate all possible choices of such circuits and use one of those that achieves
circuit-level distance $d-1$ (Appendix~\ref{app:syndrome-extraction}). With syndrome
extraction, one can transversally initialize a bicycle chain code into
$\ket{\overline{0},\dots,\overline{0}}$ and $\ket{\overline{+},\dots,\overline{+}}$
by initializing all physical qubits into $\ket{0}$ and $\ket{+}$ followed by syndrome
extraction.
To implement each scheduling step, the ancilla arrays are cyclically
shifted according to the corresponding monomial, then shuttled and entangled
with the data arrays.
The cyclic shifts of the ancilla arrays are implemented by the same AOD
reshuttling primitive as the shift automorphism.

\paragraph{Parallel STAR state injection.}
The TMR protocol is applied to each disjoint logical
representative in the same way [Fig.~\ref{fig:bicycle-chain-code}(b)]: divide
the support of a logical representative into groups; apply a $Z\otimes \cdots
\otimes Z$ rotation on each full group; then measure the syndrome and
post-select. The $Z\otimes \cdots \otimes Z$ rotation can be decomposed into
CNOT ladders with a single-qubit $R_Z$, where the CNOTs correspond to local
shuttling and entanglement within a column. Because the same operation is
applied to every disjoint column, selective AOD addressing allows parallel
resource state preparation on different logical representatives.

\section{Numerical Simulations and End-to-End Resource Estimates}
\label{sec:results}

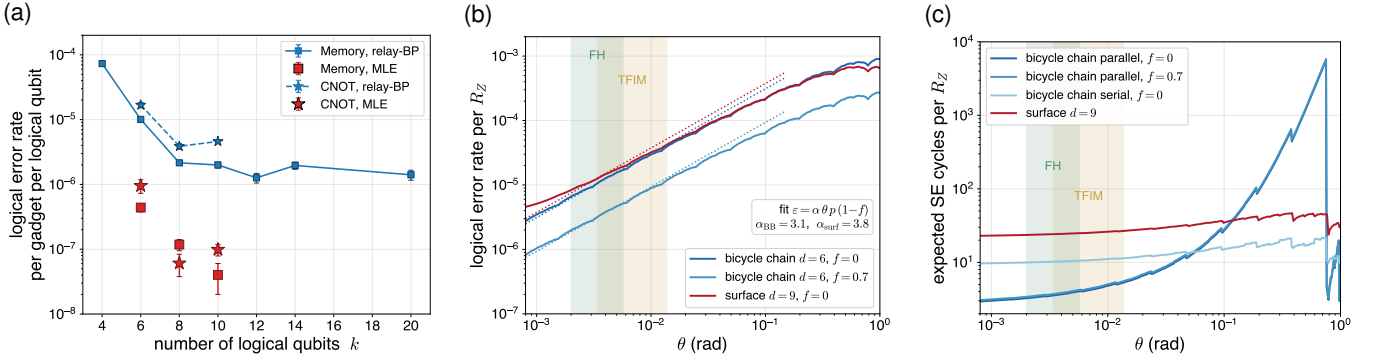
\begin{figure*}[htbp]
  \centering
  \begin{tikzpicture}
    \node[anchor=south west, inner sep=0] (pa) {\includegraphics[width=0.32\textwidth]{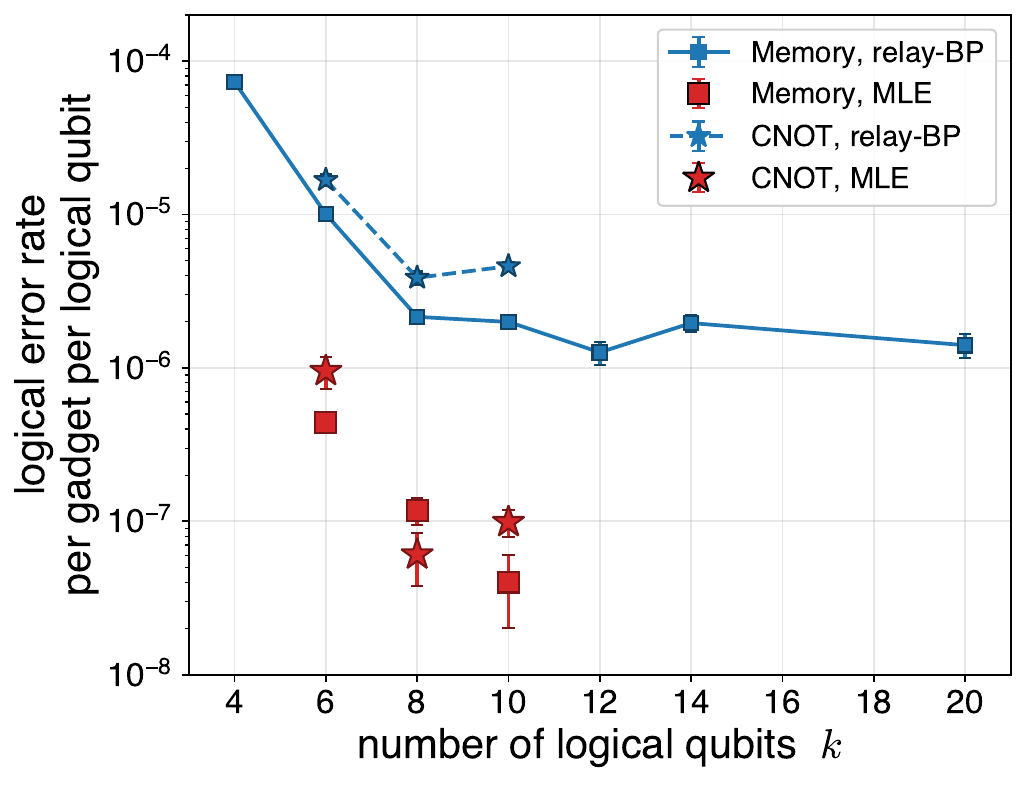}};
    \node[anchor=south west, font=\sffamily, inner sep=0] at ([yshift=1pt]pa.north west) {(a)};
  \end{tikzpicture}\hfill
  \begin{tikzpicture}
    \node[anchor=south west, inner sep=0] (pb) {\includegraphics[width=0.32\textwidth]{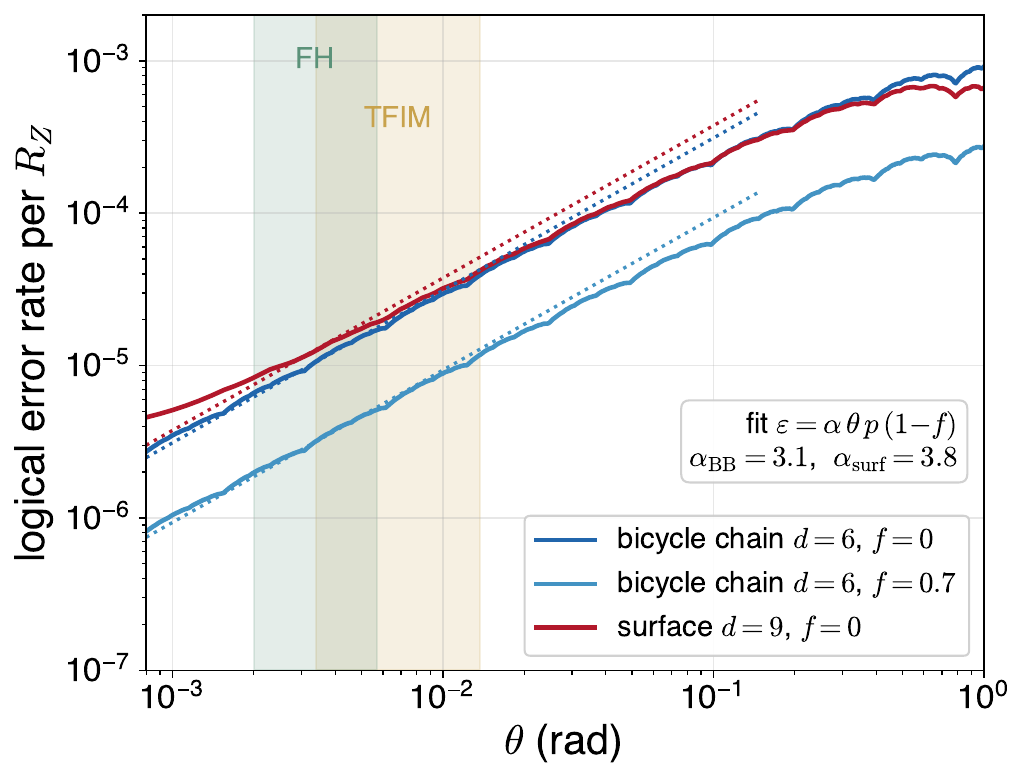}};
    \node[anchor=south west, font=\sffamily, inner sep=0] at ([yshift=1pt]pb.north west) {(b)};
  \end{tikzpicture}\hfill
  \begin{tikzpicture}
    \node[anchor=south west, inner sep=0] (pc) {\includegraphics[width=0.32\textwidth]{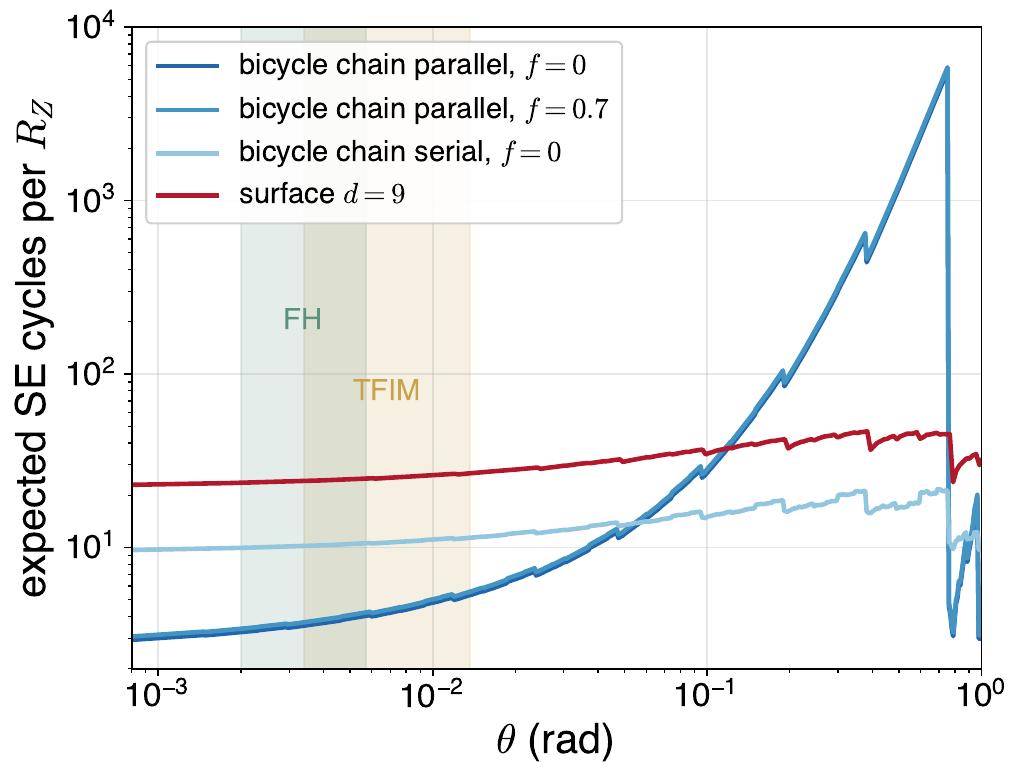}};
    \node[anchor=south west, font=\sffamily, inner sep=0] at ([yshift=1pt]pc.north west) {(c)};
  \end{tikzpicture}
  \caption{
    \textbf{Logical gadget performance at physical error rate of $p=0.1\%$.}
    (a) Per-gadget-per-logical-qubit logical error rate (LER), defined as $\mathrm{LER}/(N r)$,
    versus $k = 2\ell$ for the $\qcode{14\ell, 2\ell, 6}$ bicycle
    chain family. Here $r = 3d = 18$ syndrome rounds are applied, with $N = k$
    logical observables for the memory experiment (squares) and $N = 2k$ for the
    transversal CNOT (stars). Both are decoded by relay-BP (blue) and MLE (red).
    Error bars denote $1\sigma$ binomial confidence intervals.
    (b)--(c) Cost of the repeat-until-success (RUS)-injected $R_Z(\theta)$ gate on the $k = 8$
    bicycle chain block ($\ell = 4$, $m = 7$) versus rotation angle $\theta$,
    simulated using the stochastic RUS protocol (see Appendix ~\ref{sec:magic}).
    (b) Per-teleported-$R_Z$ logical error rate for the bicycle chain code (blue) at qubit loss error
    fraction $f = 0$ (dark) and $f = 0.7$ (light), compared with the surface
    code $d = 9$ STAR rotation (red). Dotted lines are linear fits $\varepsilon = \alpha\,\theta\,p\,(1-f)$ over the small-angle (Trotter) regime, with $\alpha_\mathrm{BB} \approx 3.1$ and $\alpha_\mathrm{surf} \approx 3.8$; the teleported error rate is linearly proportional to angle and to the Pauli (non-loss) fraction $(1-f)p$, departing from the fit for large angles.
    (c) Expected syndrome extraction cycles per $R_Z$ (blue, bicycle chain):
    parallel re-injection at $f = 0$ (dark) and $f = 0.7$ (medium), the serial
    one-$R_Z$-per-block cost (light), and the surface code $d = 9$ reference
    (red). All $k$ logical qubits are injected together and re-injected per RUS
    round, retrying only failed logical qubits. The TMR partition number $M$ is chosen adaptively from
    $\{1, 3\}$ to minimize the logical error rate. The discontinuity at large angles is the consequence of $M=1$ protocol being favorable at all teleportation stages.  Shaded bands mark the Trotter
    rotation angles used in the Fig.~\ref{fig:resources} estimates [transverse-field Ising model (TFIM) and
    Fermi--Hubbard (FH)].
  }
  \label{fig:numerics}
\end{figure*}

In this section, we report the results of detailed numerical simulations to support and validate megaquop quantum simulation capabilities based on the proposed high-rate STAR architecture. These include the simulations of circuit-level gadgets for the bicycle chain codes, with syndrome extraction, transversal Clifford gates, and TMR rotation resource state preparation.
All circuits are simulated under a depolarizing noise model with physical error rate $p$. Each preparation of
physical $\ket{0}$ or $\ket{+}$ state is followed by a single-qubit depolarizing channel of
strength $p$. The $X$- or $Z$-basis measurements report the wrong outcome
with probability $p$. Finally, one- or two-qubit gate and followed by the
corresponding one- or two-qubit depolarizing channel of total strength $p$, and
qubits that idle during entangling layers undergo single-qubit depolarizing
noise with the same strength.

For noisy Clifford circuits, detector and observable
samples are generated using Stim~\cite{gidney2021stimfast}. The TMR protocol
contains non-Clifford components, so its logical error rates are estimated using
non-Clifford simulation techniques implemented in the Clifft package~\cite{chase2026clifft}.
This is in contrast to most magic-state simulations, which until recently were done by Clifford approximate circuits~\cite{gidney2024magicstate, sahay2025foldtransversalsurface, akahoshi2024partiallyfaulttolerant,ismail2025transversalstar}. The full circuits are decoded using both relay-BP~\cite{muller2025improvedbelief}
and a mixed-integer-programming implementation of the most-likely-error (MLE) decoder~\cite{cain2024correlateddecoding,landahl2011faulttolerant,takada2024ising,gurobi},
and we report both results in Fig.~\ref{fig:numerics}(a).

\subsection{Per-gadget logical error rates}
\label{sec:gadget_ler}
For syndrome extraction, we assume that each round contributes the same
logical error rate. To eliminate boundary effects from logical initialization
and readout, we simulate a memory experiment with $r=3d$ syndrome extraction
rounds and report the average logical error rate per round.

Transversal Clifford gadgets are modeled using a similar subtraction procedure as in Ref.~\cite{ismail2025transversalstar}.
For a given gadget, we simulate initialization, $d$ memory rounds, $d$
repetitions of the transversal gadget separated by one syndrome extraction
round each, another $d$ memory rounds, and final measurement. We then subtract
the logical failure probability of the surrounding $2d$ memory rounds and
average over the $d$ internal gadget repetitions. This gives the per-gadget
logical error model used in the resource estimates. Figure~\ref{fig:numerics}(a)
reports the per-gadget-per-logical logical error rate (LER) as a function of the number of logical qubits
$k = 2\ell$ at $p = 10^{-3}$ for the memory and transversal CNOT gadgets.
Here the number of logical qubits is varied by lengthening the torus
(increasing $\ell$) within the $[[14\ell,\, 2\ell,\, 6]]$ family of
Sec.~\ref{sec:bicycle_chain_code}, holding the distance $d = 6$ and the
encoding rate $k/n = 1/7$ fixed. Both memory and transversal CNOT LERs are large at small $l$, but plateau at $l \gtrsim 4$ within the error bounds of the simulation. It appears that adding logical qubits at fixed circumference
$m$ and distance $d$ does not further improve the performance once the chain is long enough. The dependence on the physical error rate $p$ at fixed $k = 8$ is
reported in Appendix~\ref{app:bicycle-chain-codes} (Fig.~\ref{fig:ler_vs_p}).

The infidelity of TMR-generated magic states is estimated by Bell measurements together
with an ideally prepared $R_Z(-\theta)$ verification state (Appendix~\ref{sec:magic}). These Bell
measurements are implemented using transversal CNOTs and transversal measurements. Detailed
logical error rates and post-selection rates for different $p$ and $\theta$
are reported in Appendix~\ref{sec:magic}. Each TMR attempt requires only two syndrome extraction rounds (one $Z$-only round for the initialization before the rotation for codespace projection, and one full round after TMR) to saturate the TMR factory error rates, versus three for the surface code---a structural consequence of every qubit in the bicycle-chain code participating in an even number of $X$-checks  (Appendix~\ref{app:rus-clifford-floor}). Requiring fewer rounds partially compensates for the longer syndrome extraction cycle of the bicycle-chain code, whose circuit depth is $8$ versus $4$ for the surface code: although each round is roughly twice as deep, fewer of them are needed to reach the factory error floor.

In addition, we bound the effect of atom loss using heralded-erasure simulations.  We treat loss purely by post-selection: any shot with a heralded loss is
discarded, and we never decode or correct based on where or when the loss is detected. Our simulation therefore discards a run the moment a loss occurs. Although this models immediate heralding, it faithfully reproduces implementations with deferred detection---including measurement-only schemes, where loss is revealed only at readout \cite{Ma_2023Loss,Scholl_2023Loss,Chow_2024Loss}. The reason is that the location and timing of a loss are never consumed: only its occurrence matters. Consequently the accepted shots are loss-free and identical in both cases, every loss-containing trajectory is discarded in both, and the only physical difference (errors propagated by a still-undetected lost atom during the detection delay) is confined to shots that are themselves discarded, so it affects neither the
reported infidelity nor the acceptance rate. The short factory depth reinforces this by bounding the detection latency: every qubit is measured within two cycles, so loss is promptly heralded under any reliable loss-detection scheme.

Treating a fraction $f$ of the physical errors as heralded loss rather than as
Pauli noise is doubly favorable: it lowers the residual infidelity while costing
almost no acceptance. The infidelity drops because, in the STAR TMR protocol, an
undetectable logical fault requires a set of $Z$ errors on a qubit set that
exactly matches one of the chosen TMR partitions; a heralded loss never produces
such a pattern and is post-selected away, so only the remaining Pauli fraction
$1-f$ can seed a logical fault. The same fraction
of that smaller component becomes a logical error, so the TMR logical
error rate---which scales as $\alpha p|\theta|^\beta$ in the small-angle
regime~\cite{toshio2024practicalquantum, ismail2025transversalstar}---is reduced
to approximately $(1-f)\alpha p|\theta|^\beta$, consistent with the numerical
simulations reported in Appendix~\ref{sec:magic}.

The acceptance, by contrast, suffers only minimally, because most of the
factory's Pauli events are already post-selected. A fraction $\gamma \approx
0.93$ of physical Pauli events trigger a state-prep detector and are discarded
regardless (Sec.~\ref{app:tmr_characterization}); converting that fraction to
always-heralded loss therefore removes only the additional $\approx (1-\gamma)$
that would otherwise have been accepted. This is reflected in the factorized
acceptance rate of applying TMR in parallel on $N$ logical qubits,
$p_\mathrm{init}(\ell,m,f,p) \cdot p_\mathrm{TMR}(\theta;M)^N \cdot
p_\mathrm{loss}(\ell,m,f,p)$, where the Pauli survival
$p_\mathrm{init} \propto \exp[-\gamma N_\mathrm{loc}(1-f)p]$ carries the detector
fraction $\gamma$ while the loss survival
$p_\mathrm{loss} = \exp[-f N_\mathrm{loc}p]$ carries the full $f$; their product
differs from the all-Pauli ($f=0$) baseline only by $\exp[-f(1-\gamma)
N_\mathrm{loc}p]$. Here $M$ is the TMR partition count [see Appendix~\ref{app:M}
and Fig.~\ref{fig:compilation}], while $N_\mathrm{loc} = 32\ell m + 13$ is the number of possible noise-locations per factory run,  and the prefactor fits are extracted in
Appendix~\ref{app:tmr_characterization}.

Finally, to complete the numerical model of parallel $R_Z$ magic performance, we show that logical errors on distinct
logical qubits in a bicycle chain block are independent in explicit simulations (Appendix~\ref{sec:magic}). This
simplification stems from the disjoint logical representative basis. The dominant
logical error contribution comes from the TMR step, which scales as
$O(p\theta^{2(1-1/M)})$~\cite{choi2023faulttolerantnoncliffordstate, toshio2024practicalquantum} and is applied independently to different logical
qubits. Correlations between logical qubits are introduced only through
syndrome extraction and therefore enter at higher order,
$\bigO{p^{\lfloor (d+1)/2 \rfloor}}$. Combining the detailed CNOT and TMR error and postselection rate models,
the teleported $R_Z$ gate LER and expected number of retries are estimated using a stochastic RUS simulation,
as shown in Fig.~\ref{fig:numerics}(b-c).  The teleported error rate is linearly proportional to angle, $\theta$, and the Pauli (non-loss) fraction of the logical error rate at small angles, $(1-f)p$. While small-angle TMR state preparation has better-than-linear-scaling, $O(p\theta^{2(1-1/M)})$ with $M=3$, the teleported LER is ultimately limited by the large angle teleportation steps where $M=1$ is optimal, leading to the observed linear scaling \cite{toshio2024practicalquantum}, with fitted coefficient $\alpha \approx 3.1$ for the bicycle chain ($\alpha \approx 3.8$ for the surface code) at $p = 0.1\%$ [Fig.~\ref{fig:numerics}(b)].

The results imply that the $[[14l, 2l, 6]]$ bycicle chain code family is megaquop-capable. The teleported $R_Z$ gate LER without loss heralding is competitive with surface code version and is expected to be largely independent of the code, so long as the memory error (logical error of one syndrome extraction cycle in the code) is below the $R_Z$ error at the angles of interest. The Clifford error
rate, meanwhile, is firmly in megaquop territory: as a reference, surface-code
transversal STAR~\cite{ismail2025transversalstar} requires $d=9$ to reach Clifford
logical error rates comparable to the bicycle chain code at $d=6$. We caution,
however, that the bicycle chain results are decoded with the optimal MLE decoder,
whereas the surface-code results used the suboptimal MWPF
decoder~\cite{wu2025mwpfdecoder}; whether the $d=6$ versus $d=9$ gap persists once
the surface code is decoded optimally remains to be determined.

Finally, two distinct mechanisms drive the cycle-count gains for parallel bicycle STAR magic. First, the reduced code distance: $d=6$ rather than $d=9$ means fewer noise locations per shot, hence fewer post-selection retries per magic gate.
More importantly, parallel injection amortizes the dominant cost, the post-selection of the
block initialization ($p_\mathrm{init}$), across all rotated logical qubits: it is
paid once per block rather than once per logical, at the price of only $\sim\log_2(k)/k$ RUS overhead per logical [Eq.~\eqref{eq:rmax_mean}]. This amortization is most effective precisely in the small-angle regime relevant
to quantum simulation [$10^{-3}\lesssim\theta\lesssim10^{-1}$, shaded in
Fig.~\ref{fig:numerics}(c)]. There the per-logical TMR acceptance
$p_\mathrm{TMR}(\theta;M)\approx 1$, so the joint acceptance $p_\mathrm{TMR}^N$ of
injecting all $N$ logicals together stays high and the saving on $p_\mathrm{init}$
is realized in full. Climbing the RUS ladder doubles the angle at each step but
re-injects only the failed logicals, so the growing per-step post-selection cost
is paid on a shrinking set. Serial injection (rotating one logical per
block) instead pays $p_\mathrm{init}$ for every logical but avoids the
$p_\mathrm{TMR}^N$ joint-acceptance penalty; it wins only when the target angles
are large enough [$10^{-1}\lesssim\theta\lesssim 1$] that parallel post-selection
becomes prohibitive already at the early RUS steps. Those angles lie outside the
quantum-simulation regime of interest here.

In that simulation-relevant regime, the reduced distance and the amortized
parallel injection together cut the expected number of logical cycles per magic
gate by up to $\sim10\times$ relative to the $d=9$ surface code and up to
$\sim5\times$ relative to serial high-rate injection. Loss heralding can reduce
the small-angle magic LER still further, since the error scales as $(1-f)$ and the
loss fraction on current hardware is consistent with a $1-f\approx 0.3$
reduction~\cite{evered2026highfidelityentanglinggatesnonlocal}, at negligible time
cost.

\subsection{End-to-end resource estimates for megaquop quantum simulation}
\label{sec:e2e_resources}

We now proceed to evaluate the consequences of the improved encoding rate and the logical gadget performance and parallelism on the physical resources for megaquop quantum simulation. As prototypical applications, we use second-order Trotterized simulation of
the transverse-field Ising model (TFIM) and the Fermi--Hubbard (FH) model with a
local compact encoding~\cite{derby2021compact,jafarizadeh2024recipelocal} (Appendix~\ref{app:fermion-new}). For each simulation, we set the largest step size
$\delta t^*$ to satisfy the second-order Trotter error to $\epsilon_{\mathrm
{trot}} \leq 0.01$ at any given simulation time $T$. As representative examples of
megaquop simulation space-time volume, we consider a $8\times 8$ TFIM and a spinful single-band
FH model at simulation time $T^*=2.0\,(zJ)^{-1}$ and $2.0\,(zt)^{-1}$, respectively, where $z=4$ is
the lattice coordination number. This sets a comparison point across all architectures characterized with  $\delta t^*=0.0274\,(zJ)^{-1}$ and $73$ Trotter steps for the TFIM
($J=g=1$, $L=8$), and $\delta t^*=0.0230\,(zt)^{-1}$ and $87$ steps for
Fermi--Hubbard ($U=4$, $t=1$, $L=8$; see Appendix~\ref{app:trotter} for
Trotter error analysis). Finally, for $T$-based architectures, a rotation synthesis error budget of $\epsilon_{\mathrm{syn}} = \epsilon_{\mathrm{RZ}}/3$ following Ref.~\cite{beverland2022assessingrequirements} (Appendix~\ref{app:rz-to-t}), corresponds to $\sim\!2.1\times 10^5$ and
$\sim\!1.4\times 10^6$ equivalent $T$ gate counts for the TFIM and Fermi--Hubbard comparison points, respectively.

\begin{figure}[htbp]
  \centering
  \includegraphics[width=\columnwidth]{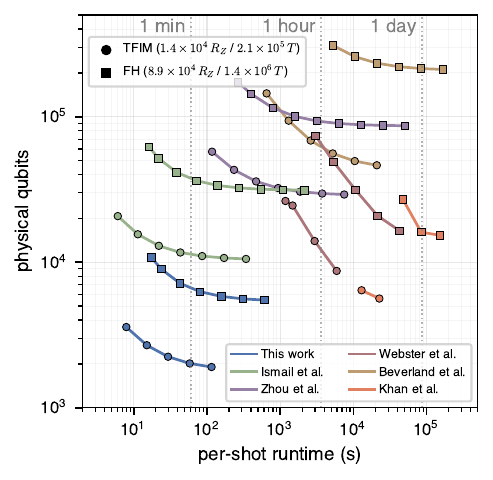}
  \caption{
    \textbf{End-to-end resource cost of bicycle chain STAR versus alternative
    architectures for prototypical megaquop simulation tasks.} Physical qubit footprint versus per-shot runtime trade-off is shown for the
    $8\times 8$ TFIM (circles) and Fermi--Hubbard model (squares). All six
    architectures implement the identical Trotterized circuit at a common
    simulation time ($T^*=2.0$ in Hamiltonian units; $73$ TFIM and $87$ FH
    Trotter steps), at heralded-atom-loss fraction $f=0$. Points on each curve
    sweep the number of magic-state factories. Bicycle chain STAR runs two
    parallel instances (left/right logical halves), halving its per-shot runtime.
    The Khan et al.~\cite{khan2026architecting}\ extractor is drawn along its space--time Pareto frontier (the
      runtime-floor tail, where added factories only grow the qubit count, is
    omitted).
    Accumulated gate errors (TFIM\,/\,FH) are: this work $0.43/0.65$, Ismail et al.~\cite{ismail2025transversalstar}, Zhou et al.~\cite{zhou2025resourceanalysis}~$0.32/0.31$, Webster et al.~\cite{webster2026pinnaclearchitecture}~$0.16/0.09$, Beverland et al.~\cite{beverland2022assessingrequirements}~$0.64/0.35$, and Khan et al.~\cite{khan2026architecting}~$0.05/0.09$. Error
    comprises gate/QEC, magic-state, and for Clifford$+T$ baselines, $R_z$-to-$T$
    synthesis error. Code distances and
    benchmark references are given in Sec.~\ref{sec:e2e_resources} and
    Table~\ref{tab:arch-axes}.
  }
  \label{fig:resources}
\end{figure}

Under the same Trotter error allocation and as comparable as possible circuit error, Fig.~\ref{fig:resources} summarizes the estimated runtime and physical qubit footprint for six architectures across varying number of factories: bicycle chain STAR ($d=6$), surface code transversal STAR ($d=9$)~\cite{ismail2025transversalstar}, a surface code transversal architecture with fold-transversal cultivation ($d=15$)~\cite{zhou2025lowoverheadtransversal,sahay2025foldtransversalsurface},
surface code lattice surgery with 15-to-1 $T$ factories of Beverland et al. [$d=15$ for TFIM and $d=19$ for FH, with factory tuple $(d_X,d_Z,d_m)=(15,5,5)$ denoting the factory's asymmetric patch distances and
measurement distance as covered in Appendix~\ref{app:beverland}]~\cite{beverland2022assessingrequirements}, the Webster et al. Pinnacle generalized bicycle (GB) architecture ($d=16$)~\cite{webster2026pinnaclearchitecture}, and the Khan et al.\ extractor-based architecture on Two-Gross bicycle codes ($d=18$)~\cite{khan2026architecting}. The physical error rate is assumed to be $p=0.1\%$ under the standard depolarizing
error model without atom loss error for all architectures.
The syndrome extraction cycle time is set to 1$\,$ms, 1.5$\,$ms,
and 2$\,$ms for the surface code, BB/GB code, and bicycle chain code,
respectively, reflecting their weight-four, weight-six, and weight-eight
stabilizers. Table~\ref{tab:arch-axes} in
Appendix~\ref{app:resource-estimation} summarizes the corresponding host code,
Clifford backbone, rotation strategy, and resulting space--time bottleneck for
each architecture.
Since estimating observable expectation values requires many repeated shots,
and two parallel simulations run concurrently on the bicycle chain STAR
architecture, enabled by the left/right logical halves of the bicycle chain code, we report its total per-shot runtime, defined as the total
circuit execution time divided by two. Note that these architectures support parallel magic factories to accelerate the computation in principle. We sweep the number of factories in each architecture to evaluate the trade-off between space and execution time.

Bicycle chain STAR requires $1{,}904$--$3{,}584$ physical qubits for the TFIM
($8$--$116\,$s per-shot) and $5{,}488$--$10{,}752$ for Fermi--Hubbard
($18\,$s--$10.2\,$min), achieving circuit errors $\epsilon_{\mathrm{gate}}=0.43$
and $0.65$, composed of logical Clifford errors and rotation resource state errors.
The higher FH error reflects its Clifford-heavier Trotter step
(roughly six transversal Cliffords per rotation versus two for the TFIM).
Surface STAR achieves $\epsilon_{\mathrm{gate}}=0.46$ (TFIM) and $0.84$ (FH)
at $5.5\times$ and $5.7\times$ the qubit count. Compared to fully fault-tolerant architectures, bicycle chain STAR achieves
$100-1000\times $ speedup over code-surgery approaches (Beverland
et al.~\cite{beverland2022assessingrequirements}, Webster et al.~\cite{webster2026pinnaclearchitecture}, and Khan et al.~\cite{khan2026architecting}) and ${>}10\times$ speedup over the
transversal surface code (Zhou et al.~\cite{zhou2025resourceanalysis}). This speed up is reported already at reduced space, as surface code
architectures require $15$--$75\times$ more physical qubits; BB/GB
architectures are closer but still $\sim\!2.8$--$13\times$ more bulky due to
surgery and factory overheads.

For comparison, the fully fault-tolerant architectures achieve the following circuit errors (TFIM/FH), including logical
gate errors, magic state errors, and rotation synthesis errors: Zhou et al.~\cite{zhou2025resourceanalysis}\ $0.32$/$0.31$, Beverland et al.~\cite{beverland2022assessingrequirements}\ $0.64$/$0.35$,
Webster et al.~\cite{webster2026pinnaclearchitecture}\ $0.16$/$0.09$, and Khan et al.~\cite{khan2026architecting}\ $0.05$/$0.09$. These errors are dominated
by synthesis and can be lowered by increasing code distances, using more
accurate synthesis, or using higher-fidelity magic-state factories. STAR-magic exclusive
architectures, by contrast, are limited by the fidelity of their rotation
resource states~\cite{chung2026partiallyfaulttolerantquantumcomputation}: there is no synthesis error to reduce, and the rotation
error is set by the physical gate fidelity and the TMR protocol. We note that while our comparison settings were chosen to represent the closest to equivalent comparison for the same task across all architectures, the resulting total circuit error rates for Webster et al.~\cite{webster2026pinnaclearchitecture} and Khan et al.~\cite{khan2026architecting} end up being notably lower than the rest of the architectures. This is the consequence of the limited number of code instances accessible to these architectures -- the decrease of the code distance to the next smaller code instances would make the task non-viable.

\begin{figure*}[!t]
  \centering
  \includegraphics[width=0.9\textwidth]{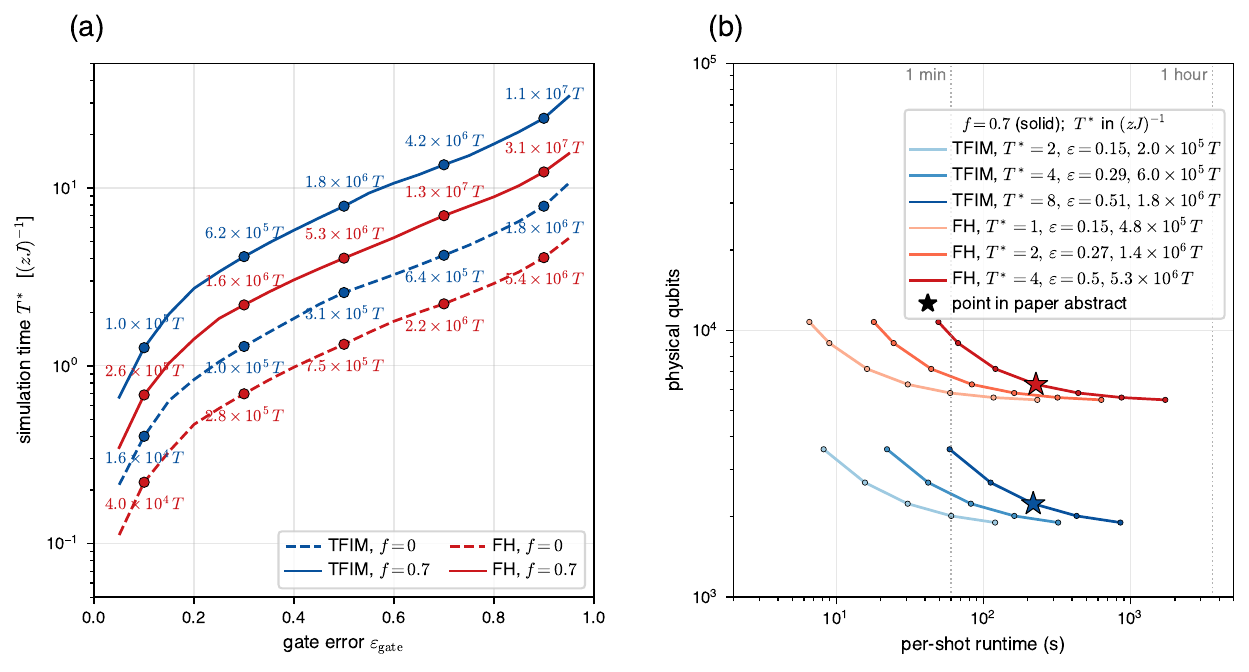}
  \caption{
    \textbf{Simulation reach of bicycle chain STAR.}
    (a) Reachable simulation time $T^*$ versus achieved gate error at
    heralded-atom-loss fractions $f=0$ (light) and $f=0.7$ (dark) for both TFIM and FH
    models. Post-selecting atom loss lowers the $R_Z$ error rates by
    $\sim\!(1-f)$, extending $T^*$. Points are labeled with their equivalent $T$-count, i.e., the number of $T$ gates needed to synthesize the rotations with synthesis error as $1/3$ of the $R_Z$ error (the same count is listed in the (b) legend).
    (b) Physical qubits versus per-shot runtime, swept over
    magic-state factory count, for both models at $f=0.7$. Three $T^*$ values
    are shown per model (different shades illustrate different $T^*$); the legend
    lists, per operating point, $T^*$, the achieved gate-error budget $\varepsilon$,
    and the equivalent $T$-count. Two points at
    $T^*\approx 8\,(zJ)^{-1}$ TFIM, and $\approx 4\,(zt)^{-1}$ FH marked as stars are the ones reported in the manuscript abstract: \headlineQubits{}-qubit / $\sim\!200\,$s (TFIM)
    and \headlineQubitsFH{}-qubit / \headlineTimeFH{}$\,$s (FH). They represent the limits of possible high-rate STAR simulation volume.
  }
  \label{fig:reach}
\end{figure*}

On neutral-atom hardware, the fidelity of STAR resource states can be improved
by detecting and post-selecting atom loss. From our simulation of the heralded atom loss (Appendix~\ref{app:loss}), the logical error rates of the $R_Z$ decrease linearly as the atom loss error fraction $f$ of the total physical errors (Fig.~\ref{fig:reach}).
At $f=0.7$, the TFIM reaches $T^*\approx 8\,(zJ)^{-1}$---equivalent to $\sim\!2\times 10^6$ $T$ gates---at \headlineQubits{} physical
qubits and a $\sim\!200\,$s per-shot runtime, while Fermi--Hubbard reaches
$T^*\approx 4\,(zt)^{-1}$ ($\sim 5 \times 10^6$ $T$ gates) at \headlineQubitsFH{} physical qubits and
a \headlineTimeFH{}$\,$s per-shot runtime. These loss-post-selected points---marked
as stars on the space--time tradeoff of Fig.~\ref{fig:reach}(b)---are the ones
highlighted in the manuscript abstract, and represent effectively limits of STAR-magic-exclusive simulation volume.

\section{Discussion and Conclusion}
\label{sec:discussion}

We present an operationally parallel, high-rate STAR architecture for megaquop-scale
local lattice Hamiltonian simulation, following the symmetry-driven co-design
principle. The translation symmetry of the target Hamiltonian is matched
to the cyclic shift automorphism of the bicycle chain codes, achieving
high encoding rates and efficient transversal computation with parallel
STAR injections. On neutral-atom hardware, the same symmetry maps onto
global AOD shifts, making all the key operations hardware-native.

End-to-end resource estimates supported by circuit-level simulations confirm the
practical advantage of this architecture: for an $8\times 8$ lattice
model, it achieves a \fullSpaceReduction{} reduction in physical qubit count relative to a surface code transversal STAR
baseline~\cite{ismail2025transversalstar} while preserving its speed, and is two to three orders of magnitude
faster than code-surgery architectures. These gains bring megaquop-scale simulation of both the
transverse-field Ising and Fermi--Hubbard models into a regime accessible with a few thousand physical qubits and a few hundred seconds of per-shot runtime, opening a low-resource pathway into the early fault-tolerant era.
The significance of this result is not only the numerical resource reduction, but that the high-rate encoding advantage is retained end to end: the qubit savings are not paid back through slower logical operations, complex surgery primitives, or serialized magic-state preparation. To our knowledge, this is among the first fully evaluated high-rate architectures to achieve this combination of space savings, runtime competitiveness, and operational simplicity.

On the error-correction side, our work opens several direct avenues of exploration.
Notably, all numerical results in this work assume a simple circuit-level depolarizing
noise model. Realistic hardware may exhibit structured noise bias that
could be exploited by tailored decoding strategies~\cite{roffe2023biastailoredquantum,bonillaataides2021xzzxsurface}.
For atom loss, we estimated the performance impact using heralded-erasure
simulations; a detailed analysis of syndrome extraction under loss~\cite{baranes2025leveragingatom,yu2025locatingrydberg,yu2025processingdecoding,liu2026achieving} remains to be explored in future work. Further studies call for additional simulation techniques that
can combine the current non-Clifford methods~\cite{haenel2026tsim,chase2026clifft} and atom loss.
As for decoding, our resource
estimates rely on the MLE decoder, which provides near-optimal error
suppression but does not scale to large circuits in real time. Recent
advances in machine-learning decoders~\cite{ataides2025neuraldecoders,gu2026scalableneural}
suggest that near-MLE accuracy can be achieved at latencies compatible
with real-time syndrome processing. Demonstrating this for bicycle chain
codes under realistic noise is an important practical milestone. We remark that the
architecture itself is not limited to neutral atoms: any platform with
the long-range connectivity required for bicycle chain code syndrome extraction,
including trapped-ion and spin qubits, could host the same logical structure.

On the algorithm side, although this work focuses on dynamical simulation of
local lattice Hamiltonians, the framework extends naturally to broader settings.
For example, one can simulate Hamiltonians beyond native lattice models, as long as the geometric
locality is present, by embedding them into the lattice grid, selectively
controlling the strength of terms at each site while preserving the lattice backbone. This also motivates exploring STAR beyond
  deterministic Trotterized Hamiltonian simulation. Randomized or sampling-based
  time-evolution protocols~\cite{childs2019fasterquantum,campbell2019randomcompiler},
  real-time response measurements~\cite{wecker2015solvingstrongly,delre2024robustmeasurements},
  and other algorithms that extensivley query local time-evolution primitives. The key determinant of applicability is
 the preservation of dominant global lattice structure in the algorithms that can benefit from parallel
STAR injection and efficient global Clifford gateset. In addition, one can generalize this framework
to quantum phase estimation~\cite{kitaev1995quantum,lu2021algorithms} or Hadamard-test-based algorithms~\cite{Lin2022, Wan2022, Epperly2022, Moreno2025, Yoshioka2025}, where
controlled-$e^{-iHt}$ is required. One approach to implementing the controlled
evolution is to use a GHZ register to keep the operation structural and global.
Preparing GHZ states across bicycle chain code blocks requires additional
one-time cost, which could be tackled by distilling the Clifford resource states or with a code surgery on high-rate codes~\cite{baspin2025fastsurgery,cowtan2025fastfaulttolerant,zhang2025timeefficientlogical,zheng2025highratesurgery, gu2026qgpu}.

Looking further ahead, the $R_Z$ gate in STAR achieves logical error
rates scaling as $\alpha p|\theta|$. This bounds the total simulatable
rotation budget to $\sim 1/(\alpha p)$, or, for $N$ logical qubits, a total
simulatable angle of $\sim 1/(\alpha p N)$, limiting the current architecture
at the megaquop scale. This rotation-error ceiling is visible in
Fig.~\ref{fig:reach}(a), where the reachable $T^*$ saturates as the tolerated
gate error approaches one. Although recent results suggest that such fidelity
bounds may be loose for local observables~\cite{flannigan2022propagationerrors},
pushing far beyond the direct STAR rotation-budget regime will likely require
either improving the STAR resource states or hybridizing with higher-fidelity
discrete-magic architectures~\cite{gidney2024magicstate,sahay2025foldtransversalsurface,menon2025magictricycles}. Such a hybrid need not discard high-rate STAR. Even in a fully fault-tolerant
setting, there may be an architecture-dependent angle threshold below which
direct high-rate STAR magic is preferable to synthesizing the rotation entirely
from discrete magic states. The resulting space-time advantage would not be the
full advantage of the direct STAR architecture studied here, since high-distance
codes, discrete-magic factories, and arithmetic subroutines may dominate other
parts of the computation. Nevertheless, a smooth interpolation between
STAR-based analog rotations and Clifford+$T$ synthesis is viable:
arithmetic-heavy simulation algorithms~\cite{Kivlichan2020improvedfault, campbell2022earlyfaulttolerant, Su2021, spagnoli2025quantumsimulationnucleardynamics}
could use conventional discrete magic for arithmetic primitives, while the
remaining small-angle rotations are supplied by STAR. This includes both the
small rotation components appearing in improved angle-dependent synthesis
schemes~\cite{kliuchnikov2022shorter} and the rare non-Clifford residuals in
recent quasi-probability approaches to small-angle rotation
synthesis~\cite{bothe2026moreefficient}. Such a hybrid would allow high-rate STAR approach to
remain useful even inside fully fault-tolerant architectures with high-distance
codes and discrete magic factories.

Complementary to extending the high-rate STAR utility by hybrid strategies is exploiting the logical multiplicity already present in the construction. The two independent left/right logical subsets of the bicycle chain code double the minimum footprint if viewed as a single simulation, but for observables estimated from many shots they can instead be used to run independent replicas of the same simulation instance, converting this multiplicity into sampling throughput. This suggests a broader route toward high-throughput logical simulation: ultra-high-rate codes~\cite{kasai2026breakingorthogonality, zhao2026ultrahighrate} whose logical qubits decompose into $N_{\mathrm{sub}}$ internally cyclic subsets could run $N_{\mathrm{sub}}$ replicas in parallel while amortizing syndrome extraction and global transversal gates across all of them. The main technical tension is with STAR injection, since denser logical packing may preclude fully disjoint physical representatives and require injections to be batched or supplemented by additional angle factories. Identifying high-rate code families with this subset-cyclic logical structure, and characterizing their injection, post-selection, and decoding behavior, is a promising direction for scaling high-rate STAR architectures speed.

\begin{acknowledgments}
  We thank Qian Xu for the insightful discussion at the early stage of this project.
  We acknowledge stimulating discussions with Andrew Sornborger, Casey Duckering, Shayan Majidy, and Samuel Tan.
  We thank Mikhail Lukin for his comments on the draft of this manuscript. This work was supported under the Entangled Logical Qubits program (Cooperative Agreement Number W911NF-23-2-0219), the DARPA MeasQuIT program (HR0011-24-9-0359), and the Innovate UK grant (10179725) as part of the QNABLE project. The numerical studies were performed on the high-performance computing system Perlmutter, a NERSC resource, using NERSC award DDR-ERCAP0030190.
\end{acknowledgments}

\bibliography{ref}
\input{appendix}

\end{document}

%% file: appendix.tex
\appendix
\section{Logical Compilation for Fermionic Hamiltonian Simulation via local encoding}
\label{app:fermion-new}

In Sec.~\ref{sec:gadgets} of the main text, we described how the local spin lattice Hamiltonian simulation is implemented in the bicycle chain STAR architecture. In this appendix, we present how to generalize it to fermionic lattice models.

Standard Jordan--Wigner (JW) encoding maps $N$ fermionic modes to $N$
qubits but represents each hopping term by a Pauli string whose length grows
with the JW ordering, up to $\mathcal{O}(N)$. These long strings are
geometrically non-local and incompatible with our transversal gate set.
Geometrically local fermion-to-qubit mappings instead trade a constant qubit
overhead for $\mathcal{O}(1)$-weight operators that respect the planar
lattice~\cite{derby2021compact,jafarizadeh2024recipelocal}.

We adopt the Derby--Klassen compact encoding~\cite{derby2021compact}. It
places one vertex qubit at each site (one per fermionic mode) and one
face (ancilla) qubit on the odd faces of the square lattice in a
checkerboard pattern, for an asymptotic overhead of $1.5$ qubits per mode: an
$L\times L$ lattice uses $L^2$ vertex and $\tfrac12 L^2$ face qubits~\cite{derby2021compact}. The encoded number
operator is the weight-one $\tilde n_j = \tfrac12(1 - Z_j)$, and each encoded
hopping operator is a Pauli string of weight at most three, supported on the two
vertex qubits of a bond and the single intervening face qubit; vertical and
horizontal bonds differ only in the face-qubit Pauli, of the form
$X_i Y_j X_f$ and $X_i Y_j Y_f$ respectively~\cite{derby2021compact}.

\begin{figure}[t]
  \centering
  \includegraphics[width=1.0\linewidth]{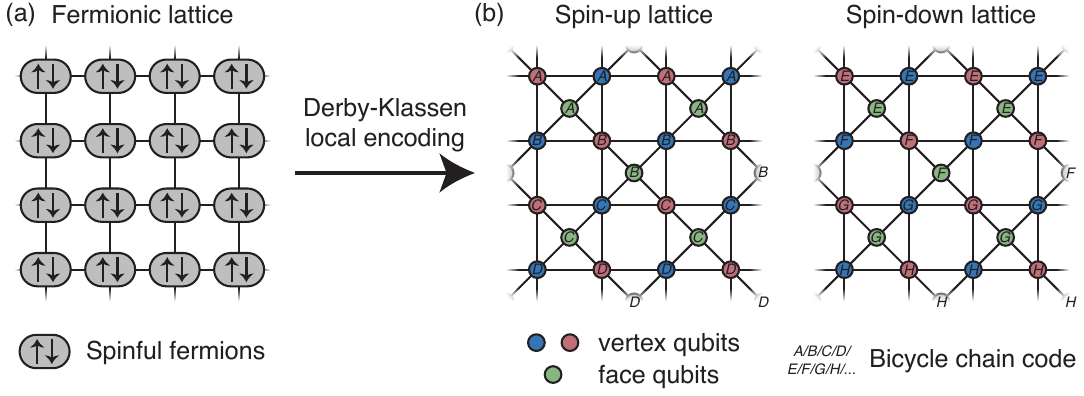}
  \caption{
    \textbf{Fermionic compilation via the Derby--Klassen local
    encoding.}
    (a)~\emph{Fermionic lattice.}
    A $4\times 4$ square lattice of spinful fermionic modes with nearest-neighbor hopping bonds.
    (b)~\emph{Spin lattice after compact encoding.}
    The Derby--Klassen mapping adds an ancilla qubit on every odd face in a
    checkerboard pattern; these ancilla qubits absorb the fermionic exchange
    statistics so that all encoded hopping operators remain weight-three and
    geometrically local. Different colors and letters indicate the different bicycle chain code blocks encoding all logical qubits on the full lattices. For example, all qubits labeled as $A$ in green belong to a half of one bicycle chain code block.
  }
  \label{fig:fermion_new}
\end{figure}

For the spinful Fermi--Hubbard model on an $L\times L$ lattice,
\begin{equation}
  H_\mathrm{FH} = -t\!\!\sum_{\langle i,j\rangle,\sigma}
  \bigl(c^\dagger_{i\sigma} c_{j\sigma} + \mathrm{h.c.}\bigr)
  + U\sum_i n_{i\uparrow} n_{i\downarrow},
  \label{eq:FH_new}
\end{equation}
we apply one
Derby--Klassen layer per spin species, following the explicit recipe of
Ref.~\cite{jafarizadeh2024recipelocal}. Each $L\times L$ lattice then carries
$2L^2$ modes encoded into $\sim\!3L^2$ qubits --- three per site: a spin-$\uparrow$
and a spin-$\downarrow$ vertex qubit plus, amortized over the checkerboard, one
face qubit. The kinetic term factors into weight-three rotations
\begin{align}
  H_V &= -\tfrac{t}{2}\bigl(X_R X_B X_G + Y_R X_B Y_G\bigr),
  \label{eq:fh_hop_V_new} \\
  H_H &= -\tfrac{t}{2}\bigl(X_R Y_B X_G + Y_R Y_B Y_G\bigr),
  \label{eq:fh_hop_H_new}
\end{align}
for vertical ($V$) and horizontal ($H$) bonds, where $R, B$ are the
vertex qubits and $G$ the face qubits (Figure~\ref{fig:fermion_new}(b)), while the on-site Coulomb
term is the weight-two
\begin{equation}
  H_I = \tfrac{U}{4}\bigl(Z_\uparrow Z_\downarrow - Z_\uparrow - Z_\downarrow\bigr),
  \label{eq:fh_onsite_new}
\end{equation}
acting on the two vertex qubits of a site (cf.\ Appendix~\ref{app:trotter},
where these terms drive the Childs--Su error bound).

Compilation of the encoded Hamiltonian requires no primitive beyond those
established for spin models. A weight-three Pauli rotation
$e^{-i\frac{\theta}{2} P_1 P_2 P_3}$ is conjugated to a $Z_R Z_B Z_G$ rotation by the
single-qubit Cliffords that rotate each factor into the $Z$ basis ($H$ for $X$,
$SH$ for $Y$), then synthesized by a CNOT ladder that computes the three-qubit
parity onto one qubit, applies a single $R_Z$, and uncomputes. The on-site
$Z_\uparrow Z_\downarrow$ term is the two-body rotation, and the linear $Z_\uparrow$, $Z_\downarrow$ pieces
are single-logical $R_Z$ rotations.

As shown in Fig.~\ref{fig:fermion_new}, the encoded graph remains translation-invariant: the vertex and face qubits
each tile the torus periodically, so the lattice translation
$\ZZ_L\times\ZZ_L$ still acts as a shift and is matched to the bicycle chain
shift automorphism exactly as in the spin case. Assigning the vertex qubits of
one row in each color, together with their face qubits, to the disjoint logical qubits of a code
block places every weight-three bond rotation either within one block
(intra-row bonds and their face qubit) or across two adjacent blocks (inter-row
bonds), both reachable by the native transversal CNOT and shift gadgets. The
only structural changes relative to the spin case are the constant $1.5\times$
qubit overhead and the appearance of weight-three rather than weight-two
rotations; neither requires a new logical gadget, so the Fermi--Hubbard model
falls within the same compilable class as the spin models of
Sec~\ref{sec:gadgets}.

\section{Bicycle chain codes}
\label{app:bicycle-chain-codes}

This appendix supplements the bicycle chain code construction of
Section~\ref{sec:bicycle_chain_code} with the formal details needed to
reproduce the code instances and their syndrome extraction circuit. The family,
its four co-design gadgets, and the parameter summary
(Table~\ref{tab:bicycle-chain-instances}) are given in the main text and are not repeated
here.

\subsection{Construction and parameters}
\label{app:bicycle-chain-construction}

Working in the group algebra $\ZZ_2[\ZZ_\ell \times \ZZ_m]$ of bivariate
polynomials modulo $x^\ell - 1$ and $y^m - 1$, the bivariate bicycle (BB) code
of two low-weight polynomials $a(x,y)$, $b(x,y)$ has parity-check matrices
$H_X = [\,A \;\; B\,]$ and $H_Z = [\,B^\top \;\; A^\top\,]$, where $A$ and $B$
are the block-circulant matrices of $a$ and $b$~\cite{bravyi2024highthresholdlowoverhead}.
The $n = 2\ell m$ data qubits split into a left half ($L$, the first
$\ell m$ columns) and a right half ($R$); both are indexed by a cell
$(i, j)$ of the $\ell \times m$ torus, and a monomial $x^p y^q$ acts on the data
qubits as the cyclic shift $(i, j) \mapsto (i + p,\, j + q)$.

We take the self-dual instances $b = a^\dagger$, where the group-algebra
adjoint sends $x^i y^j \mapsto x^{-i} y^{-j}$; equivalently $B = A^\top$, which
makes $H_X = H_Z$~\cite{liang2025selfdualbivariate}. Throughout this work, we fix the
weight-four polynomial
\begin{equation}
  a(x,y) = 1 + y^{3} + x\,y^{2} + x\,y^{4},
  \label{eq:qian_poly}
\end{equation}
so that $b(x,y) = 1 + y^{-3} + x^{-1} y^{-2} + x^{-1} y^{-4}$. Every stabilizer
generator then has weight $\mathrm{wt}(a) + \mathrm{wt}(b) = 8$ (four supports on
$L$, four on $R$); since $8 \equiv 0 \pmod 4$ the code is doubly even, the
property invoked for transversal $S$ below.

The construction decouples the two code parameters. The distance $d$ is set by
$a$ together with the circumference $m$, while the encoded count $k = 2\ell$ is
set by the chain length $\ell$, with $n = 2\ell m$. The chain is therefore
lengthened---adding logical qubits to match a larger target lattice---by
extending the torus in the $x$ direction without changing $d$. At $\ell = 3$, the
instances of Table~\ref{tab:bicycle-chain-instances} are the explicit codes
$\qcode{42,6,6}$ ($m=7$), $\qcode{66,6,8}$ ($m=11$), and $\qcode{78,6,10}$
($m=13$), matching Table~I of Ref.~\cite{xu2025batchedhighrate}. The native logical gate set follows from the code structure as summarized in
Section~\ref{sec:bicycle_chain_code}: transversal CNOT between blocks (any CSS
code), transversal Hadamard (self-duality, $H_X = H_Z$), and transversal phase
$S$ (self-duality together with the doubly-even property established above).

\subsection{Syndrome extraction}
\label{app:syndrome-extraction}
Each syndrome cycle measures all $\ell m$ $X$-checks and all $\ell m$ $Z$-checks
using $n = 2\ell m$ ancillas, one per check. Since every check has weight eight,
each ancilla executes eight CNOTs, and the shortest cycle that does so would have depth
eight. We obtain such optimal-depth schedule by the search methodology of
Ref.~\cite{bravyi2024highthresholdlowoverhead}, which realizes a depth-seven
cycle for their weight-six codes; we apply the same procedure to our
weight-eight checks and find depth-8 cycle. In brief, one enumerates the candidate CNOT orderings
permitted by the schedule grammar and keeps those in which every ancilla cleanly
measures its intended weight-eight stabilizer with no residual back-action on the
data block (verified by symbolic propagation over $\ZZ_2$); we then rank the
survivors by their circuit-level fault distance, the minimum number of
independent circuit faults producing an undetected logical error. We refer the reader to
Ref.~\cite{bravyi2024highthresholdlowoverhead} for details on the search methodology.
For the $d = 6$ code, the correctness filter leaves $212$ schedules. Ranked by
circuit-level fault distance, $192$ achieve $d_\mathrm{fault} = 4$ and $20$
achieve $d_\mathrm{fault} = 5$. We use one of these $20$ distance-optimal schedules throughout this
work; we list our choice explicitly in
Table~\ref{tab:se_schedule}. Because the grammar depends only on the check weights
$\mathrm{wt}(a) = \mathrm{wt}(b) = 4$ and not on the specific monomials, the same
schedule applies unchanged across all chain lengths $\ell$ and circumferences $m$
in Table~\ref{tab:bicycle-chain-instances}.

\paragraph{Explicit schedule.}
Table~\ref{tab:se_schedule} gives the chosen order schedule. We label the four
monomials of $a$ by $A_1,\dots,A_4$ and those of $b$ by
$B_1, \dots, B_4$, with the dictionary
$A_1 = x y^4,\ A_2 = x y^2,\ A_3 = y^3,\ A_4 = 1$. Each of the eight rounds is
one CNOT layer. In round $t$, every $X$-ancilla applies a CNOT (as control) onto
the data qubit of the listed half ($L$ or $R$) displaced by the monomial $p_X$,
while every $Z$-ancilla simultaneously receives a CNOT (as target) from the
opposite-half data qubit displaced by the transposed monomial $p_Z^\top$. Since the two ancilla
types always act on opposite data halves, both syndrome types extract
concurrently within the same eight layers. Ancillas are initialized
($X$-ancillas in $\ket{+}$, $Z$-ancillas in $\ket{0}$) before round~1 and
measured after round~8.

\begin{table}[t]
  \centering
  \caption{\textbf{Depth-eight syndrome extraction schedule.}
    Each row is one CNOT layer. The ``$X$-check'' column gives the data half and the
    monomial of the $X$-ancilla coupling; the ``$Z$-check'' column gives the
    opposite half and the transposed monomial of the simultaneous $Z$-ancilla
    coupling. $A_i$ and $B_i$ denote the monomials of $a$ and $b$;
  the schedule is polynomial-independent given the weight-four supports.}
  \label{tab:se_schedule}
  \begin{ruledtabular}
    \begin{tabular}{ccc}
      Round & $X$-check (half, monomial) & $Z$-check (half, monomial) \\
      \midrule
      1 & $L$, $A_1$ & $R$, $A_4^\top$ \\
      2 & $R$, $B_1$ & $L$, $B_4^\top$ \\
      3 & $L$, $A_2$ & $R$, $A_3^\top$ \\
      4 & $R$, $B_2$ & $L$, $B_2^\top$ \\
      5 & $R$, $B_3$ & $L$, $B_3^\top$ \\
      6 & $L$, $A_3$ & $R$, $A_2^\top$ \\
      7 & $R$, $B_4$ & $L$, $B_1^\top$ \\
      8 & $L$, $A_4$ & $R$, $A_1^\top$ \\
    \end{tabular}
  \end{ruledtabular}
\end{table}

\subsection{Logical error rate scaling}
\label{app:bicycle-chain-ler}

\begin{figure}[t]
  \centering
  \includegraphics[width=\columnwidth]{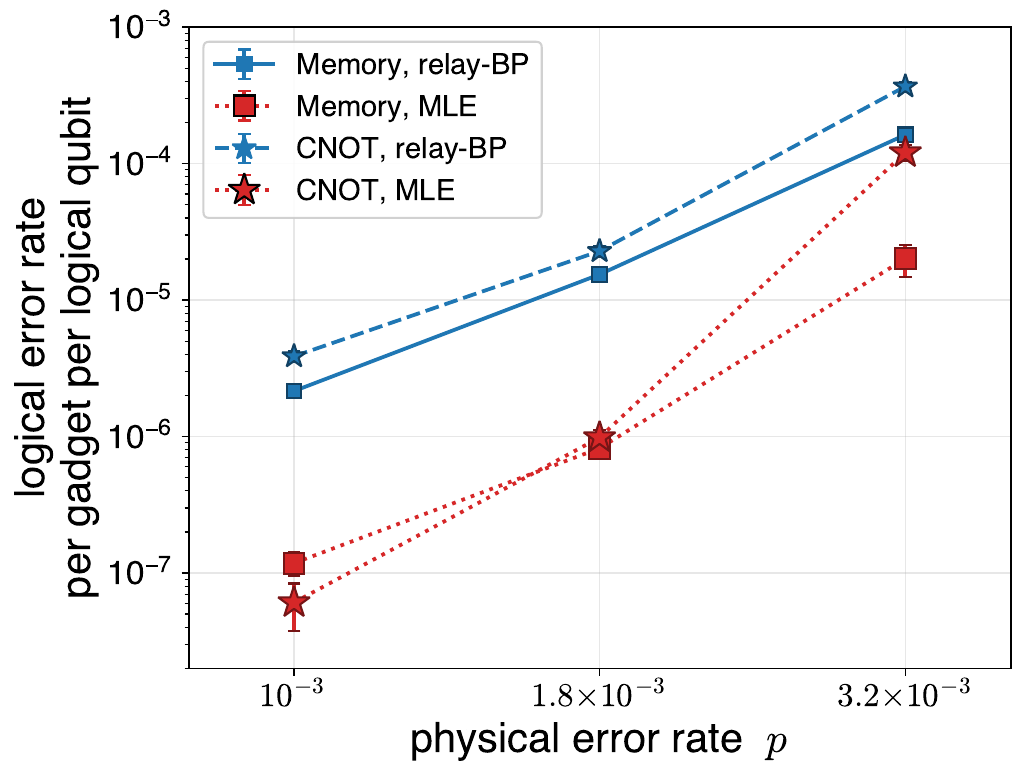}
  \caption{
    \textbf{Logical error rate scaling with the physical error rate.}
    Logical error rate per gadget per logical qubit, $\mathrm{LER}/(N r)$
    ($N = k$ logical observables for the memory experiment and $N = 2k$ for the
    transversal CNOT; $r = 18$ syndrome extraction rounds), versus the
    physical error rate $p$ at fixed $k = 8$ ($\ell = 4$, $m = 7$). Memory
    (squares) and transversal CNOT (stars) are decoded by relay-BP (blue) and
  the MLE decoder (red); error bars are $1\sigma$ binomial.}
  \label{fig:ler_vs_p}
\end{figure}

Figure~\ref{fig:ler_vs_p} shows how the logical error rate per gadget per
logical qubit scales with the physical error rate $p$ at fixed $k = 8$
($\ell = 4$), complementing the $k$-dependence reported in the main text
[Fig.~\ref{fig:numerics}(a)]. For both the memory and transversal CNOT circuits, the MLE decoder performs up to two orders of magnitude better than
relay-BP~\cite{muller2025improvedbelief} at every $p$.

\section{Small-Angle Rotations via STAR-based gate teleportation on qLDPC codes}
\label{sec:magic}

\begin{figure}[t]
  \centering
  \begin{quantikz}[row sep={1.1cm,between origins}, column sep=0.7cm]
    \lstick{$\ket{\psi}_L^{\mathrm{data}}\;/^{k}$}
    & \targ{}  & \meter{Z^{\otimes k}} \vcw{1}
    & \rstick{$\vec m\in\{0,1\}^{k}$} \\
    \lstick{$\ket{m_{\vec\theta}}_L\;/^{k}$}
    & \ctrl{-1} &  \gate{\hat X_L^{\vec m}}
    & \rstick{$\bigotimes_{j} \hat R_{z,L}\!\bigl((-1)^{m_j}\theta_j\bigr)\ket{\psi}_L$}
  \end{quantikz}
  \caption{\textbf{Rotation gate teleportation for a qLDPC
    block.} The ancilla block holds the resource state
    $\ket{m_{\vec\theta}}_L = \bigotimes_{j=1}^{k}\ket{m_{\theta_j}}_L$
    (control) and the data block holds $\ket{\psi}_L$ (target); the two
    are coupled by a transversal CNOT. A transversal $\hat Z$ measurement
    on the data block yields one outcome $m_j$ per logical qubit. Conditional on
    $\vec m$, a feed-forward logical correction $\hat X_L^{\vec m} = \bigotimes_j
    \hat X_{L,j}^{m_j}$ is applied on the ancilla. This gadget teleports the     rotated state $\bigotimes_j \hat R_{z,L}((-1)^{m_j}\theta_j)  \ket{\psi}_L$ onto the ancilla block. Outcomes $m_j = 1$ rotate by the wrong sign and
  are corrected by re-teleporting at $2\theta_j$ ~\cite{toshio2024practicalquantum}.}
  \label{fig:rotation_teleportation}
\end{figure}
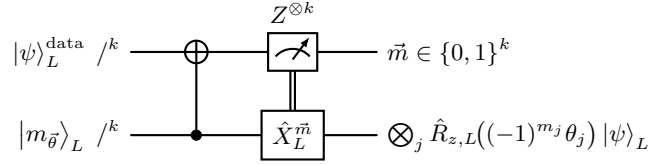

We implement a logical rotation gate $\hat R_{z,L}(\theta) = e^{-i\theta \hat Z_L/2}$, $|\theta| \ll \pi/2$, on logical qubits encoded in a qLDPC block using gate teleportation \cite{Gottesman_1999_gate_teleportation}. In this approach, the desired gate is applied indirectly: one first prepares an appropriate logical resource state on an ancilla, entangles the ancilla with the data block, measures, and then applies a feed-forward correction conditioned on the measurement outcome (Fig.~\ref{fig:rotation_teleportation}).

For the rotation gate, the required resource state is $\ket{m_{\theta}}_L  \;\equiv\; \hat R_{z,L}(\theta)\,\ket{+}_L    = \cos(\theta/2) \ket{+}_L - i\sin(\theta/2) \ket{-}_L$. In our setting, the ancilla is itself a qLDPC code patch containing multiple logical qubits, so we prepare the corresponding product resource state across all $k$ logical qubits:
\begin{equation}
  \ket{m_{\vec\theta}}_L \;\equiv\; \bigotimes_{j=1}^{k}
  \ket{m_{\theta_j}}_L.
  \label{eq:resource_block}
\end{equation}
The angles $\theta_j$ are chosen independently; setting $\theta_j = 0$ leaves the $j$-th logical qubit unrotated and effectively idles it through the gadget.

The teleportation gadget (Fig.~\ref{fig:rotation_teleportation}) enacts $\bigotimes_j \hat R_{z,L}\!\bigl(
(-1)^{m_j}\theta_j\bigr)$ on the teleported state, where $m_j \in \{0,1\}$ is the transversal-$\hat Z$ outcome on the support
of $j$th logical qubit. Wrong outcomes ($m_j = 1$) differ from the target by $\hat R_{z,L}(-2\theta_j)$. They are therefore corrected by another teleportation step using an angle $2 \theta_j$. Repeating this procedure gives the repeat-until-success (RUS) $\theta_j \to 2\theta_j \to 4\theta_j \to \cdots$ on each logical independently.

With $k$ RUS ladders running in parallel, one might worry that the
block-level process stalls until the unluckiest ladder
succeeds, with this straggler eating through the advantage of
teleporting all $k$ logical qubits using a single $\ket{m_{\vec\theta}}_L$
injection. Geometric suppression prevents this. A single
ladder survives past round $r$ only with probability $2^{-r}$, so the
slowest of $k$ ladders lands where this tail thins to $\sim\!1/k$,
i.e.\ where $2^{-r}\approx 1/k$, giving a worst-case depth of only
$\sim\!\log_2 k$.

The block-level expected depth, the round at which every
logical qubit has terminated, $R_{\max}=\max_j R_j$, is therefore
\begin{equation}
  \begin{split}
    \bar R_{\max}(k) &\;\equiv\; \mathbb{E}[R_{\max}]
    \;=\; \sum_{r=0}^{\infty}\bigl[1 - (1 - 2^{-r})^{k}\bigr] \\
    &\;=\; \sum_{j=1}^{k}\binom{k}{j}\,\frac{(-1)^{j+1}}{1 - 2^{-j}} \\
    &\;\xrightarrow{k \to \infty}\;
    \log_2 k \;+\; \frac{\gamma}{\ln 2} \;+\; \frac{1}{2}
    \;+\; o(1),
  \end{split}
  \label{eq:rmax_mean}
\end{equation}
growing only as $\log_2 k$. At the block size $k=8$ used here,
Eq.~\eqref{eq:rmax_mean} gives $\bar R_{\max}(8)\approx 4.42$---about
twice the per-logical mean $\mathbb{E}[R_j]=2$, not eight times it,
since most logical qubits terminate early and only the slowest tail fixes
$R_{\max}$.

\subsection{Resource state preparation}
\label{sec:resource_prep}
For the preparation of the resource state, we leverage the transversal
multi-rotation (TMR) protocol of Ref.~\cite{toshio2024practicalquantum}. The
TMR protocol synthesizes a logical $Z$-rotation out of physical $Z$-rotations
applied to subsets of the logical support. We partition the support of
$\hat{Z}_L$ into $M$ disjoint subsets $\{c_j\}$, writing $\hat{Z}_L = \prod_{j} Z_{c_j}$ with
$Z_{c_j} \equiv \prod_{i \in c_j} Z_i$, and on each subset we apply the joint rotation
$e^{-i\theta_{c_j} Z_{c_j}/2}$. The partition is otherwise free, subject only to the condition that no $Z_c$, nor any product of
them, be a stabilizer.  Different choices of partition trade post-selection overhead against achievable fidelity, with the best choice also depending on the target angle, a tradeoff we examine later in this section. The angles $\theta_c$ are not free individually: they must be chosen so that the
protocol realizes the target logical rotation by $\theta$. Many assignments meet
this requirement, and we specialize to the symmetric one, taking a common angle
$\theta_c = \theta^\ast$ across all $M$ subsets; as shown below, $\theta^\ast$ is
then fixed by $\theta$ and $M$.

Acting on $\ket{+}_L$, each factor expands as
$e^{-i\theta^\ast Z_c/2} = \cos(\theta^\ast/2) - i\sin(\theta^\ast/2)\,Z_c$, so the full
product is a sum of $2^{M}$ terms. A proper, nonempty product of the $Z_c$ anticommutes with at least one
stabilizer and therefore carries a nontrivial syndrome; only the identity term
and the full product $\prod_c Z_c = \hat{Z}_L$ are syndrome-free.
Post-selecting on the trivial syndrome of a subsequent syndrome extraction round
keeps just these two and, using $\hat{Z}_L\ket{+}_L = \ket{-}_L$, leaves the
(unnormalized) logical state
\begin{equation}
  \cos^{M}\!(\theta^\ast/2)\,\ket{+}_L
  + (-i)^{M}\sin^{M}\!(\theta^\ast/2)\,\ket{-}_L .
  \label{eq:tmr_postselected_state}
\end{equation}
Comparing Eq.~\eqref{eq:tmr_postselected_state} with the target
$\hat{R}_{z,L}(\theta)\ket{+}_L = \cos(\theta/2)\,\ket{+}_L - i\sin(\theta/2)\,\ket{-}_L$
fixes the per-subset angle. With $j \equiv \lfloor M/2 \rfloor$, the prepared
state is, up to normalization, the resource state $\ket{m_{(-1)^{j}\theta}}_L$
for odd $M$ and the dressed state
$\hat{R}_{x,L}(\pi/2)\,\ket{m_{(-1)^{j}\theta}}_L$ for even $M$, reproducing
Eq.~(5) of Ref.~\cite{toshio2024practicalquantum}. The angle magnitudes obey
\begin{equation}
  \lvert\tan(\theta/2)\rvert = \tan^{M}\!(\theta^\ast/2), \qquad \theta^\ast/2 \approx (\theta/2)^{1/M}
  \label{eq:tmr_angle_relation}
\end{equation}
For even $M$ the phase $(-i)^{M}$ is real, so the prepared state additionally
carries a fixed logical Clifford $\propto \sqrt{X}_L$, independent of the angle.
Because the self-dual codes used here admit a global transversal
$\sqrt{X}$ acting on all logical qubits of a block, this residue is removed by a single
transversal application (of the appropriate chirality, or equivalently by
preparing with a sign-adjusted $\theta$): every rotated logical qubit carries the same
Clifford and is thereby corrected, while every idle logical qubit sits in
$\ket{+}_L$, a $+1$ eigenstate of $\hat{X}_L$, and is left invariant up to a
global phase. Both parities are thus usable, the only requirement being that all
rotated logical qubits in a block share the parity of $M$.

In the noise-free case, the post-selection succeeds with probability
\begin{equation}
  \begin{split}
    p_\mathrm{TMR}(\theta; M) &=
    \bigl[\sin^{2/M}(\theta/2) + \cos^{2/M}(\theta/2)\bigr]^{-M} \\
    &\simeq 1 - M\, (\theta/2)^{2/M} + \mathcal{O}\!\big(\theta^4\big).
    \label{eq:tmr_pideal}
  \end{split}
\end{equation}
While the prepared state infidelity of this protocol (in the noisy case) scales as $\theta^{\,2(1 - 1/M)}\,p$~\cite{toshio2024practicalquantum}. We numerically study how these two quantities change for BB codes under a standard depolarizing error model in the subsequent discussion.

Because the self-dual BB codes considered here admit a logical basis in which each logical operator is supported on a disjoint subset of qubits, the TMR protocol can be applied to all $k$ logical qubits simultaneously, preparing the full block resource state $\ket{m_{\vec\theta}}_L$ of Eq.~\eqref{eq:resource_block}. As we show below, the resulting logical infidelities are mutually independent, and the post-selection rate factors into a part that is independent of the number of rotated logical qubits---and hence amortized across the block---and a part that grows with it. In practice, $\ket{+}_L$ is itself prepared by transversally initializing the code qubits in $\ket{+}$ and then performing one round of $Z$ syndrome extraction. A Pauli-$X$ correction conditioned on the measured $Z$ syndromes projects the state back into the code space, resetting all $Z$ syndromes to $0$. This correction requires no decoding: any two corrections differing by $\hat X_L$ act identically on $\ket{+}_L$, since $\hat X_L\ket{+}_L = \ket{+}_L$.

\subsection{Fidelity estimation method}

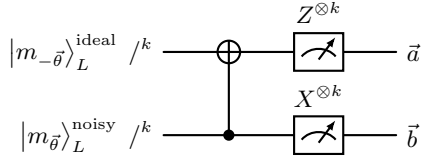
\begin{figure}[t]
  \centering
  \begin{quantikz}[row sep={1.1cm,between origins}, column sep=0.7cm]
    \lstick{$\ket{m_{-\vec\theta}}_L^{\mathrm{ideal}}\;/^{k}$}
    & \targ{}   & \meter{Z^{\otimes k}} & \rstick{$\vec a$} \\
    \lstick{$\ket{m_{\vec\theta}}_L^{\mathrm{noisy}}\;/^{k}$}
    & \ctrl{-1} & \meter{X^{\otimes k}} & \rstick{$\vec b$}
  \end{quantikz}
  \caption{\textbf{Teleportation-based fidelity estimator.} The noisily
    prepared resource state $\ket{m_{\vec\theta}}_L$ (control) is entangled
    with an ideally prepared inverse-rotation state $\ket{m_{-\vec\theta}}_L$
    (target) by a noise-free transversal CNOT, followed by noise-free
    transversal $Z$ and $X$ measurements.
    From these, an MLE decoder~\cite{gurobi} infers the logical eigenvalues
    $\hat Z_L$ and $\hat X_L$ up to correctable errors. The teleported rotation
    has random sign: the outcome $\hat Z_L=+1$ applies $+\vec\theta$, canceling
    the inverse rotation so that an ideal resource gives $\hat X_L=+1$, while
    $\hat Z_L=-1$ applies $-\vec\theta$ and is discarded. Restricted to
    $\hat Z_L=+1$ shots, the fraction with $\hat X_L=-1$ estimates the
  infidelity of the prepared state.}
  \label{fig:fidelity_estimation}
\end{figure}

Because the preparation protocol contains non-Clifford gates, standard Clifford simulators do not apply; we simulate it end-to-end using two recently introduced simulators~\cite{haenel2026tsim, chase2026clifft}. To estimate the fidelity of the prepared resource state, we reuse the teleportation gadget of Fig.~\ref{fig:rotation_teleportation}, now with the data block prepared ideally (noise-free) in the inverse-rotation state $\ket{m_{-\vec\theta}}_L$ (Fig.~\ref{fig:fidelity_estimation}). The noisy resource is coupled to this ideal inverse, and a transversal $Z$ measurement teleports the result. The sign of the applied rotation is random: per logical qubit, the outcome $\hat Z_{L,j}=+1$ (half the shots) applies $+\theta_j$, so the two rotations cancel and the teleported logical qubit is ideally in $\ket{+}_L$, whereas $\hat Z_{L,j}=-1$ applies $-\theta_j$ and is discarded. Restricted to $\hat Z_{L,j}=+1$, a transversal $X$ measurement returns $+1$ deterministically for an ideal resource, so the fraction of these shots returning $-1$ is an unbiased estimator of the prepared-state infidelity. The raw transversal $Z$ and $X$ outcomes are decoded jointly with a most-likely error (MLE) decoder (implemented with Gurobi~\cite{gurobi}), so the estimate is taken up to decoder-correctable errors.

Although this estimator may look unnecessarily indirect, we adopt it for two practical reasons. The most naive alternative would prepare the (noisy) state, apply a noise-free round of syndrome extraction, perform an ideal noise-free inverse rotation, and then measure. Making this fault-tolerant---i.e., recovering the prepared state up to correctable errors before the inverse rotation---requires live decoding of the noise-free syndrome extraction round together with a feed-forward correction conditioned on it. The simulators we use do not support live decoding with mid-circuit feed-forward, so we instead defer all correction to the offline MLE decoding of the final transversal measurements in our chosen fidelity estimation gadget. A second alternative, full state tomography of the prepared resource, requires far more shots than our single-observable estimator to reach a given relative uncertainty.

\subsection{Characterizing resource-state preparation}
\label{app:tmr_characterization}

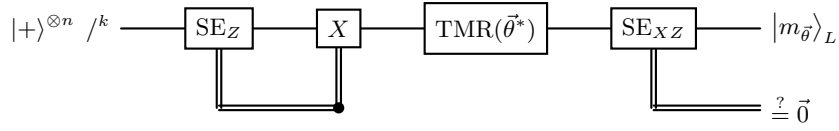
\begin{figure*}[htbp]
  \centering
  \begin{quantikz}[row sep={1.05cm,between origins}, column sep=0.85cm]
    \lstick{$\ket{+}^{\otimes n}\;/^{k}$}
    & \gate{\mathrm{SE}_{Z}}\vcw{1}
    & \gate{X}
    & \gate{\mathrm{TMR}(\vec\theta^{\ast})}
    & \gate{\mathrm{SE}_{XZ}}\vcw{1}
    & \rstick{$\ket{m_{\vec\theta}}_L$} \\
    \setwiretype{n}
    &
    & \control{}\vcw{-1}\cw
    &
    &
    & \rstick{$\stackrel{?}{=}\vec 0$}\cw
  \end{quantikz}
  \caption{\textbf{Parallel TMR resource-state preparation.}
    The $n$ physical qubits of a $k$-logical block are initialized
    transversally in $\ket{+}$. A first $Z$-stabilizer syndrome extraction
    round used only for a classical Pauli-$X$ correction (a
    frame update requiring no decoding); it does \emph{not} contribute to
    post-selection. The transversal multi-rotation
    $\mathrm{TMR}(\vec\theta^{\ast})=\prod_c e^{-i\theta^{\ast}Z_c/2}$ is then
    applied across all rotated logical qubits, followed by a full ($X$ and $Z$)
    syndrome extraction round. Post-selecting on the
    trivial syndrome $s_{XZ}=\vec 0$ leaves the block resource state
  $\ket{m_{\vec\theta}}_L$.}
  \label{fig:state_prep}
\end{figure*}

Using the teleportation-based fidelity estimator of Fig.~\ref{fig:fidelity_estimation}, we
characterize the parallel TMR preparation. We adopt a standard depolarizing noise model for our numerics, with a physical error rate of $p=10^{-3}$. Before sweeping parameters, we fix
the syndrome extraction round structure of the preparation protocol, shown in
Fig.~\ref{fig:state_prep}. Numerically, the schedule that maximizes the
post-selection rate without sacrificing fidelity is: (i) initialize all
physical qubits transversally in $\ket{+}$; (ii) perform one (half) round of
$Z$-stabilizer syndrome extraction; (iii) apply the TMR; and (iv) perform one round of full ($X$ and $Z$) syndrome extraction.
The initial $Z$ round (ii) serves only to project back into the code space through a classical Pauli-$X$ correction. This is a frame update: it requires no decoding and could be implemented in practice through classical control feedback that changes the signs of the physical angles used in TMR. It does not contribute to post-selection. Post-selection is enforced solely on the trivial syndrome of the final round (iv). We adopt this schedule for all numerical experiments reported below.

With the schedule of Fig.~\ref{fig:state_prep} fixed, we characterize the
parallel TMR preparation through six sweeps that vary, in turn, the rotation
angle $\theta$, the partition count $M$, the number of simultaneously rotated
logical qubits $N$ (out of the $k$ logical qubits in the block), the torus dimensions
$(\ell, m)$, the atom-loss fraction $f$, and the physical error rate $p$
itself (Sec.~\ref{app:p}). Unless noted, all sweeps fix $m = 7$ and, except
in the $p$-sweep, $p = 10^{-3}$. We organize the characterization around a
closed-form factorization for the prepared state infidelity and for the
post-selection success rate. The forms below are motivated by simple
structural considerations and the few constants in them are fixed by fitting to
the numerics. We first state the formulas, then pin down one factor at a time,
showing for each swept parameter that both the infidelity and the success rate
follow the predicted dependence.

\paragraph{Closed-form factorization.}
We find that the success rate can be written as
\begin{equation}
  s_\mathrm{total} = p_\mathrm{init}\,
  p_\mathrm{TMR}(\theta; M)^{N}\, p_\mathrm{loss},
  \label{eq:s_total}
\end{equation}
with
\begin{equation}
  \begin{aligned}
    p_\mathrm{TMR}(\theta; M) &=
    \bigl[\sin^{2/M}(\theta/2) + \cos^{2/M}(\theta/2)\bigr]^{-M}, \\
    p_\mathrm{init} &\approx
    \exp\!\bigl[-\gamma\, N_\mathrm{loc}(\ell, m)\,(1-f)\,p\bigr], \\
    p_\mathrm{loss} &=
    \exp\!\bigl[-f\, N_\mathrm{loc}(\ell, m)\,p\bigr],
  \end{aligned}
  \label{eq:s_total_factors}
\end{equation}
and a per-logical infidelity that collapses, across all partition
counts, onto a single theory-constrained family
\begin{equation}
  \varepsilon(\theta, f) = C(M)\,p\,(1-f)\,\theta^{\,2(1 - 1/M)},
  \qquad
  \varepsilon_\mathrm{any} = 1 - (1 - \varepsilon)^{N},
  \label{eq:epsilon_decomposition}
\end{equation}
whose exponent $2(1 - 1/M)$ is fixed by the TMR partition count $M$
(Ref.~\cite{toshio2024practicalquantum}), leaving a coefficient $C(M)$ that is given by
\begin{equation}
  \begin{array}{c|ccccccc}
    M    & 1    & 2    & 3            & 4            & 5            & 6            & 7 \\ \hline
    C(M) & 0.47 & 0.50 & 0.68         & 1.66         & 2.95         & 5.28         & 6.79 \\[2pt]
  \end{array}.
  \label{eq:C_table}
\end{equation}
$N_\mathrm{loc}(\ell, m) = 32\ell m + 13$ is the per-shot noise-location count,
and $\gamma = 0.93$ is the fraction of physical Pauli events that trigger a
state-prep post-selection detector. The $C(M)$ are the only infidelity
constants; they are extracted in Sec.~\ref{app:M} by fitting
Eq.~\eqref{eq:epsilon_decomposition}. The
$M = 1$     infidelity is the flat floor
$\varepsilon = C(1)\,p\,(1-f) \approx 4.7\times10^{-4}\,(1-f)$ at $p = 10^{-3}$.

Two features of this factorization fix the order of the discussion that follows.
First, the per-logical infidelity $\varepsilon$ in
Eq.~\eqref{eq:epsilon_decomposition} depends only on the rotation angle $\theta$
and the partition count $M$; it carries no dependence on the code dimensions
$(\ell, m)$ or on the number of rotated logical qubits $N$. The only $N$-dependence of
the error is collective, through the any-logical rate
$\varepsilon_\mathrm{any} = 1 - (1-\varepsilon)^N$, which is the statement that
errors on distinct rotated logical qubits are independent to leading order. Second,
the success rate $s_\mathrm{total}$ separates into three pieces with clean
operational meanings:
\begin{itemize}
  \item $p_\mathrm{TMR}(\theta; M)$ is the ideal TMR acceptance---the
    fraction of weight the partitioned injection leaves inside the code space at
    $p = 0$. It is the $p \to 0$ limit of $s_\mathrm{total}$
    (where $p_\mathrm{init}, p_\mathrm{loss} \to 1$), and it is raised to the
    $N$th power because each of the $N$ rotated logical qubits contributes an
    independent acceptance factor.
  \item $p_\mathrm{init}$ is the state-prep PS survival of the logical
    $\ket{+}_L$ block. It is the $\theta \to 0$ limit of the acceptance
    (where $p_\mathrm{TMR} \to 1$), i.e.\ the cost of preparing and projecting the
    encoded plus state independent of the rotation.
  \item $p_\mathrm{loss}$ is the heralded-loss survival, present when the
    physical noise budget contains a detectable (erasure) component of weight
    $f$, discussed in Sec.~\ref{app:loss}.
\end{itemize}
We now verify each factor in turn.

\paragraph{Verification protocol.}
The infidelity is read out with the teleportation-based estimator
of Fig.~\ref{fig:fidelity_estimation}, run on two bicycle-chain patches
occupying the two halves of one atom array: patch~$A$ holds the noisy resource
state from the full TMR circuit, and patch~$B$ holds an ideal reference at angle
$-\theta$. Each resource state is accepted with probability $\sim\!1/2$, and the
accepted shots split into two equally likely branches; the cancel branch
estimates $\langle X_L\rangle$ on each rotated
logical~\cite{xu2025batchedhighrate, ismail2025transversalstar}.

\subsubsection{Partition count $M$}
\label{app:M}

\begin{figure}[t]
  \centering
  \includegraphics[width=\columnwidth]{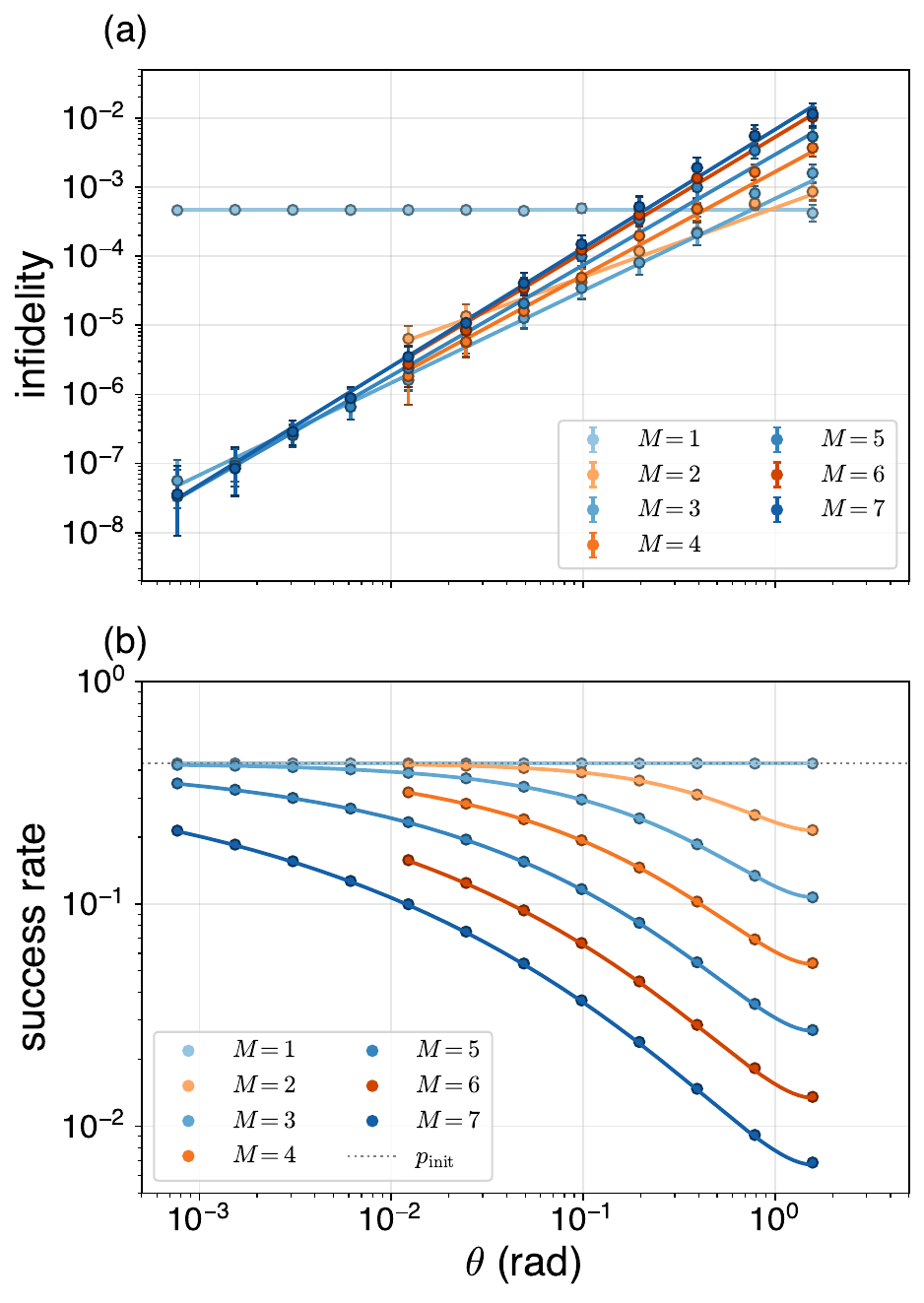}
  \caption{\textbf{Partition-count sweep: Infidelity of preparing resource state using various $M$.}
    (a) Infidelity $\varepsilon$ of injecting one logical rotation vs $\theta$ at $\ell = 4$, $m = 7$,
    $N = 1$, $p = 10^{-3}$, for $M \in \{1, \dots, 7\}$ (odd $M$ blue, even $M$
    orange). Solid curves are the constrained family
    $\varepsilon = C(M)\,p\,\theta^{\,2(1 - 1/M)}$ of Eq.~\eqref{eq:C_table},
    with $C(M)$ fitted at the fixed exponent on the $\theta/\pi \le 1/16$ subset
    (drop-2) (b) Measured success rate vs the master
    $p_\mathrm{init}\,p_\mathrm{TMR}(\theta; M)$ [Eq.~\eqref{eq:s_total}],
    agreeing to within $1.3\times10^{-3}$ across all $(M, \theta)$; the dotted
    line marks $p_\mathrm{init}$. Even-$M$ data are taken after the global
  $\sqrt{X}$ residue correction (Sec.~\ref{sec:resource_prep}).}
  \label{fig:tmr_M}
\end{figure}

The partition count $M$ sets both the ideal acceptance and the angle scaling of
the infidelity. We sweep $M \in \{1, \dots, 7\}$ at $\ell = 4$, $m = 7$,
$N = 1$, $p = 10^{-3}$, spanning both parities: odd $M$, where Eq.~(5) of
Ref.~\cite{toshio2024practicalquantum} cancels with no logical residue, and even
$M$, where the logical Clifford residue is removed by the global transversal
$\sqrt{X}$ available on the block (Sec.~\ref{sec:resource_prep}).
Fitting Eq.~\eqref{eq:epsilon_decomposition} at the fixed exponent
$n = 2(1 - 1/M)$ extracts one coefficient $C(M)$ per partition count
(Fig.~\ref{fig:tmr_M}); the constrained fits track the data with
$R^2_{\log} > 0.98$ for every $M \ge 2$ across both parities, so the single
family of Eq.~\eqref{eq:C_table} describes the full sweep. Leaving the exponent
free instead improves $R^2_{\log}$ by less than $10^{-3}$, so we adopt the constrained family.

The two ends of the sweep bracket the behaviour. For $M = 1$, the injection is a
single block on the full logical-$Z$ support, $p_\mathrm{TMR}(\theta; 1) = 1$
leaks nothing outside the code space, and the infidelity is the flat,
$\theta$-independent floor $\varepsilon = C(1)\,p\,(1-f) \approx
4.7\times10^{-4}\,(1-f)$ ($n = 0$). For $M \ge 2$, the acceptance acquires the
angle dependence $p_\mathrm{TMR}(\theta; M) < 1$ and the infidelity rises as
$\theta^{\,2(1 - 1/M)}$.  Which partition count is preferable therefore depends on the target angle.
Equating the $M = 1$ floor with the $M = 3$ power law,
$C(3)\,p\,\theta^{4/3} = C(1)\,p$, places the crossover at
$\theta/\pi \approx 0.24$ at $p = 10^{-3}$. 
Consequently, the preferred partition count over our sweep is
\begin{equation}
M^\star(\theta) =
\begin{cases}
1, & \theta/\pi \in \{1/4,\,1/2\} \quad (T,\,S\ \text{rotations}),\\[4pt]
3, & \theta/\pi \le 1/8.
\end{cases}
\end{equation}
Increasing $M$ beyond $3$ lowers $\varepsilon$
only for angles well below the smallest we consider; because the realized
rotation is synthesized by repeat-until-success, its fidelity is dominated
by the larger-angle resource states consumed in the later RUS stages, so this
small-angle improvement does not improve the synthesized gate (Sec.~\ref{sec:gadget_ler}). 

The infidelity exponent $2(1 - 1/M)$ is the one derived in
Ref.~\cite{toshio2024practicalquantum} (their Eq.~15), and our constrained fits confirm it against simulation results. The reference further estimates the prefactor $C(M)$ to be linear in the partition count, $C(M) \sim M\,p$, whereas our extracted $C(M)$ grows faster: a power law $C(M) \simeq A\,M^{B}$ with $A = 0.061$ and $B = 2.44$ fits the $M \ge 2$ values at $R^2 = 0.99$ on the linear scale, against $R^2 = 0.73$ for the linear form.

\subsubsection{Number of simultaneously rotated logical qubits $N$}
\label{app:N}

\begin{figure*}[htbp]
  \centering
  \includegraphics[width=\textwidth]{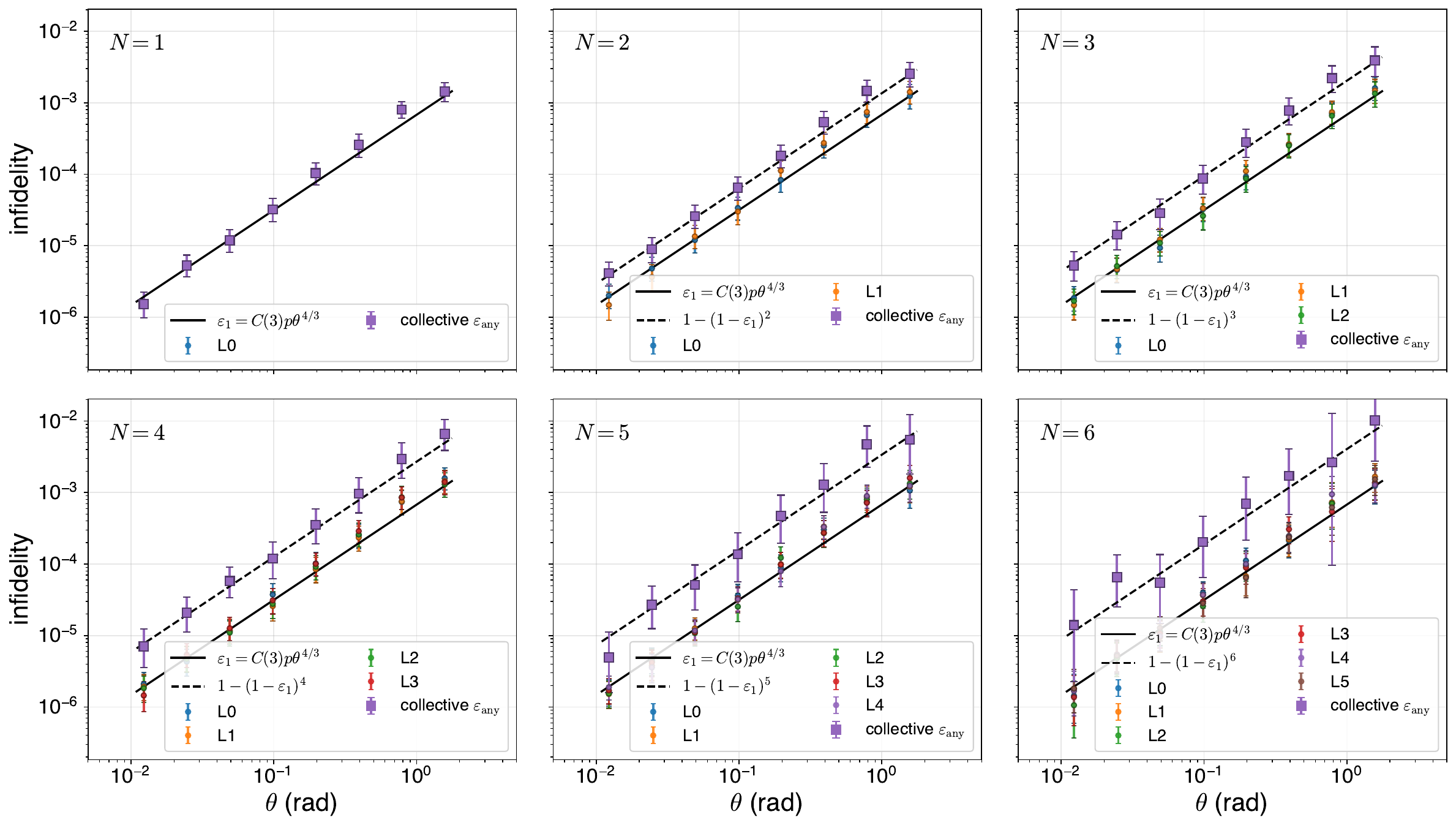}
  \caption{\textbf{Per-logical infidelity independence of the number of simultaneously rotated logicals, $N$.}
    Infidelity at $\ell = 4$, $m = 7$, $M = 3$, $f = 0$, one panel per number of
    simultaneously rotated logical qubits $N \in \{1, \dots, 6\}$. Coloured points are
    the per-logical $\varepsilon(\theta)$; purple squares are the collective
    any-logical rate $\varepsilon_\mathrm{any}$. The solid black curve is the
    constrained $N = 1$ member $\varepsilon_1 = C(3)\,p\,\theta^{4/3}$ of
    Eq.~\eqref{eq:C_table}; the dashed curve is the parameter-free independence
    prediction $\varepsilon_\mathrm{any} = 1 - (1 - \varepsilon_1)^{N}$. The
    per-logical collapse across $N$ and the agreement of
    $\varepsilon_\mathrm{any}$ with the independence curve confirm that errors on
  distinct rotated logical qubits are independent to leading order.}
  \label{fig:tmr_independence}
\end{figure*}

Fixing $M = 3$, we sweep the number of simultaneously rotated logical qubits
$N \in \{1, \dots, 6\}$ at $\ell = 4$, $m = 7$, $f = 0$. The per-logical
infidelity $\varepsilon(\theta)$ is statistically indistinguishable across $N$,
and the collective failure rate tracks the independence prediction
$\varepsilon_\mathrm{any} = 1 - (1-\varepsilon)^{N}$ within the binomial
confidence bands (Fig.~\ref{fig:tmr_independence}). Over the same sweep, the
$\langle X_L\rangle$-flip rate on the un-rotated logical qubits sits below the
binomial floor of $6\times10^{-9}$ (zero errors in $1.24\times10^{9}$ shots),
so the cascade touches only its targeted logical qubits. This is the numerical content
of the $N$-independence of $\varepsilon$ asserted after
Eq.~\eqref{eq:epsilon_decomposition}, and it validates the parallel-$R_Z$
logical-error model of Sec.~\ref{sec:gadget_ler}.

\begin{figure}[htbp]
  \centering
  \includegraphics[width=\columnwidth]{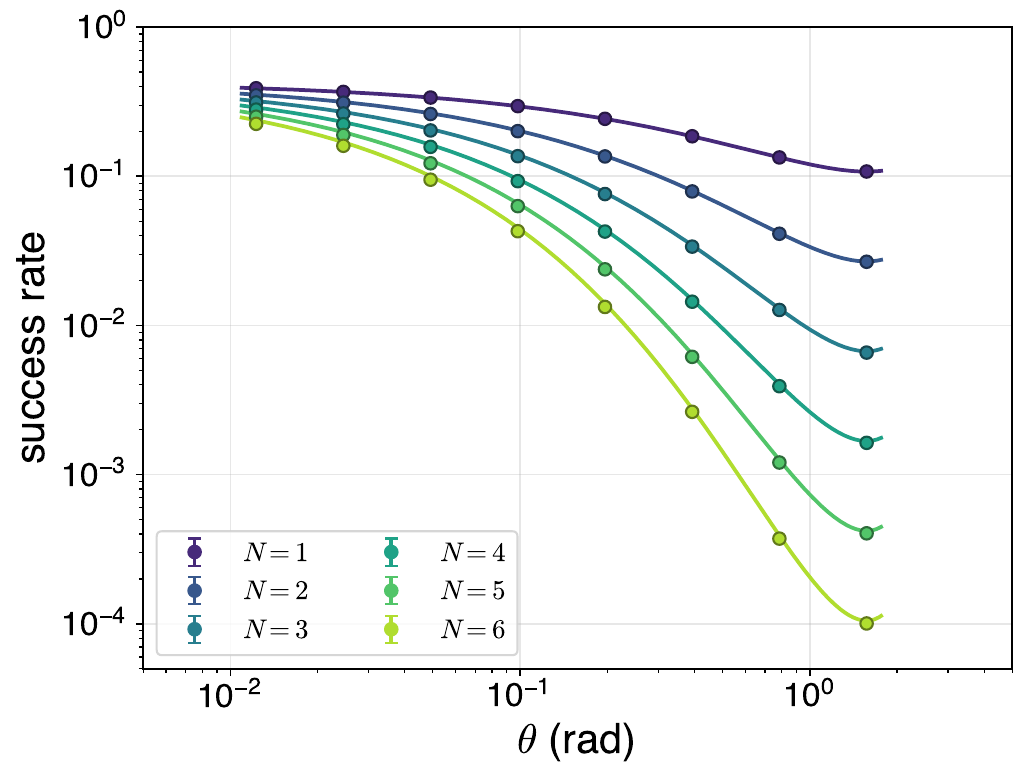}
  \caption{\textbf{Success rate vs the number of simultaneously rotated logical qubits $N$.}
    Measured success rate at $\ell = 4$, $m = 7$, $M = 3$, $f = 0$ for
    $N \in \{1, \dots, 6\}$, overlaid with
    $s_N = p_\mathrm{init}\,p_\mathrm{TMR}(\theta; 3)^{N}$
    (Eq.~\eqref{eq:s_total}): each additional rotated logical multiplies in one
    more ideal-acceptance factor $p_\mathrm{TMR}(\theta; 3)$, while
  $p_\mathrm{init}$ is paid once per block.}
  \label{fig:tmr_independence_succ}
\end{figure}

On the success-rate side, the $N$-dependence enters only through the
$p_\mathrm{TMR}(\theta; M)^{N}$ factor of Eq.~\eqref{eq:s_total}: each additional
rotated logical multiplies in one more ideal-acceptance factor
(Fig.~\ref{fig:tmr_independence_succ}). The state-prep
survival $p_\mathrm{init}$, by contrast, is a single block-level factor that is
amortized across all $k$ logical qubits in the patch and does not scale with $N$. Thus
rotating more logical qubits in parallel costs acceptance only through
$p_\mathrm{TMR}^{N}$, while the code-size cost $p_\mathrm{init}$ is paid once per
block regardless of how many of its logical qubits are rotated.

\subsubsection{Code dimensions $(\ell, m)$}
\label{app:l}

\begin{figure}[htbp]
  \centering
  \includegraphics[width=\columnwidth]{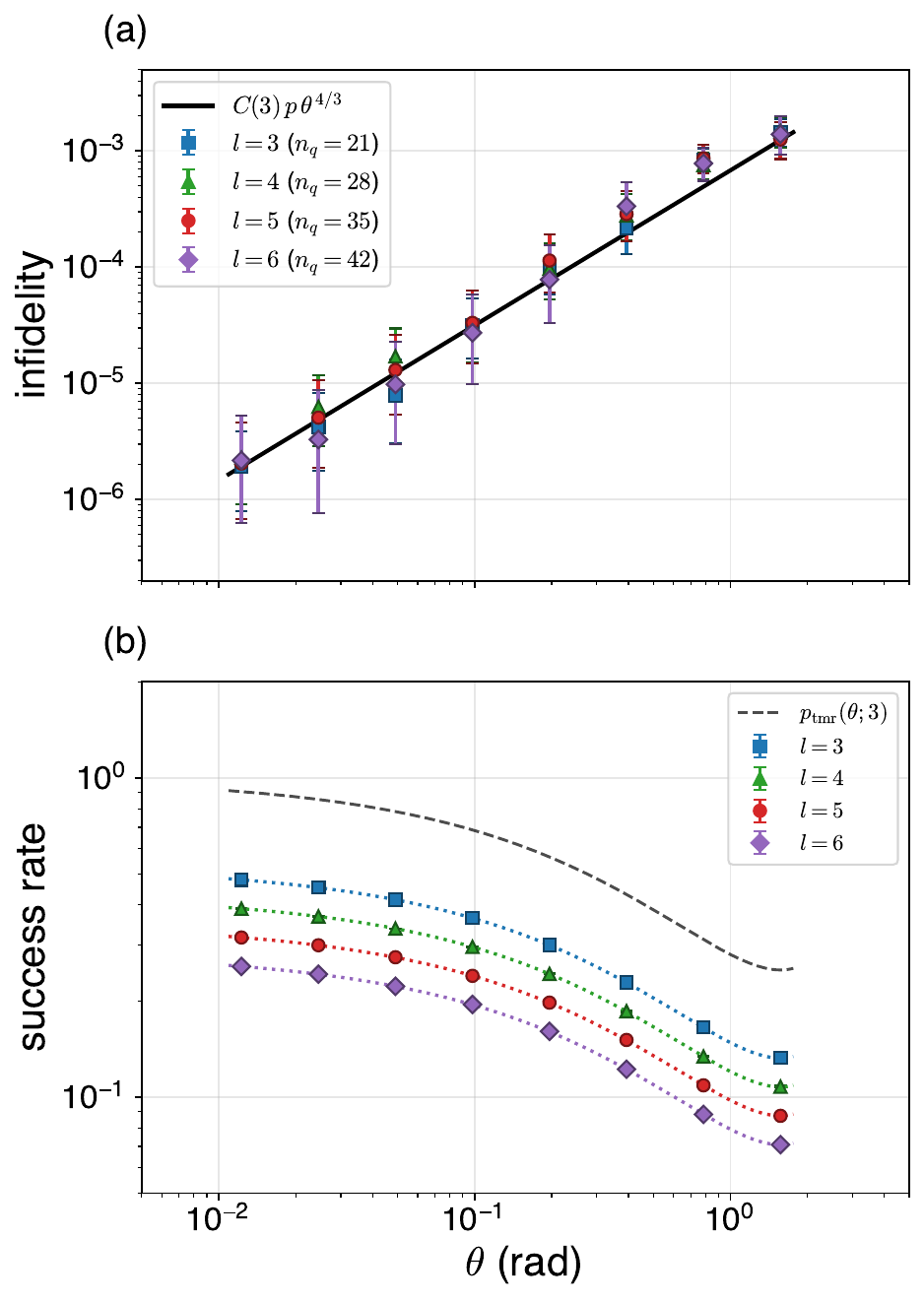}
  \caption{\textbf{Infidelity of preparing resource state is independent of code dimensions; success rate is not.}
    Resource-state verification at $M = 3$, $N = 1$, $f = 0$, sweeping
    $\theta/\pi$ from $1/2$ to $1/256$ at $\ell \in \{3,4,5,6\}$, $m = 7$.
    (a) Per-logical infidelity $\varepsilon(\theta)$ collapses to one
    curve, overlaid with the constrained $M = 3$ member
    $C(3)\,p\,\theta^{4/3}$ of Eq.~\eqref{eq:C_table}. (b) Success rate
    per $\ell$ fitted to
    $s = p_\mathrm{init}(\ell)\,p_\mathrm{TMR}(\theta; 3)$ with one free
    $p_\mathrm{init}$ per $\ell$; the extracted $p_\mathrm{init}(\ell)$ match the
    prediction $\exp[-\gamma\,(32\ell m + 13)\,p]$ at $\gamma = 0.93$, fixing the
    calibration of $\gamma$ in Eq.~\eqref{eq:s_total_factors} (cross-checked
  against the loss sweep in Sec.~\ref{app:loss}).}
  \label{fig:tmr_vary_l}
\end{figure}

Still at $M = 3$, $f = 0$, we sweep the code width $\ell \in \{3,4,5,6\}$ at
$m = 7$. The per-logical infidelity collapses across all four widths onto the
constrained $M = 3$ member $\varepsilon = C(3)\,p\,\theta^{4/3}$ of
Eq.~\eqref{eq:C_table} (Fig.~\ref{fig:tmr_vary_l}a), confirming the $(\ell, m)$-independence of
$\varepsilon$: the error is set by the TMR cascade rather than by the
syndrome extraction floor $O\!\bigl(p^{\lfloor(d+1)/2\rfloor}\bigr)$, which is
subdominant at $p = 10^{-3}$.

The state-prep survival, on the other hand, depends on $(\ell, m)$ exactly as
predicted, and through a single channel: the per-shot noise-location count
$N_\mathrm{loc}(\ell, m) = 32\ell m + 13$. The two SE rounds contribute
$20\ell m$ two-qubit-gate locations, the verification-patch round $10\ell m$, the
codespace projection $\sim\!2\ell m$, and the TMR cascade a size-independent
constant $\sim\!13$. Fitting $p_\mathrm{init} = \exp[-\gamma N_\mathrm{loc}(1-f)p]$
at $\ell = 4$ gives a single free parameter $\gamma = 0.93$, which then
reproduces the independently measured $p_\mathrm{init}$ at
$\ell \in \{3, 5, 6\}$ to four decimals (Fig.~\ref{fig:tmr_vary_l}b and
Sec.~\ref{app:loss}). The
larger acceptance cost of larger codes is thus entirely the cost of more error
locations, as encoded in $N_\mathrm{loc}$; $\gamma$ is $\ell$- and
$M$-independent in the regime studied.

\subsubsection{Heralded atom loss error fraction $f$}
\label{app:loss}

Finally we incorporate atom loss, modeled as a heralded erasure: a fraction $f$ of each per-location physical-error budget $q_\mathrm{loc}$ is a loss event that is heralded and post-selected away (the entire shot is
discarded) leaving the residual Pauli budget $(1-f)\,q_\mathrm{loc}$ as the only error able to seed a logical fault on the accepted shots. Our simulation realizes this by aborting and discarding a run the moment a loss is heralded. As argued in the main text (Sec.~\ref{sec:gadget_ler}), because loss is handled by pure
post-selection (its location and timing is never used) this immediate-heralding model is also a good proxy for other implementations with delayed loss detection.

\begin{figure}[htbp]
  \centering
  \includegraphics[width=\columnwidth]{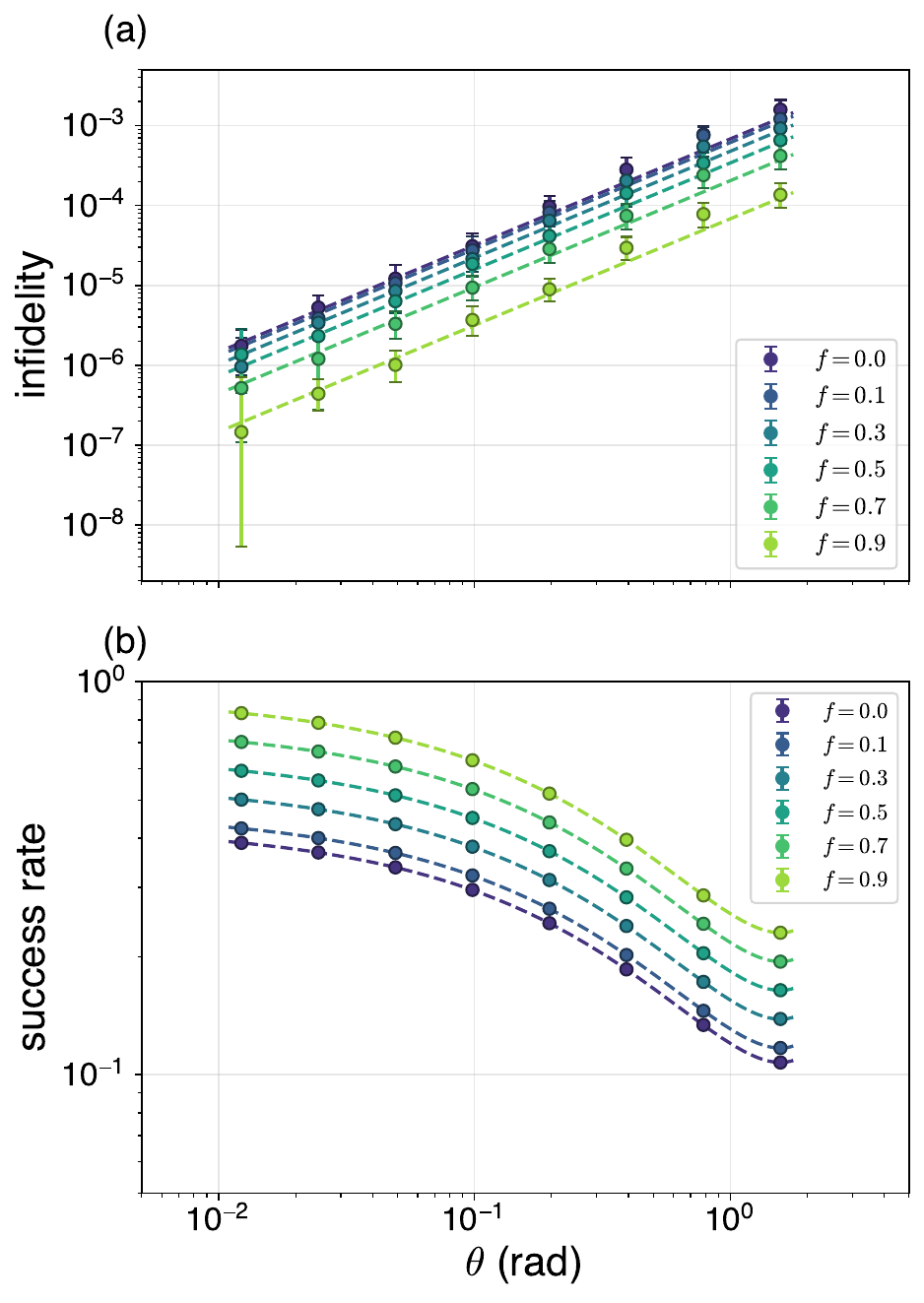}
  \caption{\textbf{Scaling of infidelity of preparing resource state under heralded loss.}
    $\ell = 4$, $m = 7$, $M = 3$, sweeping $f \in [0, 0.5]$ (colour).
    (a) Infidelity vs $\theta$; dashed curves are the constrained
    reference $\varepsilon = C(3)\,p\,(1-f)\,\theta^{4/3}$ of
    Eq.~\eqref{eq:epsilon_decomposition}, one per $f$. The clean $(1-f)$
    collapse confirms that the heralded-erasure rescaling is structurally exact
    and not a small-$f$ approximation. (b) State-prep success rate per $f$
    vs $p_\mathrm{init}(p_\mathrm{eff})\,p_\mathrm{TMR}(\theta; 3)$, with
    $p_\mathrm{init}$ rescaled by the surviving Pauli rate $p_\mathrm{eff} =
  (1-f)\,p$.}
  \label{fig:tmr_loss_eps}
\end{figure}

Under this model both the infidelity and the acceptance rescale by the surviving Pauli rate. Since only the residual Pauli weight $(1-f)\,q_\mathrm{loc}$ can produce a logical error, the infidelity scales as $\varepsilon \propto (1-f)$, giving $C(M)\,p\,(1-f)\,\theta^{\,2(1 - 1/M)}$ for every $M$ in
Eq.~\eqref{eq:epsilon_decomposition}. Fig.~\ref{fig:tmr_loss_eps} verifies the $(1-f)$ collapse across $f \in [0, 0.9]$. The same split sends $p \to (1-f)p$ in the state-prep exponent, so $p_\mathrm{init} = \exp[-\gamma N_\mathrm{loc}(1-f)p]$, while the detected loss is post-selected through $p_\mathrm{loss} = \exp[-f N_\mathrm{loc}p]$. The two PS exponents thus share the common budget $N_\mathrm{loc}$ and differ only in their trigger fraction, $\gamma$ for state-prep PS and $f$ for loss heralding, which is the structural cross-check verified in Sec.~\ref{app:l}.

\subsubsection{Physical error rate $p$}
\label{app:p}

\begin{figure}[htbp]
  \centering
  \includegraphics[width=\columnwidth]{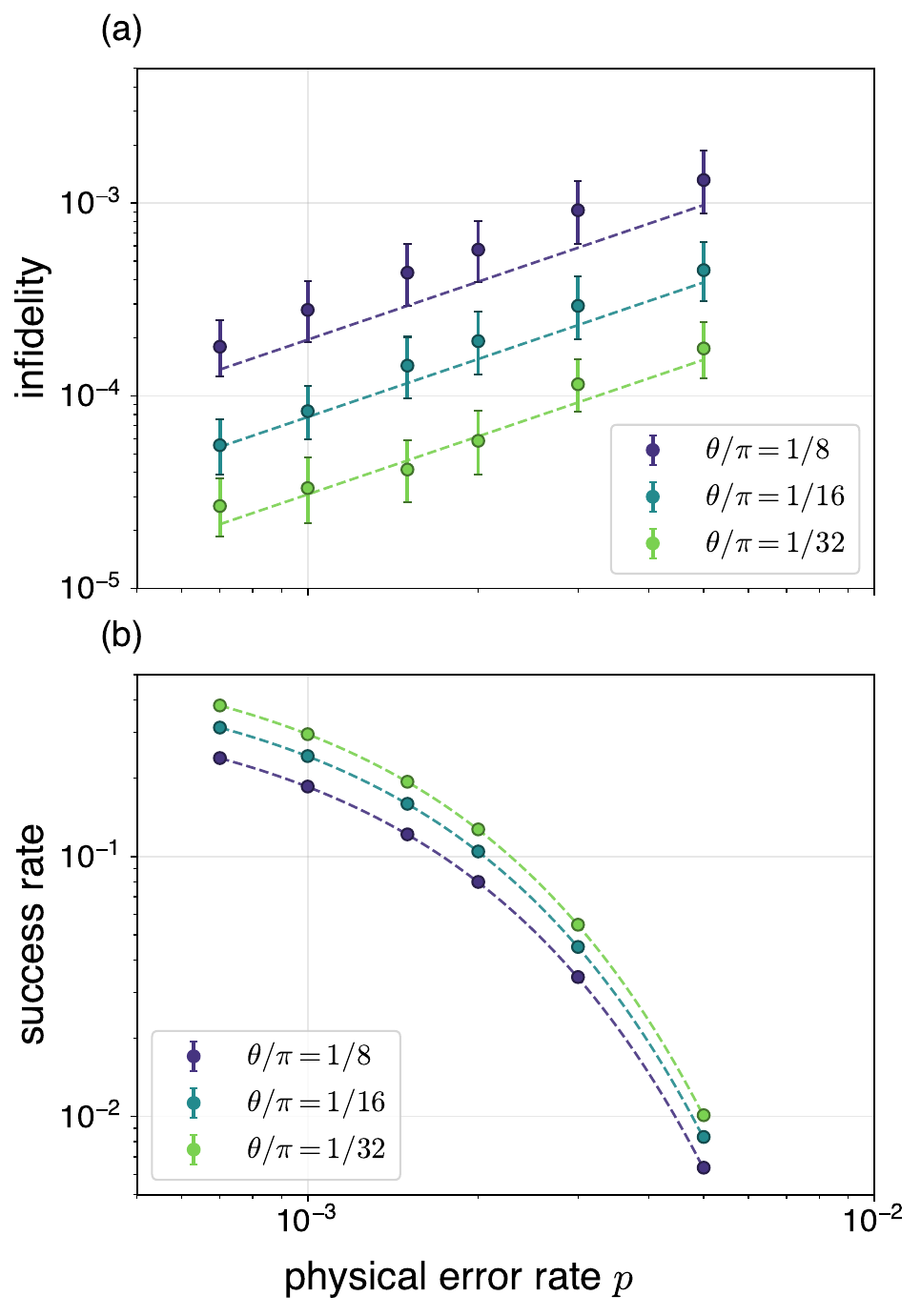}
  \caption{\textbf{Dependence on physical error rate for infidelities and success rates.}
    Resource-state cost model ansatz verification at $M = 3$, $N = 1$, $f = 0$, $\ell = 4$,
    $m = 7$, sweeping the physical error rate $p \in [7\times10^{-4},\,
    5\times10^{-3}]$ at $\theta/\pi \in \{1/8, 1/16, 1/32\}$ (colour).
    (a) Per-logical infidelity $\varepsilon$ vs $p$ (log--log); points are
    data with binomial confidence intervals, dashed lines the constrained
    $M = 3$ family $C(3)\,p\,\theta^{4/3}$ of Eq.~\eqref{eq:C_table} with no
    re-fit. (b) State-prep success rate $s$ vs $p$ (log--log); points are
    data, dashed lines the fitted ansatz
    $s = p_\mathrm{init}(p)\,p_\mathrm{TMR}(\theta; 3)$ of Eq.~\eqref{eq:s_total}
    with $p_\mathrm{init} = \exp[-\gamma\,N_\mathrm{loc}\,p]$ ($\gamma = 0.93$).
  }
  \label{fig:tmr_vary_p}
\end{figure}

The code-width sweep (Sec.~\ref{app:l}) and the loss-fraction sweep
(Sec.~\ref{app:loss}) both fixed $p = 10^{-3}$ while varying a structural
parameter. The factorized $p$-dependence written into the model, a single
power of $p$ in the infidelity [Eq.~\eqref{eq:epsilon_decomposition}] and an
exponential $p_\mathrm{init} = \exp[-\gamma\,N_\mathrm{loc}\,(1-f)\,p]$ in the
success rate [Eq.~\eqref{eq:s_total_factors}], was therefore assumed but not
tested. We now verify it directly by sweeping $p$ at fixed $M = 3$,
$N = 1$, $f = 0$, $\ell = 4$, $m = 7$, over
$p \in [7\times10^{-4},\, 5\times10^{-3}]$ and three angles
$\theta/\pi \in \{1/8, 1/16, 1/32\}$ (Fig.~\ref{fig:tmr_vary_p}).

As shown in Fig.~\ref{fig:tmr_vary_p}(a), the infidelity is linear in $p$ at every angle. A free power-law fit $\varepsilon \propto p^{\,n}$ returns
$n = 1.02$, $1.08$, and $1.01$ for $\theta/\pi = 1/8$, $1/16$, $1/32$, all
consistent with the single power of $p$
in Eq.~\eqref{eq:epsilon_decomposition}. The three curves separate only by the
$\theta^{4/3}$ prefactor of the constrained $M = 3$ family
$C(3)\,p\,\theta^{4/3}$, which tracks each curve with no re-fit.

As shown in Fig.~\ref{fig:tmr_vary_p}(b), the success rate follows the
factorization ansatz at every angle. The
measured $s(p)$ is reproduced across all three angles by
$s = p_\mathrm{init}(p)\,p_\mathrm{TMR}(\theta; 3)$ with the exponential
state-prep survival $p_\mathrm{init}(p) = \exp[-\gamma\,N_\mathrm{loc}\,p]$ of
Eq.~\eqref{eq:s_total_factors} and no parameter re-fit
(Fig.~\ref{fig:tmr_vary_p}b). Here $\gamma = 0.93$ and
$N_\mathrm{loc}(\ell, m) = 32\ell m + 13 = 909$ are fixed by the code-width
sweep of Sec.~\ref{app:l}. The only angle dependence enters through the constant
acceptance prefactor $p_\mathrm{TMR}(\theta; 3)$, so the three curves share a
single exponential rate; an unconstrained fit of that rate returns $844$, within
$1\%$ of the predicted $\gamma\,N_\mathrm{loc} = 845$. This confirms the
$p$-dependence of $p_\mathrm{init}$ that the earlier sweeps took as given.

\paragraph{Regime of validity.}
The fits hold at small $p$ and small $\theta$. The trigger fraction
$\gamma = 0.93$ is calibrated at $\ell = 4$ and verified through $\ell = 6$, but
not toward $\ell \to \infty$. The $C(M)$ coefficients are fitted at the theory
exponent $n = 2(1 - 1/M)$ on the $\theta/\pi \le 1/16$ subset, since the power
law saturates at larger $\theta$. And the
exponential acceptance factors approximate the exact per-location product
$\prod_\mathrm{loc}(1 - q_\mathrm{loc})$, agreeing to $10^{-6}$ at
$p \le 10^{-3}$ but not to be extrapolated past $p \gtrsim 10^{-2}$ without
re-checking.

\subsection{Repeat-until-success gate cost and the Clifford floor}
\label{app:rus-clifford-floor}

\begin{figure}[htbp]
  \centering
  \includegraphics[width=\columnwidth]{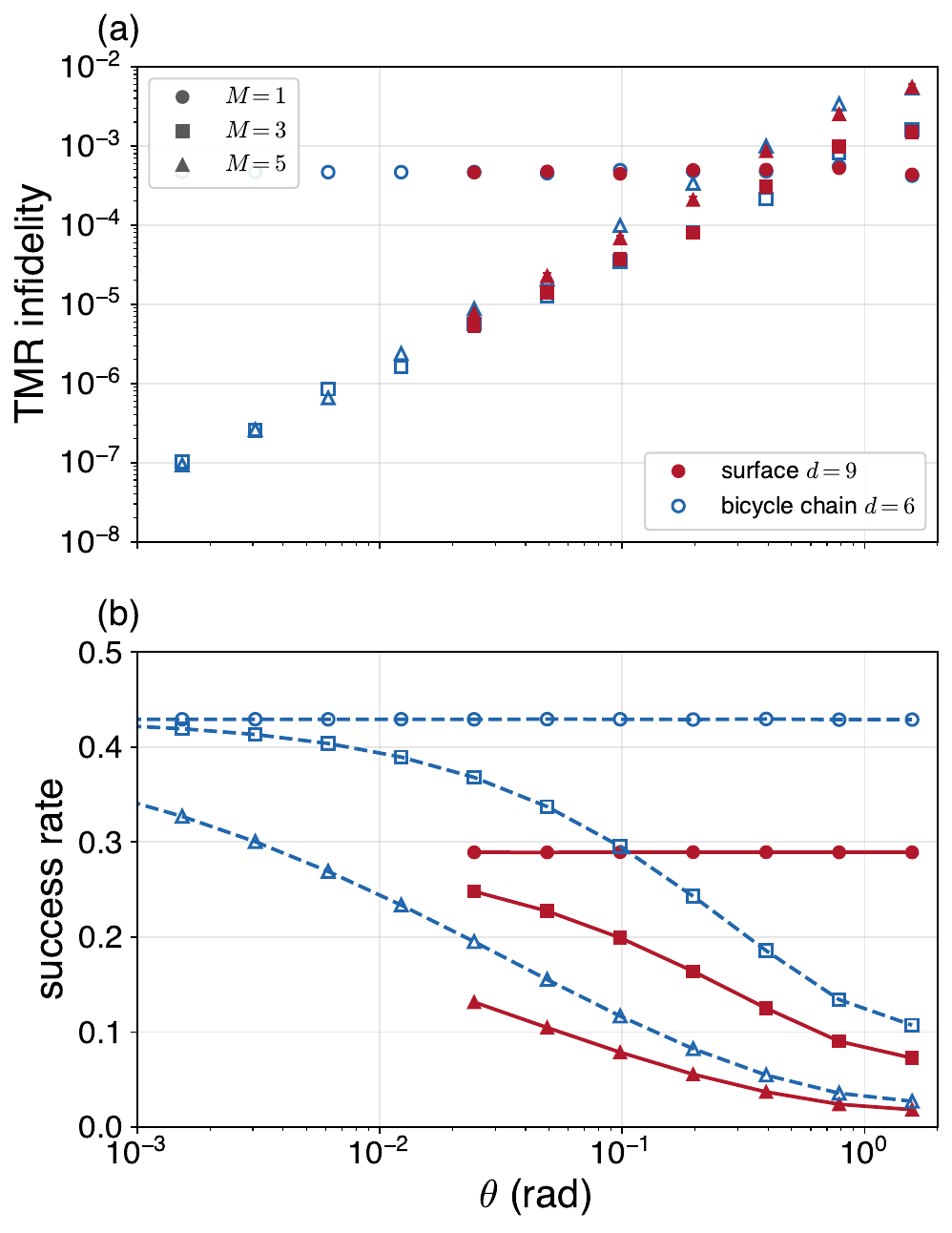}
  \caption{
    \textbf{Comparison of rotated surface code versus the bicycle chain for angle resource state preparation.} All numbers are obtained with MLE decoder at $p = 10^{-3}$, with each code's optimal syndrome extraction
    schedule (surface: one $Z$-only round before the rotation and two after;
    bicycle chain: one $Z$-only before and one after). Different colors represent different code: surface $d = 9$ (red) and bicycle chain $d = 6$ (blue); the different markers represent the partition counts $M \in \{1, 3, 5\}$. (a) Per-injection infidelity versus rotation angle $\theta$: simulation
    data only (surface filled with $1\sigma$ binomial error bars, bicycle chain
    open). The $M = 3$ data of the two codes matches, and the $M = 1$ floor
    $\varepsilon \approx 0.47\,p$ is code- and distance-independent.
    (b) State-preparation post-selection rate $s$ versus $\theta$: the
    surface code accepts at $p_\mathrm{init} \approx 0.29$ ($d = 9$), below the
    bicycle chain's $p_\mathrm{init} \approx 0.43$, the combined cost of its larger patch and
  its extra post-selection round.}
  \label{fig:surface_tmr_numerics}
\end{figure}

\begin{figure}[htbp]
  \centering
  \includegraphics[width=\columnwidth]{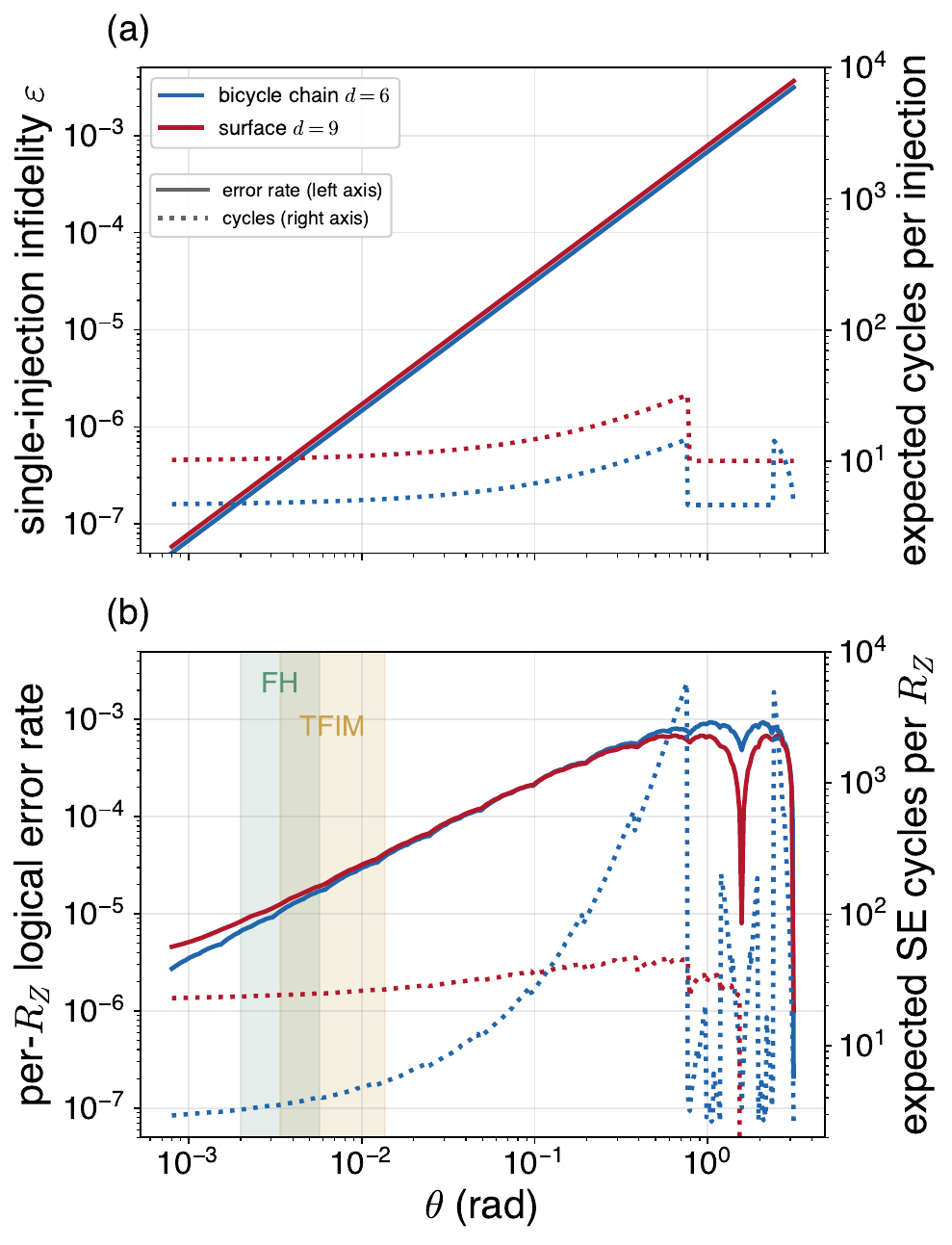}
  \caption{
      \textbf{$R_Z$ gate via RUS teleportation: single injection versus full
      gate.} Bicycle-chain code at $d=6$ and rotated surface code at $d=9$, both
      fitted from circuit-level simulations at $p = 10^{-3}$. In both panels,
      solid curves show logical error rate (left axis) and dotted curves show
      the expected number of syndrome-extraction cycles (right axis). (a) Single
      TMR resource-state infidelity: the two codes have similar fitted
      infidelities, $0.68\,p\,\theta^{4/3}$ for the bicycle chain and
      $0.79\,p\,\theta^{4/3}$ for the surface code at $M=3$. The higher expected cycles curve for the surface code is likely  partially due to its higher distance compared to bicycle chain code and partially due to its three post-selection cycles per
      attempt, compared with two for the bicycle chain. (b) Full RUS gate. Shaded bands mark the Trotter rotation angles of the
  Fig.~\ref{fig:resources} estimates.}
  \label{fig:rus_clifford_floor}
\end{figure}

The preceding subsections characterized a single TMR injection
(Appendix~\ref{app:tmr_characterization}). The logical gate the architecture
actually executes is the full RUS chain of
Fig.~\ref{fig:rotation_teleportation}: on a wrong-sign outcome, the angle doubles
and is re-injected, and each level consumes one teleportation transversal CNOT
in addition to the TMR resource state. We first present that the infidelities
of single-injection resource states are similar in both surface code at $d=9$
and the bicycle chain at $d=6$ (Fig.~\ref{fig:surface_tmr_numerics}), and then
show that the two architectures differ once the full RUS procedure is
considered because of different Clifford error rates and post-selection rates
(Fig.~\ref{fig:rus_clifford_floor}).

Circuit-level simulations of a single TMR injection show that the infidelity follows the TMR ansatz $\varepsilon = C(M)\,p\,\theta^{2(1-1/M)}$. For the $d=6$ bicycle-chain code we find $C(1) = 0.47$ and $C(3) = 0.68$; for the $d=9$ rotated surface code we find the same $C(1) = 0.47$ and a slightly larger $C(3) = 0.79$.  Thus the per-injection infidelities of the two codes are essentially the same [Fig.~\ref{fig:surface_tmr_numerics}(a)] and mainly depends on the angle and the number of partitions $M$ as formula ~\eqref{eq:epsilon_decomposition} suggests.

The architectures differ instead in the number of syndrome extraction
cycles each TMR attempt must post-select on. The rotated surface code requires
three noisy cycles per attempt --- one $Z$-only round before the rotation, which
fixes the random $\ket{+}_L$ $Z$-syndrome by feed-forward and is not itself
post-selected, and two rounds after, whereas the bicycle chain needs only two
(one before and one after). The second post-rotation round is needed only for the surface code, and the reason is structural. A noise-free TMR already leaves the block in a superposition of the intended rotation acting on a state in the codespace and branches with a wrong logical rotation acting on states in orthogonal subspaces and therefore have a nontrivial syndrome. (Physical noise during TMR enters separately, at order $p$: it either produces an
unavoidable logical error that sets the $O(p)$ floor and is suppressed at small
rotation angle $\theta$, or leaves a nontrivial syndrome and is post-selected
away.) A single perfect syndrome round would reject every nontrivial branch. A
noisy round is dangerous only if a measurement error makes a nontrivial syndrome
read as trivial, which can happen only when the incoming syndrome has weight one,
so that a single flipped check masks it. Only weight-one syndromes matter here: the
TMR error rate is already $O(p)$, so an event needing two measurement faults is
higher order and irrelevant.

The nontrivial TMR syndromes are exactly those of $Z$ acting on subsets of the
rotated logical (set by the TMR partition choice). For the surface code the logical-$Z$ representative is a
geometric string that terminates on a boundary qubit checked by a single
$X$-stabilizer; that endpoint carries a weight-one syndrome, which a single
measurement error can hide, so the surface code requires the confirming second
round. For the bicycle chain every data qubit is touched by four $X$-stabilizers,
an even number, so any subset of qubits is touched by an even number of $X$-checks
in total. No subset of $Z$ blocks can therefore produce an odd-weight
syndrome, and in particular none can have weight one; a single measurement error
can never mask such a syndrome, and the bicycle chain needs no second post-rotation
round to guard against single measurement faults.

The success rate factorizes as $s = p_\mathrm{init}\,p_\mathrm{TMR}(\theta; M)$,
with the ideal acceptance $p_\mathrm{TMR}$ fixed by $M$ and $\theta$ and a
code-dependent state-preparation survival $p_\mathrm{init}$. We measure $p_\mathrm{init} = 0.29$ for the $d = 9$ surface code and $0.48$ for $d = 7$, both
reproduced by $p_\mathrm{init} = \exp(-\gamma\,N_\mathrm{loc}\,p)$ with
$N_\mathrm{loc} = N_\mathrm{cyc}\,d^2$, $N_\mathrm{cyc} = 3$, and $\gamma = 5.0$,
while the bicycle-chain block accepts at $\sim\!0.43$ with only
$N_\mathrm{cyc} = 2$ [Fig.~\ref{fig:surface_tmr_numerics}(b)]. The surface
code's lower acceptance is the combined cost of its larger patch and the extra
post-selection round.

Beyond the per-attempt magic state, the full $R_Z$ gate at the small
Trotter angles of the resource estimates ($\theta \sim 10^{-2}$\,rad) is
dominated by the teleportation Clifford, not by the magic state. The
bicycle-chain transversal block CNOT has logical error $\sim\!10^{-7}$ per
logical qubit (simulated, $k = 8$ MLE, $d = 6$, $p = 10^{-3}$), whereas the
surface-code transversal/lattice-surgery CNOT at $d = 9$ is
$\sim\!4\times10^{-6}$. The full per-$R_Z$ gate therefore floors at
$\sim\!10^{-7}$ for the bicycle chain versus $\sim\!10^{-5}$ for the surface
code [Fig.~\ref{fig:rus_clifford_floor}(b)]: in the relevant small-angle regime,
the bicycle-chain gate has no error floor, while the surface-code gate saturates
well above it. Together with the extra post-selection cycle, this is a
key advantage of the co-design approach: compact, low-cycle transversal qLDPC
post-selection and Cliffords make every RUS attempt cheap, so the rotation gate
is limited by the magic state rather than by the surrounding Clifford in practice.

The RUS chain folds the doubled angle to its nearest multiple of $\pi$ for
the bicycle chain (terminus $R_Z(\pi) = Z$, a free Pauli-frame update) and to its
nearest multiple of $\pi/2$ for the surface code (terminus $R_Z(\pi/2) = S$).
Because $R_Z(\pi/2) = S$ is implemented transversally and deterministically
on the surface code, its terminus error is that of a transversal Clifford,
$\sim\!10^{-6}$ at $d = 9$ --- comparable to the teleportation CNOT and already
included in every RUS cost reported here, in contrast to the bicycle chain whose
$R_Z(\pi) = Z$ terminus is a virtually free Pauli-frame update. All RUS costs
reported here include the Clifford error of every teleportation CNOT.

\section{Resource Estimation Methodology}
\label{app:resource-estimation}

In this Appendix, we provide more details on the estimation of the physical cost of simulating 2D lattice Hamiltonians
(transverse-field Ising model and Fermi--Hubbard model with local compact
encoding) under second-order Trotterization.  The estimation pipeline
determines the optimal simulation parameters, decomposes the circuit into
architecture-specific primitives, and compares six fault-tolerant
architectures.  All formulas below are implemented in the open-source
package accompanying this work.

\subsection{Simulation parameters and operating point}
\label{app:crossing}

We fix the simulation time to a representative operating point $T^*=2.0\,(zJ)^{-1}$ (Section~\ref{sec:e2e_resources}). Given a
Trotter error budget $\epsilon_{\mathrm{trotter}}$, we choose the largest Trotter step size $dt^*$ that satisfies the second-order Trotter bound at $T^*$ via bisection:
\begin{equation}
  \mathcal{E}_{\mathrm{trotter}}\bigl(dt,\, \lceil T^*/dt \rceil\bigr)
  \;\le\; \epsilon_{\mathrm{trotter}}.
  \label{eq:trotter-constraint}
\end{equation}
At the resulting operating point $(T^*,\, dt^*)$ we build the Trotterized evolution circuit and evaluate the gate infidelity $\epsilon_{\mathrm{gate}}$ (Section~\ref{app:gate-error-budget}). The same $(T^*,\, dt^*)$ is applied to every architecture, so the reported gate errors all refer to the same simulation task.

\subsection{Trotter error bounds}
\label{app:trotter}

We use the Childs--Su spectral-norm bounds~\cite{childs2021theory} for
product-formula error.  Throughout, $n_{\mathrm{steps}} = \lceil T/dt \rceil$
denotes the number of Trotter steps.

\paragraph{TFIM.}
The Hamiltonian is split as $H = A + B$ with
$A = -g \sum_v X_v$ and $B = -J \sum_{\langle u,v\rangle} Z_u Z_v$.
On a 2D square lattice with periodic boundaries (degree $\delta = 4$)
containing $N_s$ sites and $N_b$ bonds:

\smallskip
\noindent\emph{First order (Lie--Trotter):}
\begin{equation}
  \mathcal{E}_1
  \;=\; n_{\mathrm{steps}} \,\frac{dt^2}{2}\,
  4\, N_b\, |g|\,|J|.
\end{equation}

\noindent\emph{Second order (symmetric Strang):}
\begin{equation}
  \mathcal{E}_2
  \;=\; n_{\mathrm{steps}} \left[
    \frac{dt^3}{12}\, 4\,\delta^2 N_s\,|g|\, J^2
    + \frac{dt^3}{24}\, 16\, N_b\, g^2\, |J|
  \right].
\end{equation}

\paragraph{Fermi--Hubbard.}
For the local compact encoding the Hamiltonian is split into five layers
$\{V_0, V_1, H_0, H_1, I\}$, where each hopping layer is a sum of
3-qubit terms
\begin{align}
  H_V &= -\tfrac{t}{2}\bigl(X_R X_B X_G + Y_R X_B Y_G\bigr), \\
  H_H &= -\tfrac{t}{2}\bigl(X_R Y_B X_G + Y_R Y_B Y_G\bigr),
\end{align}
and the on-site interaction is implemented in its
chemical-potential-shifted single-$ZZ$ form
$H_I = \tfrac{U}{4}\,Z_\uparrow Z_\downarrow$~\cite{campbell2022earlyfaulttolerant}.
This differs from the unshifted $\tfrac{U}{4}(Z_\uparrow Z_\downarrow - Z_\uparrow - Z_\downarrow)$
by a number-operator term that commutes with both hopping and interaction, so it
only shifts the energy by a known per-particle-number constant while removing two
of every three on-site rotations.
The subscripts $R, G$ denote primary (data) qubits and $B$ the secondary
(auxiliary) qubit at the plaquette center.

On a 2D lattice with periodic boundaries ($N_s = L_x L_y$ sites, $s$
spin species), the Childs--Su bound is an exact polynomial in
$(|t|, |U|)$ with coefficients that depend only on $N_s$ and~$s$.  The
per-overlap Pauli commutator norms that determine the coefficients are:

\begin{center}
  \begin{tabular}{lc}
    \toprule
    Overlap topology & $\|\,[\cdot,\cdot]\,\|$ \\
    \midrule
    $V$--$H$ sharing only auxiliary $B$ ($X_B$ vs $Y_B$) & 2 \\
    Any pair sharing one primary qubit ($R$ or $G$) & 1 \\
    $V$--$V$ or $H$--$H$ sharing only $B$ (same Pauli) & 0 \\
    Hop--$I$ sharing both $R$ and $G$ & 0 \\
    $I$--$I$ sharing one qubit & 0 \\
    \bottomrule
  \end{tabular}
\end{center}

Counting overlapping bond pairs/triples on the translation-invariant
lattice and weighting by these norms gives the first-order bound
\begin{equation}
  \mathcal{E}_1
  \;=\; n_{\mathrm{steps}} \,\frac{dt^2}{2}\,
  \bigl(6\,s\, N_s\, t^2
  + 4\, N_s\, |tU|\,\delta_{s,2}\bigr).
\end{equation}

For the second-order estimate, we adapt Campbell's split-operator bound for
  the Fermi--Hubbard model~\cite{campbell2022earlyfaulttolerant}. Let
  $H_h=V_0+V_1+H_0+H_1$ denote the full hopping Hamiltonian and
  $\mathcal{L}_{\mathrm{hop}}=\{V_0,V_1,H_0,H_1\}$ its four implementable hopping
  layers. We use a Strang step in which the interaction brackets the hopping
block,
\begin{equation}
  U_{\mathrm{step}}
  =e^{-iH_I\,dt/2}\,S_2^{\mathrm{hop}}(dt)\,e^{-iH_I\,dt/2},
\end{equation}
where $S_2^{\mathrm{hop}}$ is the symmetric second-order product formula for
  the four hopping layers. This separates the error into Campbell's
hopping--interaction contribution and a residual pure-hopping contribution:
\begin{equation}
 \mathcal{E}_2
  \;=\; n_{\mathrm{steps}}\,dt^3\,\bigl(W_{\mathrm{SO2}} + W_{\mathrm{hop}}\bigr),
\end{equation}
where:
\begin{align}
  W_{\mathrm{SO2}} &= \frac{U\,t^2}{6}\,N_s\,(\sqrt{5}+8)\,\delta_{s,2}
  + \frac{U^2}{24}\,\|H_h\|\,\delta_{s,2},\\
  W_{\mathrm{hop}} & = \tfrac{1}{12}
    \bigl\|{\textstyle\sum_{a,b,c\in\mathcal{L}_{\mathrm{hop}}}}
  [[H_c,H_b],H_a]\bigr\|,
\end{align}
and $\|H_h\|=\|R\|_1$ is the trace norm of the $N_s\times N_s$ single-particle hopping matrix, while $W_{\mathrm{SO2}}$ is the Campbell hopping--interaction split-operator
  contribution, adapted to our encoded hopping matrix $H_h$. The factor
  $\delta_{s,2}$ removes the respective terms in the spinless case. The term
  $W_{\mathrm{hop}}$ accounts for the residual error from implementing $H_h$ as
  four hopping layers; we evaluate this free-fermion single-particle norm with
  the layer sums kept inside the norm, retaining cancellations between bonds.
  Both terms are evaluated directly at the target $L$, with no finite-size
extrapolation, as in~\cite{campbell2022earlyfaulttolerant}.


\subsection{Gate-level error budget}
\label{app:gate-error-budget}

The Trotterized circuit on $n_{\mathrm{logical}}$ logical qubits decomposes
into $n_{\mathrm{Cliff}}$ Clifford gates and $M_R$ non-Clifford
$R_Z(\theta)$ rotations.  We assign each gate a logical error probability
$p_i$ and spacetime volume $v_i$; errors compose under independent failure:
\begin{equation}
  p_{\mathrm{total}}
  \;=\; 1 - \prod_i (1 - p_i),
  \qquad
  V_{\mathrm{total}} \;=\; \sum_i v_i.
\end{equation}

The elementary gates fall into two categories:
\begin{itemize}
  \item \textbf{Clifford gates} ($H$, $S$, CNOT and their global transversal
    variants). The per-gate logical error depends on the QEC code.
  \item \textbf{Non-Clifford rotations} ($R_Z(\theta)$). The per-gate
    cost depends on the magic-state resource model: direct resource-state
    injection for STAR ansätze (Section~\ref{app:resources}), or $T$-gate
    synthesis for the other architectures (Section~\ref{app:rz-to-t}).
\end{itemize}

Summing the Clifford gate errors yields $p_{\mathrm{Clifford}}$; the
remaining gate-infidelity budget available for rotations is
\begin{equation}
  \epsilon_{\mathrm{RZ}}
  \;=\; \epsilon_{\mathrm{gate}} - p_{\mathrm{Clifford}}.
  \label{eq:rz-budget}
\end{equation}
This decomposition constrains all six architectures: the STAR ansätze
consume $\epsilon_{\mathrm{RZ}}$ through per-gate spacetime volume, while
the $T$-based architectures convert it into a $T$-gate requirement
(Section~\ref{app:rz-to-t}).

\subsection{Rotation synthesis cost}
\label{app:rz-to-t}

The $T$-based architectures (Beverland et al., Webster et al., and Zhou et al.) implement each
$R_Z(\theta)$ rotation by decomposing it into a Clifford+$T$ gate
sequence. The Trotterized evolution circuits in this work contain only
Clifford gates and arbitrary-angle rotations (no native $T$ or Toffoli
gates), so the total $T$-count is determined entirely by the rotation
synthesis.

Following Beverland et al.~\cite{beverland2022assessingrequirements} (Eq.~D4)
and Kliuchnikov et al.~\cite{kliuchnikov2022shorter}, the $T$-count per
rotation at synthesis error budget $\epsilon_{\mathrm{syn}}$ allocated
across $M_R$ rotations is
\begin{equation}
  n_{T,\mathrm{rot}}
  \;=\; \left\lceil
  -0.53\,\log_2\!\left(\frac{\epsilon_{\mathrm{syn}}}{M_R}\right) + 5.3
  \right\rceil,
  \label{eq:t-per-rotation}
\end{equation}
and the total $T$-count is $n_T = M_R \cdot n_{T,\mathrm{rot}}$. We note that the recently improved estimates for small-angle synthesis under limited precision~\cite{bothe2026moreefficient} do not apply as the scale of the quantum simulations and the optimal angle sizes we operate at are beyond the applicable low precision regime.

The synthesis error budget $\epsilon_{\mathrm{syn}}$ is drawn from the
overall error budget differently in each architecture:
\begin{itemize}
  \item \textbf{Beverland et al.:}
    $\epsilon_{\mathrm{syn}} = \epsilon_{\mathrm{RZ}} / 3$, with the
    remaining two-thirds allocated to distillation and data storage
    (Section~\ref{app:beverland}).
  \item \textbf{Webster et al. (Pinnacle):}
    $\epsilon_{\mathrm{syn}} = \epsilon_{\mathrm{gate}} / 3$, using the
    full gate-infidelity budget because the Webster et al. failure model
    jointly accounts for data and $T$-state failure
    (Section~\ref{app:pinnacle}).
  \item \textbf{Zhou et al.:}
    $T$-count is obtained from the Beverland et al. model at budget
    $\epsilon_{\mathrm{RZ}}$; the Clifford budget is handled separately
    for distance selection (Section~\ref{app:zhou}).
\end{itemize}

Because Webster et al. frame-track their Cliffords at zero error, they synthesize rotations
against the full gate-infidelity budget, whereas Zhou et al. and Beverland et al.
must reserve part of the budget for Clifford error and therefore use a smaller
$\epsilon_{\mathrm{syn}}$. Through Eq.~\eqref{eq:t-per-rotation} the smaller
budget raises their $T$-per-rotation. The difference is negligible for the TFIM
(all three round to the same $\sim\!1.6\times10^6$ $T$ states) but visible for
the Clifford-heavy Fermi--Hubbard model, where Webster et al. synthesize
$\sim\!8.0\times10^5$ $T$ states against the larger $\sim\!9.6\times10^5$ of Zhou et al. and Beverland et al.

The synthesis-approximation error $\epsilon_{\mathrm{syn}}$ is itself a real
contribution to each $T$-based architecture's achieved error, not just a sizing
device, so it is composed into the total achieved error reported in
Fig.~\ref{fig:resources} alongside the QEC and distillation terms. We recover it
from the architecture's actual $T$-count by inverting
Eq.~\eqref{eq:t-per-rotation},
$\epsilon_{\mathrm{syn}} = M_R\,2^{-(n_{T,\mathrm{rot}} - 5.3)/0.53}$ with
$n_{T,\mathrm{rot}} = n_T / M_R$. The STAR ansätze inject each rotation directly
from an analog resource state and carry no Clifford$+T$ synthesis term (their
  injection error is already in the per-gate $R_Z$ cost of
Section~\ref{app:resources}); counting $\epsilon_{\mathrm{syn}}$ for the
$T$-based architectures therefore places all six on the similar total-error regime.

\subsection{STAR resource count}
\label{app:resources}

For the STAR-based architectures (surface code and BB code), the physical
qubit count and runtime are
\begin{align}
  Q_{\mathrm{phys}}
  &= 2\,(Q_{\mathrm{data}} + Q_{\mathrm{magic}}), \\
  t_{\mathrm{run}}
  &= \left(\frac{V_{\mathrm{RZ}}}{Q_{\mathrm{magic}}}
  + \frac{V_{\mathrm{Cliff}}}{Q_{\mathrm{data}}}\right)
  \times t_{\mathrm{cycle}},
\end{align}
where $Q_{\mathrm{data}} = n_{\mathrm{logical}} \times (n/k)$ is the
data-qubit footprint ($n/k$ being the physical-to-logical ratio of the code),
$Q_{\mathrm{magic}} = n_{\mathrm{factories}} \times n_{\mathrm{factory}}$
is the magic-state factory footprint ($n_{\mathrm{factory}}$ physical qubits
per factory), and
$V_{\mathrm{RZ}}$, $V_{\mathrm{Cliff}}$ are the rotation and Clifford
spacetime volumes.  We sweep $n_{\mathrm{factories}}$ from~1 to
$n_{\mathrm{logical}}$ to trace the space--time Pareto frontier.

For the BB code, the self-dual structure allows two independent algorithm
instances to run in parallel on the left and right halves of the same
code block.  The hardware therefore hosts $2 \times Q_{\mathrm{phys}}$
physical qubits, and the per-shot runtime is
$t_{\mathrm{run}} / 2$. Default cycle times are $t_{\mathrm{cycle}}^{\mathrm{(surf)}} = 1.0$\,msand $t_{\mathrm{cycle}}^{\mathrm{(BB)}} = 2.0$\,ms.

\subsection{Zhou et al. resource count}
\label{app:zhou}

The Zhou et al.\ architecture~\cite{zhou2025resourceanalysis} uses rotated surface-code
patches with transversal gates and $T$-state cultivation.  The $T$-count
is obtained from the Beverland et al. model (Section~\ref{app:rz-to-t}) at
budget $\epsilon_{\mathrm{RZ}}$.

Each patch occupies $2d^2 - 1$ physical qubits.  The code distance is
the smallest odd $d\geq 15$ (cultivation produces rotated surface codes with $d=15$) satisfying
\begin{equation}
  (n_{\mathrm{Cliff}} + 1.5\, n_T)\,
  p_{\mathrm{Cliff}}^{\mathrm{(surf)}}(p_{\mathrm{phys}}, d)
  \;\le\; \epsilon_{\mathrm{Cliff}},
\end{equation}
where $\epsilon_{\mathrm{Cliff}} = \epsilon_{\mathrm{gate}} -
\epsilon_{\mathrm{RZ}}$ is the Clifford error budget and
$p_{\mathrm{Cliff}}^{\mathrm{(surf)}}(p_{\mathrm{phys}}, d)$ is the
per-round surface-code logical error rate. The factor $1.5$ accounts for
$T$-gate teleportation via a transversal CNOT, which incurs a
probability-$1/2$ $S$ correction.

The runtime is
\begin{equation}
  t_{\mathrm{run}}
  \;=\; \left[
    \left\lceil \frac{n_{\mathrm{Cliff}}}{n_{\mathrm{logical}}} \right\rceil
    + \left\lceil
    \frac{n_T\,(r_{\mathrm{cult}} + 1.5)}{n_{\mathrm{factories}}}
    \right\rceil
  \right] t_{\mathrm{cycle}},
\end{equation}
where $r_{\mathrm{cult}} = \lceil V_{\mathrm{cult}} / (2d^2 - 1) \rceil$
with cultivation volume $V_{\mathrm{cult}} = 1.6 \times 10^4$
qubit-cycles and cultivation error rate
$\epsilon_{\mathrm{cult}} = 10^{-8}$ per $T$ state. The qubit-cycle is obtained from
fold-transversal surface-code cultivation~\cite{sahay2025foldtransversalsurface}
at $p_{\mathrm{phys}} = 10^{-3}$, by assuming 4 gate layers is equivalent to a cycle. Because the cultivated $T$ states already satisfy the error rate requirements, no furture distillation is needed. We sweep
$n_{\mathrm{factories}}$ from~1 to $n_{\mathrm{logical}}$.
Because the transversal Clifford only requires a single round of syndrome extraction due to algorithmic fault tolerance, rather than $d$ syndrome rounds, the per-logical-operation cost omits the
$d_t$-round temporal factor that the code-surgery architectures (Beverland et al., Webster et al., and Khan et al.) incur, giving this architecture a runtime advantage of order the code distance~\cite{zhou2025resourceanalysis}.

\subsection{Webster et al. (Pinnacle) resource count}
\label{app:pinnacle}

The Webster et al. (Pinnacle) architecture~\cite{webster2026pinnaclearchitecture} uses generalized bicycle (GB)
codes with parameters drawn from Table~\ref{tab:gb-codes}.

\begin{table}[h]
  \centering
  \caption{GB code parameters (from~\cite{webster2026pinnaclearchitecture}).
    $n_{\mathrm{cb}}$: physical qubits in the code block (data $+$ ancilla);
    $n_g$: physical qubits in the measurement gadget appended to form a
    processing block; $n_b$: physical qubits for bridging to an adjacent
  processing block.}
  \label{tab:gb-codes}
  \begin{tabular}{ccccc}
    \toprule
    $d$ & $k$ & $n_{\mathrm{cb}}$ & $n_g$ & $n_b$ \\
    \midrule
    4  & 8  & 60   & 13 & 7  \\
    6  & 10 & 124  & 19 & 11 \\
    10 & 12 & 252  & 31 & 19 \\
    16 & 14 & 508  & 57 & 31 \\
    24 & 16 & 1020 & 99 & 51 \\
    \bottomrule
  \end{tabular}
\end{table}

The Webster et al. (Pinnacle) architecture uses Pauli-based computation (PBC), where Cliffords
are frame-tracked at zero cost: a circuit with $n_T$ $T$~gates on
$n_{\mathrm{logical}}$ qubits is executed via
$n_T + n_{\mathrm{logical}}$ logical Pauli measurements.
The logical Pauli measurement error rate is fitted as
\begin{equation}
  p_L(p_{\mathrm{phys}}, k, d)
  \;=\; \frac{6.2}{k}\,
  \left(\frac{p_{\mathrm{phys}}}{0.0158}\right)^{d/2 + 0.47}.
\end{equation}

Runtime is dominated by the $T$ gates, which are produced by a magic engine
that combines fold-transversal cultivation with 15-to-1 distillation. At the
operating point used here ($p_{\mathrm{phys}} = 10^{-3}$), cultivation
($d_a = 7$, $r = 2$ rounds) lowers the injected-state error to
$\epsilon_{\mathrm{in}} \approx 10^{-4}$, and a single 15-to-1 distillation
round with output error
$\epsilon_{\mathrm{out}} = 35\,\epsilon_{\mathrm{in}}^3 \approx 10^{-9}$
produces the final $T$ states. The magic-engine footprint is
\begin{equation}
  n_{\mathrm{ME}}
  = n_{\mathrm{cb}}(d_e) + 10\,n_g(d_e)
  + 30\,(n_a + d_a - 1) + n_\alpha,
  \label{eq:pinnacle-nme}
\end{equation}
with engine distance $d_e$, ancilla patch $n_a = d_a^2$, and a
cultivation-ancilla overhead $n_\alpha$; for the present operating point
($d_e = 24$, $d_a = 7$, $n_\alpha = 750$) this evaluates to
$n_{\mathrm{ME}} = 4{,}410$ physical qubits.
The $T$-count follows from Eq.~\eqref{eq:t-per-rotation} with
$\epsilon_{\mathrm{syn}} = \epsilon_{\mathrm{gate}}/3$
(Section~\ref{app:rz-to-t}).

The physical qubit count and runtime with parallelism parameter $\rho$
are
\begin{align}
  \beta &= \left\lceil \frac{n_{\mathrm{logical}}}{\rho\, k} \right\rceil
  \rho, \\
  Q_{\mathrm{phys}}
  &= \beta \cdot n_{\mathrm{pb}}
  + \rho \cdot n_{\mathrm{ME}}, \\
  t_{\mathrm{run}}
  &= \left[\frac{n_T}{\rho\,(1 - p_{\mathrm{reject}})}
    + \left\lceil\frac{n_{\mathrm{logical}}}{\rho}\right\rceil
  \right]
  d_t \cdot t_{\mathrm{code}},
\end{align}
where $n_{\mathrm{pb}} = n_{\mathrm{cb}} + 4 n_g + 4 n_b$ is the
physical qubit count per processing block,
$n_{\mathrm{ME}}$ is the magic-engine qubit count of
Eq.~\eqref{eq:pinnacle-nme},
$p_{\mathrm{reject}}$ is the $T$-state rejection probability,
$d_t = d + 2$, and $t_{\mathrm{code}} = 1.5$\,ms by default.
We sweep $\rho$ from~1 (space-optimized) to
$\beta = \lceil n_{\mathrm{logical}} / k\rceil$ (time-optimized).

\begin{table*}[!ht]
  \centering
  \caption{\textbf{Architectures by design choice.}
      Comparison of the six architectures, organized by host code,
      Clifford backbone, rotation strategy, and the resulting dominant space and
      time costs. Entries are evaluated at $p_{\mathrm{phys}}=10^{-3}$.}
  \label{tab:arch-axes}
  \begin{ruledtabular}
    \footnotesize

    \renewcommand{\arraystretch}{1.2}
    \begin{tabular}{lcccccc}
      & \makecell{\textbf{Bicycle chain STAR}\\\textbf{(this work)}}
      & \makecell{\textbf{Surface STAR}\\\textbf{(Ismail et al.)}}
      & \makecell{\textbf{Surface transv.}\\\textbf{$+$ cultivation}\\\textbf{(Zhou et al.)}}
      & \makecell{\textbf{Pinnacle}\\\textbf{(Webster et al.)}}
      & \makecell{\textbf{Surface  surgery}\\\textbf{$+$ distillation} \\(\textbf{Beverland et al.})}
      & \makecell{\textbf{Extractor}\\\textbf{(Khan et al.)}} \\
      \midrule
      \makecell[l]{\emph{Host code}}
      & \makecell{Bicycle chain code}
      & \makecell{Rotated\\surface ($d=9$)}
      & \makecell{Rotated\\surface}
      & \makecell{Generalized\\bicycle}
      & \makecell{Rotated\\surface}
      & \makecell{Bivariate bicycle\\(Two-Gross)} \\
      \makecell[l]{Code parameters}
      & \makecell{$\qcode{14\ell, 2\ell, 6}$}
      & \makecell{$\qcode{d^2, 1, d}$}
      & \makecell{$\qcode{d^2, 1, d}$}
      & \makecell{Table~\ref{tab:gb-codes}}
      & \makecell{$\qcode{d^2, 1, d}$}
      & \makecell{$\qcode{288, 12, 18}$} \\
      \makecell[l]{Encoding rate}
      & \makecell{$1/7$}
      & \makecell{$1/81$}
      & \makecell{$1/d^2$}
      & \makecell{$3\%$--$26.7\%$}
      & \makecell{$1/d^2$}
      & \makecell{$1/24$} \\
      \makecell[l]{Cycle time\\(stab.\ wt.)}
      & \makecell{$2.0\,$ms\\(wt-8)}
      & \makecell{$1.0\,$ms\\(wt-4)}
      & \makecell{$1.0\,$ms\\(wt-4)}
      & \makecell{$1.5\,$ms\\(wt-6)}
      & \makecell{$1.0\,$ms\\(wt-4)}
      & \makecell{$1.5\,$ms\\(wt-6)} \\
      \midrule
      \multicolumn{7}{l}{\emph{Clifford backbone}} \\
      \makecell[l]{Implementation}
      & \makecell{Global Transversal}
      & \makecell{Transversal}
      & \makecell{Transversal}
      & \makecell{Frame tracking\\(PBC)}
      & \makecell{Code surgery}
      & \makecell{Code surgery} \\
      \makecell[l]{Footprint per-logical}
      & \makecell{14\\(data + ancillas)}
      & \makecell{$2d^2{-}1$\\$=161$}
      & $2d^2{-}1$
      & \makecell{$n_{\mathrm{pb}}/k$\\($\approx\!38$, $d{=}10$)}
      & \makecell{$\geq 2d^2{-}1$}
      & \makecell{$\sim\!76$ with\\adapters/LPU} \\
      \midrule
      \multicolumn{7}{l}{\emph{Rotation strategy}} \\
      \makecell[l]{Angle synthesis}
      & \makecell{Direct analog\\injection (no $T$)}
      & \makecell{Direct analog\\injection (no $T$)}
      & \makecell{Clifford$+T$\\synthesis}
      & \makecell{Clifford$+T$\\synthesis}
      & \makecell{Clifford$+T$\\synthesis}
      & \makecell{Clifford$+T$\\synthesis} \\
      \makecell[l]{Magic generation}
      & \makecell{Parallel TMR}
      & \makecell{TMR}
      & \makecell{$T$ cultivation\\($\epsilon_{\mathrm{cult}}{=}10^{-8}$)}
      & \makecell{Cultivation $+$\\15-to-1 distill.\\($\epsilon_{\mathrm{out}}{\approx}10^{-9}$,\\$n_{\mathrm{ME}}{=}4{,}410$)}
      & \makecell{15-to-1 distillation\\$(d_X,d_Z,d_m)$\\$=(15,5,5)$}
      & \makecell{$T$ cultivation\\($\epsilon_{\mathrm{cult}}{=}10^{-8}$)} \\
      \makecell[l]{Magic\\consumption}
      & \makecell{RUS teleport.\\(transversal)}
      & \makecell{RUS teleport.\\(transversal)}
      & \makecell{$T$-teleportation\\(transversal)}
      & \makecell{$T$-teleportation\\(surgery)}
      & \makecell{$T$-teleportation\\(surgery)}
      & \makecell{$T$-teleportation\\(surgery)} \\
      \midrule
      \multicolumn{7}{l}{\emph{Resulting cost}} \\
      \makecell[l]{Dominant\\space cost}
      & \makecell{Data encoding\\($k/n{=}1/7$;\\smallest)}
      & \makecell{Surface data\\patches}
      & \makecell{Surface patches\\(data $+$ cult.)}
      & \makecell{Magic engine\\($n_{\mathrm{ME}}{\approx}4{,}410$)}
      & \makecell{Data block $+$\\distill.\ factories\\(largest)}
      & \makecell{Data encoding\\($k/n{=}1/24$;\\smallest $T$-based)} \\
      \makecell[l]{Bottleneck\\(time)}
      & \makecell{Analog-$R_Z$\\budget (caps $T$)}
      & \makecell{Analog-$R_Z$\\budget}
      & \makecell{$T$-cultivation\\throughput}
      & \makecell{$T$ magic\\engine}
      & \makecell{15-to-1 distill.\\throughput}
      & \makecell{Serial rotation\\floor (factory-\\independent)} \\
    \end{tabular}
  \end{ruledtabular}
\end{table*}

\subsection{Beverland et al. resource count}
\label{app:beverland}

We use the Beverland et al.\ lattice-surgery
model~\cite{beverland2022assessingrequirements} with a 15-to-1 magic-state
distillation factory. The 15-to-1 block consumes 15 noisy input $T$ states and
outputs one distilled $T$ state. Its surface-code implementation is specified by
the factory-distance tuple $(d_X, d_Z, d_m) = (15, 5, 5)$: $d_X$ and $d_Z$ are
the asymmetric $X$- and $Z$-type distances of the factory patches, and $d_m$ is
the measurement distance, i.e.\ the number of syndrome-extraction rounds used for
each lattice-surgery measurement in the factory. These are factory parameters;
the data-code distance $d$ is selected separately below.
The $T$-count follows from Eq.~\eqref{eq:t-per-rotation}
(Section~\ref{app:rz-to-t}).

The error budget $\epsilon_{\mathrm{RZ}}$ is split into thirds:
\begin{align}
  \epsilon_{\mathrm{syn}}  &= \epsilon_{\mathrm{RZ}} / 3
  && \text{(rotation synthesis)}, \\
  \epsilon_{\mathrm{dist}} &= \epsilon_{\mathrm{RZ}} / 3
  && \text{(distillation)}, \\
  \epsilon_{\mathrm{data}} &= \epsilon_{\mathrm{RZ}} / 3
  && \text{(data storage)}.
\end{align}
The data code distance is auto-selected as the smallest odd $d$ satisfying
$Q \cdot C \cdot p_L^{\mathrm{(surf)}}(p_{\mathrm{phys}},d) \le
\epsilon_{\mathrm{data}}$, where $Q$ is the number of logical qubit tiles,
$C$ is the total number of syndrome extraction rounds, and
$p_L^{\mathrm{(surf)}}$ is the per-round surface-code logical error rate.

\subsection{Architecture comparison}
\label{app:arch-comparison}

  Table~\ref{tab:arch-axes} organizes the six architectures into four blocks:
  the host code, the Clifford backbone, the rotation
  strategy, and the resulting dominant space and time costs. The
  host-code block fixes the baseline encoding rate and syndrome-extraction cycle
  time. The Clifford-backbone block records whether logical Cliffords are applied
  transversally, tracked in a Pauli frame, or implemented by code/lattice surgery.
  The rotation-strategy block distinguishes direct STAR analog injection from
  Clifford$+T$ synthesis, together with the corresponding source and consumption
  of magic states. All entries use the common operating point
  $p_{\mathrm{phys}}=10^{-3}$ of Section~\ref{sec:results}; the formulas behind
  the table are defined earlier in this appendix.

These four blocks explain the resource trends in Fig.~\ref{fig:resources}.
  The two STAR architectures are limited in time by the cumulative
  analog-rotation error budget, not by magic-state factory throughput. The
  $T$-based architectures instead shift the
  bottleneck to how synthesized rotations are supplied: Zhou et al.\ and
  Beverland et al.\ trade runtime against the number of cultivation or
  distillation factories, Webster et al.\ concentrates the cost in a larger magic
  engine, and Khan et al.\ has a serial code-surgery rotation floor that
  additional factories cannot remove. On the space side, the high encoding rate
  of the bicycle-chain code keeps the data footprint small; architectures based
  on low-rate surface-code data blocks or large magic engines pay the
  corresponding qubit overhead.

The Khan et al.\ extractor~\cite{khan2026architecting} is a useful contrast:
  it uses a bicycle-code family, as this work does, but chooses code-surgery
  Cliffords and Clifford$+T$ rotations rather than transversal Cliffords and
  analog injection. This isolates the value of the two co-design choices used
  here.